\documentclass[pre,aps,showpacs,twocolumn,superscriptaddress]{revtex4-1}

\usepackage{graphicx,dcolumn}
\usepackage{amsfonts,amsmath,amssymb,bm,bbm, mathtools, siunitx}
\usepackage{float}
\usepackage{color}
\usepackage{physics}
\usepackage[toc,page]{appendix}
\usepackage[yyyymmdd,hhmmss]{datetime}
 
\def\be{\begin{eqnarray}}
\def\ee{\end{eqnarray}}
\def\ben{\begin{eqnarray*}}
\def\een{\end{eqnarray*}}
\def\ba{\begin{align}}
\def\ea{\end{align}}
\def\ban{\begin{align*}}
\def\ean{\end{align*}}
\def\bes{\begin{subequations}}
\def\ees{\end{subequations}}
\def\ds{\displaystyle}
\def\nn{\nonumber}

\newcommand{\wig}[1]{\mathrel{\hbox{\hbox to 0pt{\lower.6ex\hbox{$\sim$}\hss    }\raise.4ex\hbox{$#1$}}}}

\newcommand{\myFig}[7]{ %
\begin{figure}[h] 
\begin{center} 
\includegraphics[width=#1\columnwidth,height=#2\columnwidth,clip=true,keepaspectratio=#3,angle=#4]{#5}
\caption{#6} \vspace{-0.5cm} \label{#7} 
\end{center} \end{figure}}

\begin{document}

\title{Calculation of electron-ion temperature equilibration rates and friction coefficients\\
in plasmas and liquid metals using quantum molecular dynamics}

\author{Jacopo \surname{Simoni}}
\email{jsimoni@lanl.gov}
\author{J\'er\^ome \surname{Daligault}}
\email{daligaul@lanl.gov}
\affiliation{Theoretical Division, Los Alamos National Laboratory, Los Alamos, NM 87545, USA}

\begin{abstract}
We discuss a method to calculate with quantum molecular dynamics simulations the rate of energy exchanges between electrons and ions in two-temperature plasmas, liquid metals and hot solids.
Promising results from this method were recently reported for various materials and physical conditions [J. Simoni and J. Daligault, Phys. Rev. Lett. 122, 205001 (2019)].
Like other ab-initio calculations, the approach offers a very useful comparison with the experimental measurements and permits an extension into conditions not covered by the experiments.  
The energy relaxation rate is related to the friction coefficients felt by individual ions due to their non-adiabatic interactions with electrons. 
Each coefficient satisfies a Kubo relation given by the time integral of the autocorrelation function of the interaction force between an ion and the electrons.
These Kubo relations are evaluated using the output of quantum molecular dynamics calculations in which electrons are treated in the framework of finite-temperature density functional theory.
The calculation presents difficulties that are unlike those encountered with the Kubo formulas for the electrical and thermal conductivities.
In particular, the widely used Kubo-Greenwood approximation is inapplicable here.
Indeed, the friction coefficients and the energy relaxation rate diverge in this approximation since it does not properly account for the electronic screening of electron-ion interactions.
The inclusion of screening effects considerably complicates the calculations.
We discuss the physically-motivated approximations we applied to deal with these complications
in order to investigate a widest range of materials and physical conditions.
Unlike the standard method used for the electronic conductivities, the Kubo formulas are evaluated directly in the time domain and not in the energy domain, which spares one from needing to introduce an extraneous undetermined numerical parameter to account for the discrete character of the numerical density of states.
We highlight interesting properties of the energy relaxation rate not shared by other electronic properties, in particular its self-averaging character.
We then present a detailed parametric and convergence study with the numerical parameters, including the system size, the number of bands and $k$-points, and the physical approximations for the dielectric function and the exchange-correlation energy.
\end{abstract}

\pacs{} 

\date{\today\,\,at \currenttime}

\maketitle


\section{Introduction}\label{sec:1}

In a recent Letter \cite{Simoni_2019}, we presented first-principle calculations of the electron-ion temperature relaxation rate in materials under warm dense plasma and liquid metal conditions, including aluminum and several transition metals.
We used quantum molecular dynamics simulations to numerically evaluate a formal expression for the relaxation rate that is valid for physical systems ranging from hot solid metals to plasmas.
The justification and the properties of this theoretical expression were presented in detail in Ref.~\cite{Daligault_Simoni_2019}.
The goal of this companion paper is to present the approach we followed to numerically evaluate this theory with quantum molecular dynamics.

The underlying theory can be summarized as follows \cite{Daligault_Simoni_2019}.
We consider a material of volume $\Omega$ composed by a single atomic species.
We assume that the material can be described as an isolated,
homogeneous, two-temperature system comprised of ions (mass
$m_i=Am_u$, number density $n_i=N_i/\Omega$, charge $Ze$) and of electrons
(mass $m_e$, density $n_e=Zn_i$) that are characterized at all times $t$ by the temperatures $T_i(t)$ and $T_e(t)$, respectively.
Under mild assumptions suitable for physical conditions ranging from hot solid metals to plasmas, it can be shown that the temperatures evolve according to \cite{Daligault_Simoni_2019}
\be
c_i^0\frac{dT_i}{dt}=G_{ei}\,(T_e-T_i)\quad,\quad c_e\frac{dT_e}{dt}=-G_{ei}\,(T_e-T_i) \label{ttm_equations}
\ee
where $c_i^0=3n_ik_B/2$ is the kinetic contribution to the ionic heat capacity, $c_{e}$ is the specific heat capacity of electrons at constant volume, and
\be 
G_{ei}(T_e,T_i)&=& 3n_{\rm i}k_{\rm B}\Gamma(T_e,T_i) \nn
\ee
is the electron-ion coupling, which measures the rate of energy exchanges between electrons and ions.
The electron-ion coupling is related to the average friction $\Gamma$ felt by an ion as a result of its non-adiabatic interactions with the electrons.
More specifically, the average friction
\be
\Gamma(T_e,T_i)&=&\bigg\langle\frac{1}{3N_{\rm i}}\sum_{\rm a=1}^{N_{\rm i}}\sum_{x=1}^3\gamma_{\rm ax,ax}^{[{\bf R}]}(T_{\rm e})\bigg\rangle \label{Eq:Temp_rel_rate}
\ee
is given by the thermal average over ionic configurations ${\bf R}=({\bf R}_1,\dots,{\bf R}_{N_i})$ at temperature $T_i$ of the sum over all ions and spatial dimensions of the electron-ion friction coefficient $\gamma_{\rm ax,ax}^{[{\bf R}]}(T_e)$ felt by ion ${\rm a}$ along the ${\rm x}$-direction as a result of non-adiabatic interactions with the electrons.
The friction coefficients satisfy the Kubo relation
\be 
  \gamma_{\rm ax,by}^{[{\bf R}]}(T_{\rm e}) = \frac{1}{2m_ik_{\rm
      B}T_{\rm e}}{\rm Re}\int_0^\infty
  dt\,\big\langle\delta\hat{\mathcal{F}}_{\rm ax}(t)\delta\hat{\mathcal{F}}_{\rm
    by}(0)\big\rangle_{\rm e}, \label{Eq:friction_tensor}
\ee
where $\left\langle \dots\right\rangle_{\!e}$ is the electronic thermal average at temperature $T_e$, and $\hat{\mathcal{F}}_{\rm
  ax}(t)=e^{i\hat{H}_{\rm e}^{[{\bf R}]}t/\hbar}[-\partial\hat{H}_{\rm e}^{[{\bf
    R}]}/\partial R_{\rm ax}]e^{-i\hat{H}_{\rm e}^{[{\bf R}]}t/\hbar}$ is the electron-ion force at time $t$,  where $\hat{H}_{\rm e}^{[{\bf R}]}=\sum_i\hat{\bf p}_i^2/2m_{\rm  e}+\sum_{\rm i,a}v_{\rm ie}(\hat{\bf r}_i-{\bf R}_{\rm a}) +
\sum_{i\neq j}e^2/|\hat{\bf r}_i-\hat{\bf r}_j|$ is the electronic Hamiltonian and $v_{ie}$ the electron-ion interaction potential discussed in more details below.

In this paper, we explain how the friction coefficients (\ref{Eq:friction_tensor}) and, in turn, the electron-ion coupling (\ref{Eq:Temp_rel_rate}) can be calculated using quantum molecular dynamics simulations that treat the electrons within the framework of finite-temperature density functional theory and the ions classically within the Born-Oppenheimer approximation \cite{MartinBook}.
Such simulations are widely used to evaluate the Kubo formulas related to other electronic transport properties such as the electrical or the thermal conductivity\cite{Holst_2011}.
Although we here follow an analogous approach, the calculation of friction coefficients (\ref{Eq:friction_tensor}) presents additional and non-trivial difficulties that are addressed here.

The paper is organized as follows.
In section~\ref{sec:2}, we recast the Kubo relations (\ref{Eq:friction_tensor}) in terms of the quantities directly calculated in quantum molecular dynamics simulations.
We comment on the important role played by the shielding of electron-ion interactions due to all the electrons and, as a consequence, the non applicability of the widely-used Kubo-Greenwood approximation.
In section~\ref{sec:2-2}, we present the physically-motivated approximations we developed to account for the electronic screening effects in pseudopotential calculations of electron-ion forces, including local and plane-augmented-wave (PAW) pseudopotentials.
In section~\ref{sec:3}, we discuss the method we used to evaluate the expression (\ref{Eq:friction_tensor}), which, unlike the popular method used for the electronic conductivities, does not necessitate introducing an extraneous undetermined numerical parameter to account for the discrete nature of the numerical density of states.
We highlight the self-averaging character of the friction coefficient $\Gamma$, and discuss the statistical distribution of individual friction coefficients (\ref{Eq:friction_tensor}).
In section~\ref{sec:4}, we present a detailed parametric and convergence study of the proposed method with respect to the main numerical and physical parameters, including the system size, the number of {\bf k}-points and of energy bands, the dielectric functions, etc.
For clarity, many technical details are included in the appendices; in particular, detailed formulas useful for the practical implementation of the method are given in the appendices \ref{appendix_definition_PAW} and \ref{I}.

Throughout the paper, $\hbar$ is the reduced Planck constant, $k_B$ is the Boltzmann constant, and $e^2=q_e^2/4\pi\epsilon_0$, where $q_e$ is the elementary charge and $\epsilon_0$ the vacuum permittivity. ${\bf e}_x$ ($x=1,2,3$) denote unit vectors along the three cartesian directions.
When the illustrative calculations assume $T_e=T_i$, we denote the common temperature by $T$.

\section{Friction coefficients in the Kohn-Sham Density Functional Theory framework} \label{sec:2}

We first express the friction coefficients (\ref{Eq:friction_tensor}) in terms of the basic quantities that are directly computed in a quantum molecular dynamics calculation, namely the Kohn-Sham wave functions and energies.
This exact reformulation highlights the importance of the shielding of electron-ion forces produced by all the electrons.
In practice, the inclusion of the shielding effect is challenging, and in Sec.~\ref{sec:2-2} we present the method we developed for this purpose.

\subsection{Exact reformulation} \label{sec:2-1}

We assume to work with standard Quantum Molecular Dynamics (QMD) simulations in which electrons follow adiabatically the classical motion of ions and are treated quantum-mechanically within the framework of finite-temperature Kohn-Sham (KS) Density Functional Theory (DFT).
For each instantaneous ionic configuration ${\bf R}$ along a molecular dynamics trajectory, the electronic structure is obtained from the solution of the KS equations $\big{(}\frac{\hat{{\bf p}}^2}{2m_e}+V_{KS}[\rho_e,{\bf R}]\big{)}|n\rangle=\epsilon_n |n\rangle$, where $\epsilon_n$ and $|n\rangle$ are the single-particle KS energies and states, and $V_{KS}$ is the KS potential.
The Hamiltonian is a functional of the electron density $\rho_e({\bf
  r})=2\sum_n{p_n|\Psi_n({\bf r})|^2}$, where $\Psi_n({\bf
  r})=\langle{\bf r}|n\rangle$ and
$p_n=\left(1+e^{-(\mu-\epsilon_n)/k_BT_e}\right)^{-1}$ is the
Fermi-Dirac occupation number of state $n$, the factor $2$ account for
electron spin degeneracy.
Here and in the remaining of the paper, we often omit to indicate the dependence of the quantities from the instantaneous ionic configuration $[{\bf R}]$ in order to avoid cluttering the mathematical expressions.
We indicate the dependence on $[{\bf R}]$ when it is useful to be reminded.

As shown in the companion paper \cite{Daligault_Simoni_2019}, the friction coefficient $\gamma_{\alpha\beta}$ defined by Eq.(\ref{Eq:friction_tensor}) can be exactly written in terms of the KS spectrum as follows
\begin{equation}\label{Eq:MB_friction_tensor}
  \gamma_{\alpha\beta} =
  \tilde{\gamma}_{\alpha\beta}+
  \delta\tilde{\gamma}_{\alpha\beta}.
\end{equation}
where the indices $\alpha$ and $\beta$ are of form ${\rm ax}$, where ${\rm a}=1,\dots,N_i$ labels the ions and ${\rm x}$ denotes one of the three spatial directions.
The first term in Eq.(\ref{Eq:MB_friction_tensor}) reads
\begin{equation}\label{Eq:friction_tensor_KS_DFT}
  \tilde{\gamma}_{\alpha\beta} =-\frac{\pi\hbar}{M}\sum_{n\neq
    m}\frac{p_n-p_m}{\epsilon_n-\epsilon_m}{f}_{nm}^{\alpha,\mathrm{L}}{f}_{mn}^{\beta,R}\delta(\epsilon_n-\epsilon_m).
\end{equation}
where $f_{nm}^{\alpha={\rm ax},\mathrm{L}}$ and $f_{nm}^{\alpha={\rm
    ax},\mathrm{R}}$ denote matrix elements between KS states of the
{\it screened} force along the ${\rm x}$ direction between ion ${\rm
  a}$ and an electron  at ${\bf r}$ \cite{NoteOnElectronAndKS}.
They are given by the expressions
\begin{align}
{f}_{nm}^{{\rm ax},\mathrm{L(R)}}&=\langle n|\hat{f}_{{\rm ax},\rm L(R)}|m\rangle,\nonumber\\
&={\bf e}_x\cdot\int_\Omega{d{\bf r}\,\Psi_n({\bf r})^*\,{{\bf f}}_{{\rm a},{\rm L(R)}}({\bf r})\,\Psi_m({\bf r})}\,,\label{f_nm_Ix_L}
\end{align}
where
\begin{align}
  & {\bf f}_{{\rm a},{\rm L}}^{[{\bf R}]}({\bf r}_1) = \int_\Omega d{\bf r}\,{\bf F}_{\rm a}({\bf
    r}){\varepsilon}_{\rm L}^{[{\bf R}]}({\bf r},{\bf
    r}_1,\omega=0)^{-1} && \label{Eq:forceL}
\end{align}
and
\begin{align}
  &{\bf f}_{{\rm a},{\rm R}}^{[{\bf R}]}({\bf r}_1) = \int_\Omega d{\bf r}\,{\varepsilon}_{\rm R}^{[{\bf R}]}({\bf r}_1,{\bf r},\omega=0)^{-1}
  {\bf F}_{\rm a}({\bf r}) \label{Eq:forceR} &&
\end{align}
are the effective electron-ion forces that result from the electronic shielding of the bare electron-ion force
\be
{\bf F}_{\rm a}({\bf r})=\nabla_{{\bf r}}v_{\rm ie}({\bf r}-{\bf R}_{\rm a})\,. \label{Eq:forcebare}
\ee
The dielectric functions
\begin{align}
  & {\varepsilon}_{\rm L}^{[{\bf R}]}({\bf r},{\bf r'},\omega)\nn\\
  &= \delta({\bf r}-{\bf r'}) -\int_\Omega d{\bf r}_1\chi^{[{\bf
        R}]}({\bf r},{\bf r}_1,\omega) K^{[\mathbf{R}]}(\mathbf{r}_1,\mathbf{r'},\omega) \label{Eq:12a}\\
  &{\varepsilon}_{\rm R}^{[{\bf R}]}({\bf r},{\bf r'},\omega) \nn\\
  &=\delta({\bf r}-{\bf r'}) -\int_\Omega d{\bf r}_1 K^{[\mathbf{R}]}(\mathbf{r},\mathbf{r}_1,\omega)\chi^{[{\bf R}]}({\bf r}_1,{\bf r'},\omega)\,, \label{Eq:12b}
\end{align}
accounts for the screening effect of electrons, where $\chi^{[{\bf R}]}$ is the density-density response function of the KS system in the presence of the ionic background ${\bf R}$, and $K^{[\mathbf{R}]}(\mathbf{r},\mathbf{r'},\omega) = e^2/|\mathbf{r}-\mathbf{r'}| + f_\mathrm{xc}^{[\mathbf{R}]}[\rho_\mathrm{e}](\mathbf{r},\mathbf{r'},\omega)$ is written in terms of the KS exchange-correlation kernel $f_\mathrm{xc}^{[\mathbf{R}]}$.
Note that at this level of generality, the distinction of the left (L)
and right (R) dielectric functions is needed to account of the spatial
inhomogeneity of the electronic system in the ionic configuration ${\bf R}$.
As discussed in \cite{Daligault_Simoni_2019}, the first term in Eq.(\ref{Eq:MB_friction_tensor}) prevails and is related to the time correlation function of the interaction force between an ion and a KS particle screened by the rest of KS particles,
\be \label{tildegammaalphabetaKubo}
\tilde{\gamma}_{\alpha\beta}&=&\frac{\beta_e}{2M}\mathrm{Re}\int_0^\infty dt\left\langle\,\delta\!{\hat{f}}_{\alpha,L}(t)\delta\!{\hat{f}}_{\beta,R}(0)\,\right\rangle_\mathrm{e}\nn
\ee
where ${\hat{f}}_{\alpha,L}(t)=e^{i\hat{h}_\mathrm{KS} t/\hbar} \hat{f}_{\alpha,\mathrm{L}} e^{-i\hat{h}_\mathrm{KS} t/\hbar}$ is the time-dependent screened electron-ion force and $\beta_e=1/k_BT_e$.
Alternatively, for later reference, Eq.(\ref{tildegammaalphabetaKubo})
can also be written such as
\be
\tilde{\gamma}_{\alpha\beta}&=&\frac{\beta_e}{2M}\int_{-\infty}^\infty dt\,{K_{\alpha\beta}(t)} \label{tildegammaab}\,,
\ee
in terms of the Kubo correlation function
\be
K_{\alpha\beta}(t)=\frac{1}{\beta_e}\int_0^{\beta_e}{d\lambda \left\langle\,e^{\lambda\hat{h}_{KS}}\delta\!{\hat{f}}_{\beta,R}(0)e^{-\lambda\hat{h}_{KS}}\delta\!{\hat{f}}_{\alpha,L}(t)\,\right\rangle_\mathrm{e}}\,.\nn \label{tildegammaab2}
\ee

The second term in Eq.~(\ref{Eq:MB_friction_tensor}) reads
\begin{align}\label{Eq:MBcorrection}
  &\delta\tilde{\gamma}_{\alpha\beta}^{[{\bf R}]} = -\frac{1}{M}\nonumber \\
  &\times{\rm Im}\int_\Omega\,d{\bf r}_1\int_\Omega\,d{\bf r}_2 n_\alpha'({\bf
    r}_1)\partial_\omega f_{\rm xc}^{[{\bf R}]}({\bf r}_1,{\bf
    r}_2,\omega=0)n_\beta'({\bf r}_2)
\end{align}
where $n_\alpha'({\bf r})=\int_\Omega\,d{\bf r}_1 f_\alpha({\bf r}_1)\chi_{\rm ee}^{[{\bf R}]}({\bf r}_1,{\bf r},\omega=0)$, $f_{\alpha={\rm ax}}({\bf r})=F_{\rm ax}({\bf r})$ and
\be
\lefteqn{\chi_\mathrm{ee}^{[\mathbf{R}]}(\mathbf{r}_1,\mathbf{r}_2,\omega) = \chi^{[\mathbf{R}]}(\mathbf{r}_1,\mathbf{r}_2,\omega)}&&\label{DysonchichiKS}\\
&&+\int_\Omega d{{\bf r}}\int_\Omega d{{\bf r}'}\chi^{[\mathbf{R}]}(\mathbf{r}_1,\mathbf{r},\omega)K^{[\mathbf{R}]}(\mathbf{r},\mathbf{r'},\omega)\chi_\mathrm{ee}^{[\mathbf{R}]}(\mathbf{r'},\mathbf{r}_2,\omega),\nn
\ee
is the density response function of the electronic system.
$\delta\tilde{\gamma}_{\alpha\beta}$ represents a correction due to intricate, dynamical many-body correlations not included in the first term (see the last remark below).

\subsection{Remarks} \label{sec:2-2}

\paragraph*{1.} We stress that Eq.(\ref{Eq:MB_friction_tensor}) is an exact representation of the friction coefficient Eq.(\ref{Eq:friction_tensor}) and it does not correspond to a Kubo-Greenwood approximation \cite{Greenwood_1958}.
In the Kubo-Greenwood approximation, the many-body electronic states that one should in principle use to evaluate the Kubo formulas (\ref{Eq:friction_tensor}) are approximated by Slater determinant of KS orbitals.
This leads to an expression for ${\gamma}_{\alpha\beta}$ that is analogous to Eq.(\ref{Eq:MB_friction_tensor}) but where the screened forces (\ref{Eq:forceL}) and (\ref{Eq:forceR}) are replaced by the bare forces Eq.(\ref{Eq:forcebare}).
In other words, the dielectric functions $\epsilon_L$ and $\epsilon_R$ are set to unity in this approximation.
By discarding the electronic screening, the Kubo-Greenwood approximation for $\gamma_{\alpha,\beta}$ can be shown to diverge logarithmically due to the infinite range of the electron-ion Coulomb interaction at large distances, an effect that is analogous to the well-known infrared divergence that occurs in lowest order calculations of scattering cross sections in Coulomb systems.

\paragraph*{2.} Nevertheless, the Kubo relation (\ref{Eq:friction_tensor_KS_DFT}) resembles the Kubo-Greenwood expression for the electrical conductivity\cite{Holst_2011}
\begin{equation}
\sigma_{xy} = \frac{2\pi\hbar q_e^2}{3\Omega m_e^2}\sum_{n,m}\frac{p_n - p_m}{\epsilon_m - \epsilon_n} p_{nm}^x p_{mn}^y\delta(\epsilon_n - \epsilon_m)
\end{equation}
where $p_{nm}^x=\mel{n}{\hat{p}_x}{m}$ and $\hat{p}_x$ is the $x$ component of the linear momentum operator. The equivalent of expression (\ref{tildegammaab}) for the electrical conductivity is instead
\begin{equation}
\sigma_{xy} = \frac{\beta_e}{\Omega}\int_0^\infty{dt K_{xy}^\sigma(t)}\,,
\end{equation}
with Kubo correlation function
\begin{equation}
K_{xy}^\sigma(t) = \frac{1}{\beta_e}\int_0^{\beta_e}{d\lambda\left\langle\,e^{\lambda\hat{h}_{KS}}\hat{J}_y e^{-\lambda\hat{h}_{KS}}\hat{J}_x(t)\,\right\rangle_\mathrm{e}}\,,
\end{equation}
where $\hat{\bf J}$ is the current operator. On the basis of these similarities, many features of the standard method developed to calculate $\sigma_{xy}$ can be used.
Yet, there are additional nontrivial complications arising from the necessity to account for the screening.
According to Eq.(\ref{f_nm_Ix_L}), for each ionic configurations ${\bf R}$, one should in principle first calculate the right and left inverse static dielectric functions, calculate the screened forces and only then calculate the matrix element.
The determination of the dielectric functions is well-known to be rather challenging in itself.
To address these difficulties and reduce the cost of the algorithm without compromising the physics, we developed the approximations that are presented in Sec.~\ref{sec:2-2}.

\paragraph*{3.} The practical difficulties one faces to include the screening effects strongly depend on the pseudopotential chosen to model the tightly bound, core electrons.
Below, we consider two different categories of pseudopotentials, including: the local, Embedded Core Electrons (ECE) potentials, where the combined effect of a nucleus and its core electrons is described by a local potential; the Plane-Augmented Wave (PAW) pseudopotentials, which allow for a much more detailed account of the effect of core electrons and give access to a wide range of materials far beyond the reach of the simpler ECE potentials.

\paragraph*{4.}
In practice, the evaluation of Eq.(\ref{Eq:MB_friction_tensor}) is
limited also by the approximate nature of DFT calculations.
The mapping between the real system and the KS system is known only approximately and in practice the exchange-correlation potential energy $v_{xc}^{[{\bf R}]}$ and kernel $f_{xc}^{[{\bf R}]}$ must be also approximated.
The dependence on ${\bf R}$ is poorly known and, as usual, we use expressions based on the homogeneous electron gas.
In Sec.~\ref{sec:4-6}, we discuss the effect of two common choices for $v_{xc}$, namely the local density (LDA) and the generalized gradient (GGA) approximations.

The full exchange-correlation kernel $f_{xc}({\bf r}_1,{\bf r}_2,\omega)$ remains elusive although several important properties and approximations have been recently reported.
Here, in Eqs.~(\ref{Eq:friction_tensor_KS_DFT}) and (\ref{Eq:MBcorrection}), is essentially the low frequency behavior of $f_{xc}({\bf r}_1,{\bf r}_2,\omega)$ that is needed.
Qian and Vignale \cite{Qian_2002} reported an explicit expression for $f_{xc}\approx f_{xc}^{LDA,h}$ for the homogeneous (h) electron gas and in local density approximation, which is exact at low frequencies to leading order in the Coulomb interaction.
The approximation was used by Nazarov et al. \cite{Nazarov_2005} to compute the friction felt by a single charged impurity ion $X^{+Z_{imp}}$ in a degenerate electron gas with $n_e\sim 10^{23}-10^{22}$ $\rm cm^{-3}$.
Because the frequency derivative of $f_{xc}^{LDA,h}$ is negative definite, they find that the dynamical correlation carried by $f_{xc}$ tend to systematically enhance the friction coefficients, but the enhancement due to $\delta\tilde{\gamma}_{\alpha\beta}$ remains rather small at low impurity charges $Z_{imp}$.
For instance , at aluminum density, the enhancement due to dynamical for an aluminum-like impurity $X^{+3}$ is $< 10$ $\%$ (see Fig.1(b) in Ref.~\cite{Nazarov_2005}).
In this work, we accordingly neglect the dynamical effects described by Eq.(\ref{Eq:MBcorrection}).
However, in principle, given a reliable approximation for $f_{xc}$, the correction (\ref{Eq:MBcorrection}) could be numerically evaluated and added to $\tilde{\gamma}_{\alpha\beta}$.

\subsection{On the calculation of the screened electron-ion forces} \label{sec:2-2}

We first consider the simplest situation where, in Eq.(\ref{Eq:forcebare}), a ECE local pseudopotential $v_{ie}(r)$ is used to describe the interaction between valence electrons and an ion with its unresponsive bound electrons. 
This approach is limited to simple systems and physical conditions such as Aluminum at melting or dense Hydrogen for which reliable pseudopotentials exist.
For other elements, more sophisticated descriptions are needed and we here consider, as previously mentioned, the case of PAW pseudopotentials due to their accepted suitability to warm dense matter modeling.

\myFig{1}{1}{true}{0}{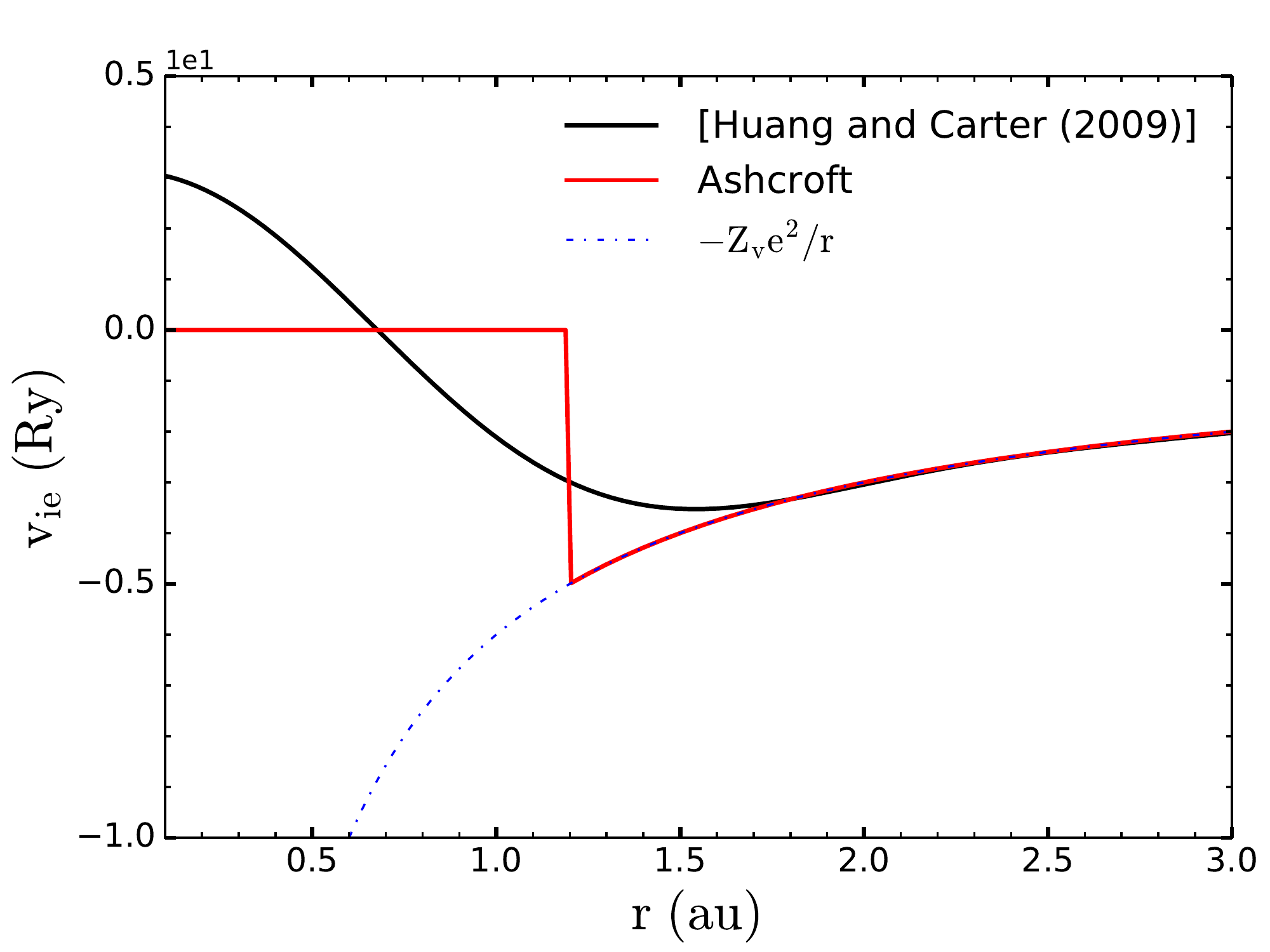}{(Color online) Local ECE pseudopotential (black line) for $\rm Al^{3+}$ used in this work (see \cite{Huang_2008}) with core radius $r_c=1.8$ $\rm a.u.\,$.
For comparison, the figure shows the celebrated Ashcroft potential \cite{Ashcroft_1966} $\ds v_{\rm ie}(r)=-\left(Z_{\rm v}e^2/r\right)\theta(r-r_c)$ with $r_c=1.2$ $\rm a.u.$ and the Coulomb potential $\ds -Z_{\rm v}e^2/r$ with $Z_{\rm v}=3$.}{Fig:15}

\subsubsection{Calculation with ECE pseudo-potentials}

In a hypothetical calculation in which all the electrons of the system are included, the KS wave functions would show very sharp features close to the nuclei in the so-called core regions since all the states are non-zero in this region and they are constrained by the requirement of orthogonality.
In contrast, outside the core region only the valence states are non-zero, resulting in much smoother wavefunctions in this interstitial region. 
The oscillatory behavior in the core regions, would require a very large set of plane waves to be described accurately.
As already mentioned, the simplest way of solving this problem consists in using a local, spherically symmetric pseudopotential in which the heterogeneous system composed by a nucleus and its tightly bound, unresponsive core electrons is described by an effective, much smoother, potential $v_{ie}(r)$ behaving as $-Z_ve^2/r$ a large distances, where $Z_v$ is the number of valence electrons per atom.
Examples of two standard ECE pseudopotentials used to describe the effective interaction betweeen the valence electrons and an $Al^{\rm 3+}$ atom with $Z_v=3$ are shown in Fig.~\ref{Fig:15}.

The theory of section \ref{sec:2-1} should be applied as follows when such a ECE local pseudopotential is used.
The KS equations are solved for the valence electrons only,
The electron density $\rho_e({\bf r})$ is the one corresponding to the $N_i\times Z_v$ valence electrons and the KS potential, for each ionic configuration ${\bf R}$, is written as
\be
V_{KS}[\rho_e]({\bf r})=\sum_{a=1}^{N_i}{v_{\rm ie}({\bf r}-{\bf R}_a)}+v_{\rm Hxc}[\rho_e]({\bf r})\,.
\ee
In addition, the dielectric functions $\tilde{\varepsilon}_{\rm L}$ and $\tilde{\varepsilon}_{\rm R}$ describe the screening power of these valence electrons only.
The shielding effect of core electrons is embraced in the pseudopotential.
Because the core electrons strongly screen the bare atomic charge and because of the Pauli principle, the valence electron-ion interaction $v_{ie}$ is often relatively weak and a reasonable approximation for $\tilde{\varepsilon}_{\rm L}$ and $\tilde{\varepsilon}_{\rm R}$ is the lowest order one given by the dielectric function of the homogenous electron gas (jellium) \cite{T7},
\bes
\be
{\varepsilon}_{\rm L}^{[{\bf R}]}({\bf r},{\bf r'},\omega)&\simeq&{\varepsilon}_{\rm eg}({\bf r}-{\bf r'},\omega),\\
{\varepsilon}_{\rm R}^{[{\bf R}]}({\bf r},{\bf r'},\omega)&\simeq&{\varepsilon}_{\rm eg}({\bf r}-{\bf r'},\omega)\,.
\ee
\label{approximationeLeReeg}
\ees
The static dielectric function $\varepsilon_{\rm eg}({\bf r},\omega=0)$ or, equivalently, its spatial Fourier transform $\varepsilon_{\rm eg}({\bf k})=\int_\Omega{d{\bf r}\,\varepsilon_{\rm eg}({\bf r},\omega=0)e^{-i{\bf k}\cdot{\bf r}}}$, is given by
\be
\frac{1}{\varepsilon_{\rm eg}({\bf k})}=1+\frac{v({\bf k})\chi_0({\bf k})}{1-v({\bf k})[1+G({\bf k})]\chi_0({\bf k})}
\ee
where $\chi_0({\bf k})$ is the density response function of the free electron gas (at $T_e$) and $G({\bf k})$ is the local field correction that accounts for exchange and correlation effects in the interacting electron gas beyond the mean field approximation, $v({\bf k})=4\pi e^2/|{\bf k}|^2$ is the Fourier transform of the Coulombic interaction.
From the approximation (\ref{approximationeLeReeg}), ${\bf f}_{a,\rm L}^{[{\bf R}]}({\bf r})={\bf f}_{a,{\rm R}}^{[{\bf R}]}({\bf r})\equiv {\bf f}_a^{eg}({\bf r})$ and 
\be
{\bf f}_a^{\rm eg}({\bf r})=\int_\Omega d{\bf r}'\,{\bf F}_a({\bf r}')\varepsilon_{\rm eg}({\bf r}-{\bf r}',\omega=0)^{-1}\,. \label{fIeg}
\ee
In practice, Eq.(\ref{fIeg}) can be efficiently computed by means of three-dimensional Fourier transform.

\myFig{1}{1}{true}{0}{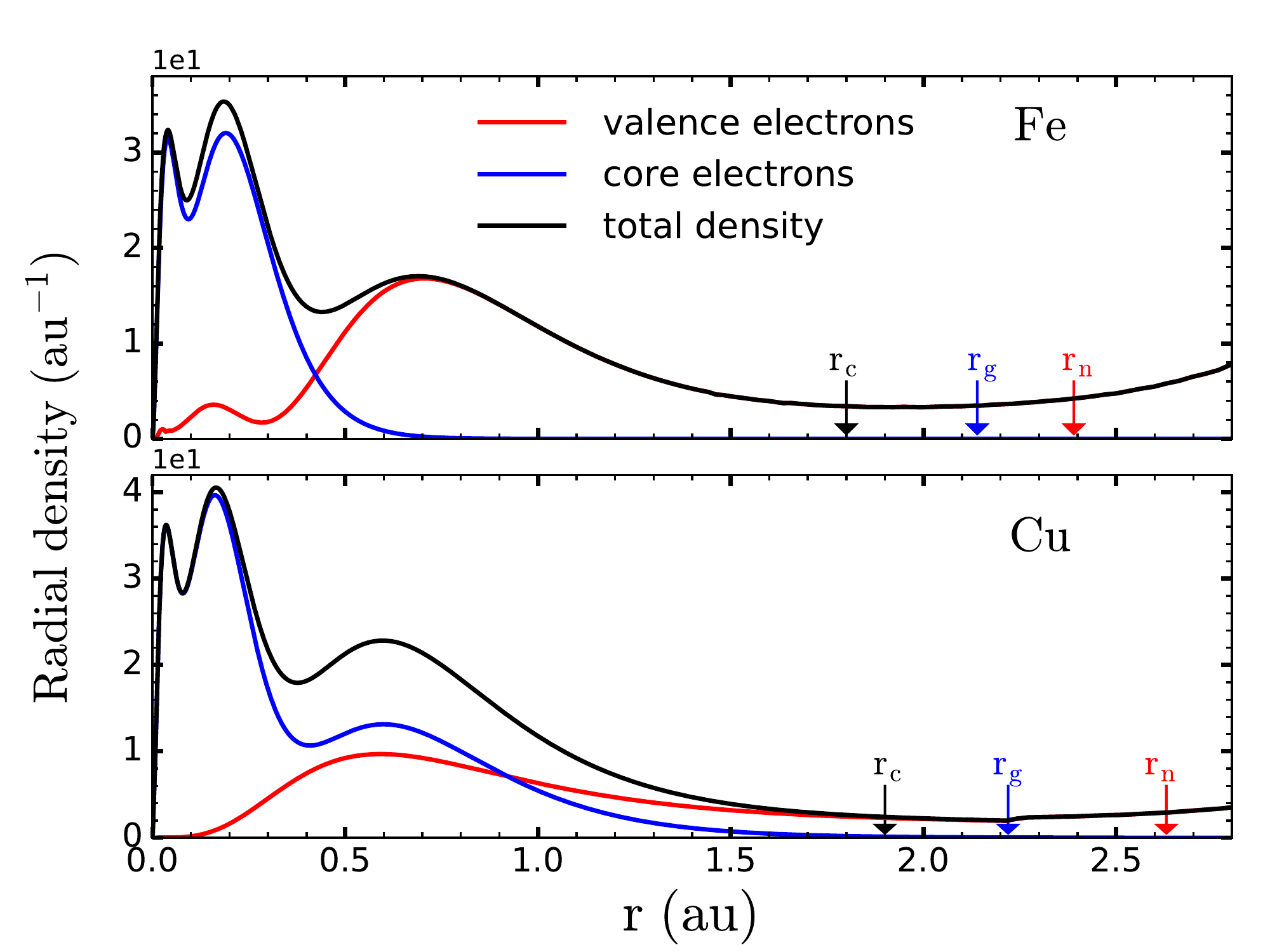}{(Color online) Radial electron density $4\pi r^2n_{\rm ie}(r)$ around a nucleus in liquid iron (upper panel) at melting temperature $T=0.156$ $\rm eV$ and solid density $\rho=7.87\,\rm{g/cm^3}$, and in liquid copper at $T=0.2$ $\rm eV$ and melting density $\rho=8.02$ $\rm g/cm^3$. In both panels, the blue line shows the radial density of core electrons, the red line shows the radial density of valence electrons, and the black line shows the total radial electron density.
The density of valence electrons  were calculated by averaging over the ions the electron density directly calculated in a quantum molecular dynamics (see the introduction of Sec.~\ref{sec:4} for details).
A PAW pseudopotential was used in each case, generated with $Z_v=16$ valence electrons and $Z_c=10$ frozen (neon-like) core electrons for iron, and $Z_v=11$ valence electrons and $Z_c=18$ frozen (argon-like) core electrons for copper.
The arrows indicate important distances discussed in the text: the PAW frozen core radius $r_c$, the neutral sphere radius $r_n$ defined such that $Z=Z_v+Z_c=\int_0^{r_n}{dr\,4\pi r^2n_{\rm ie}(r)}$, and the distance $r_g$ equal to half the average distance separating two ions.}{Fig:10}

\subsubsection{Calculation with a Plane-Augmented Wave pseudo-potential}\label{subsub:PAW}

The calculation of the matrix elements $f_{nm}^{\alpha,L}$ and $f_{nm}^{\alpha,R}$ in Eq.(\ref{f_nm_Ix_L}) from a PAW pseudopotential-based QMD simulation is more tricky.
The main reason being that all the electrons are accounted for in this approach and the screening does not affect the electrons inside the ionic cores as it does outside the cores.
In this section, we present the physically-motivated approximations that we have implemented, which result in the formula (\ref{fmnPAW}) below.
The precise definition of PAW pseudopotentials is involved and we refer the reader to the specialized literature for a detailed presentation \cite{Blochl_2003}; for convenience, important relations for the numerical implementation of Eq.(\ref{fmnPAW}) are included in appendix~\ref{appendix_definition_PAW}.
Here, we only recall the basic properties that are useful to our discussion.
\myFig{1}{1}{true}{0}{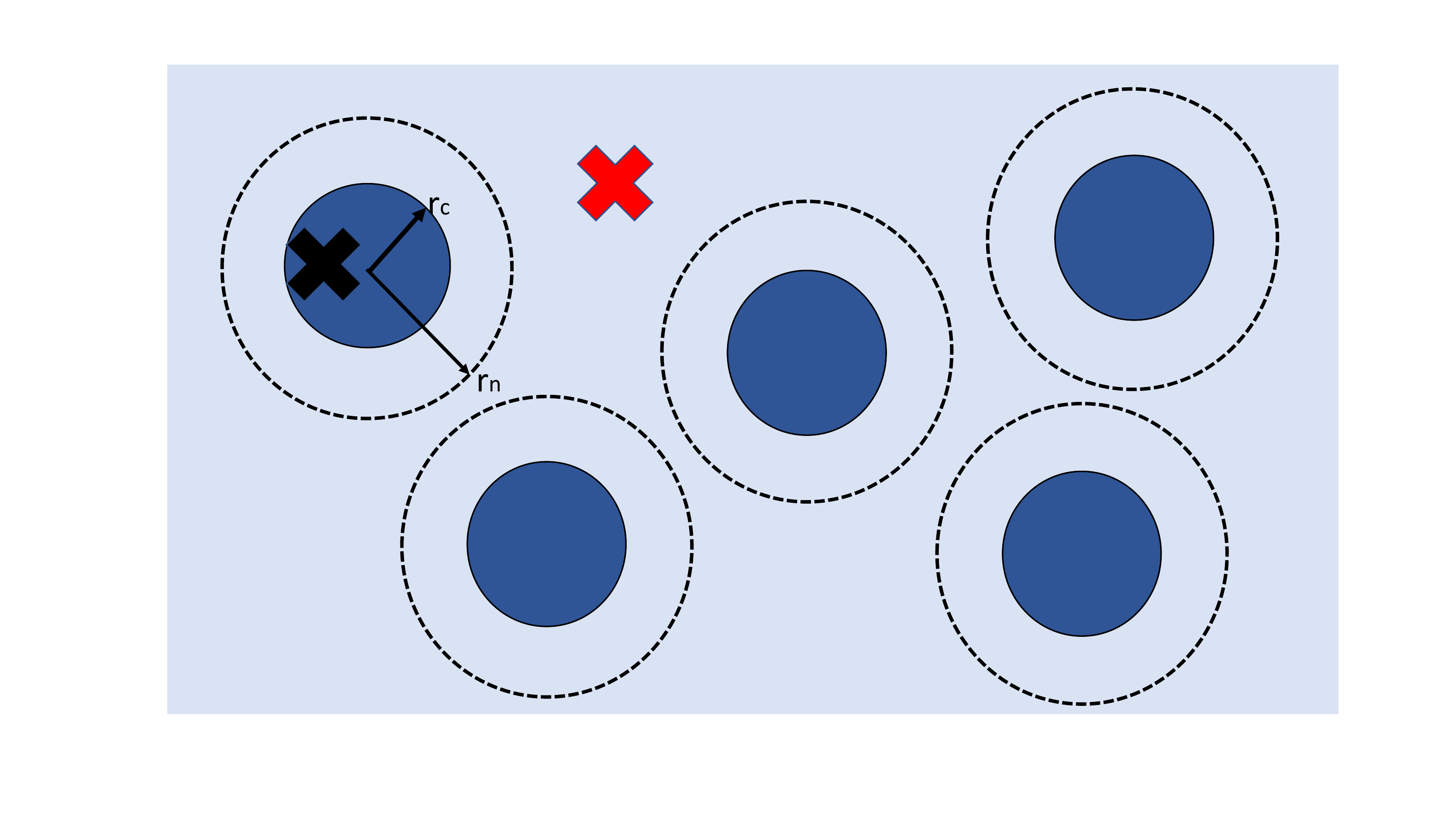}{Cartoon representation of the spatial regions that are identified in the main text in order to derive the approximated expression (\ref{fmnPAW}) of the force matrix elements with a PAW pseudopotential.
The dark spheres represent the frozen core of radius $r_c$, in which core electron states are confined.
The light blue represents the valence electrons in the interstitial region outside the frozen atomic core regions.
The dashed spheres indicate the neutral sphere of radius $r_n$ surrounding each nucleus.
Realistic electron densities found in these regions are shown in Fig.~\ref{Fig:10} for iron and copper.
For an electron at the red cross located at a distance greater than $r_n$ from any nucleus, the bare Coulomb potential $-Ze^2/r$ that it feels due to a given nucleus at a distance $r$ is first reduced to $-(Z-\bar{Z})e^2/r$ by the $\bar{Z}$ electrons located inside the core radius, and then it is further reduced by the screening effect of the delocalized valence electrons (light blue).
In this work, the screening due to delocalized electrons is modeled with the dielectric function of the homogeneous electron gas.
For an electron located at the black cross inside a core region, the previous picture fails and, in this case, we appeal to an exact sum rule to describe the shielding of electrons.}{Fig:cartoon}
Firstly, all the electrons of the system are explicitely described by Kohn-Sham wavefunctions in the PAW method.
Yet, one still conveniently distinguishes between tightly bound core states and valence states.
The distinction is an informed choice of the user based on the physical conditions under consideration.
In the standard frozen core approximation used in this work, it is assumed that the $Z_c=Z-Z_v$ core states of the isolated atoms are not affected by the surrounding particles and are identical to the isolated atomic core states.
These states are naturally localized within a sphere of radius $r_c$ around their parent nucleus.
The KS atomic core wave functions $\phi_{n}^c({\bf r})$ are calculated beforehand separately from the actual QMD simulations.
The valence electrons and their wave functions $\Psi_n({\bf r})$ are the only orbitals that are actually calculated and updated along the QMD simulation with the Kohn-Sham potential
\be
V_{\rm KS}[\rho_e]({\bf r})=V_{ei}({\bf r})+v_{\rm Hxc}[\rho_e]({\bf r})\,, \label{V_KS_PAW}
\ee
where $V_{ei}({\bf r})=-Ze^2\sum_{\rm a=1}^{N_i}{1/|{\bf r}-{\bf R}_{\rm a}|}$ is the bare electron-nulceus potential and $\rho_e({\bf r})=\rho_v({\bf r})+\sum_{\rm a=1}^{N_i}\rho_{c}^{\rm a}({\bf r})$ is the all-electron density.
The latter consists of the contribution $\rho_v$ of all valence electrons and of the localized core electron densities $\rho_{c}^{\rm a}$ around each atom ${\rm a}$.
These contributions are illustrated in Fig.~(\ref{Fig:10}) for iron and copper systems (details are in the caption) and in the cartoon shown in Fig.~(\ref{Fig:cartoon}).

Secondly, unlike with local ECE pseudopotentials, the effect of a nucleus and its core electrons on the valence electrons is not modeled by a potential $v_{\rm ie}$ but is instead directly parametrized in the wave functions.
This is accomplished mathematically with a transformation that maps the true wave functions $\ket{\Psi_n}$ with their complete nodal structure onto auxiliary smooth wave functions that have a rapidly convergent plane wave expansion,
\begin{align}
  \ket{\Psi_n} &= \hat{\boldsymbol{\tau}}\ket*{\tilde{\Psi}_n}\nonumber\\
  &= \ket*{\tilde{\Psi}_n} +\sum_{a=1}^{N_i}\sum_{i}(\ket{\phi_{ai}}-\ket*{\tilde{\phi}_{ai}})\ip*{\tilde{p}_{ai}}{\tilde{\Psi}_n}\,,
\end{align}
where the different terms, which are not essential to the present discussion, are defined in the appendix~\ref{appendix_definition_PAW}.
This transformation leads to a new set of transformed KS equations for the smooth wave functions $\hat{\boldsymbol{\tau}}^\dagger\hat{h}_{KS}\hat{\boldsymbol{\tau}}\ket*{\tilde{\Psi}_n}=\epsilon_n\hat{\boldsymbol{\tau}}^\dagger\hat{\boldsymbol{\tau}}\ket*{\tilde{\Psi}_n}$, which are actually solved by the QMD program instead of the usual set of KS equations.

Thirdly, the true valence wave functions ${\Psi}_n({\bf r})$ are identical to $\tilde{\Psi}_n({\bf r})$ ouside the core regions, i.e. $\Psi_n({\bf r})=\tilde{\Psi}_n({\bf r})$ when $|{\bf r}-{\bf R}_{\rm a}| \ge r_c^{\rm a}$ for all ${\rm a}$.
This property is conveniently written as follows 
\be
\Psi_n({\bf r})=\tilde{\Psi}_n({\bf r})\Pi^{\rm out}({\bf r})+\sum_{\rm a=1}^{N_i}\Psi_n({\bf r})\Pi_{\rm a}^{\rm in}({\bf r}) \label{Psi_n_split}
\ee
in terms of the indicator functions
\ben
\Pi^{\rm out}({\bf r})&=&\left\{
\begin{array}{l}
  1,\quad \text{if }\forall {\rm a}\,:\,|{\bf r}-{\bf R}_{\rm a}| \ge r_c^{\rm a}\\
  0,\quad \text{otherwise}
\end{array}
\right.\,,
\een
which indicates when ${\bf r}$ lies outside any ion cores, and
\ben \label{Pi_in^a}
\Pi_{\rm a}^{\rm in}({\bf r})&=&\left\{
\begin{array}{l}
  1,\quad \text{if }\,|{\bf r}-{\bf R}_{\rm a}| < r_c^{\rm a}\\
  0,\quad \text{otherwise}
\end{array}
\right.\,,
\een
which indicates instead when ${\bf r}$ lies inside the core of atom ${\rm a}$.
Equation (\ref{Psi_n_split}) implies the following decompositon of the matrix elements,
\begin{eqnarray}
{f}_{nm}^{\rm ax,L}&=&{\bf e}_x\cdot\left[\int_\Omega{d{\bf r}\,\Pi^{\rm out}({\bf r})\tilde{\Psi}_n({\bf r})^*\,{\bf{f}}_{\rm a,L}({\bf r})\tilde{\Psi}_m({\bf r})}\right.\nonumber\\
&+&\int_\Omega{d{\bf r}\,\Pi_a^{\rm in}({\bf r})\Psi_n({\bf r})^*\,{\bf{f}}_{\rm a,L}({\bf r})\Psi_n({\bf r})}\nonumber\\
&+&\left.\sum_{b=1,b\ne a}^{N_i}\int_\Omega{d{\bf r}\,\Pi_b^{\rm in}({\bf r})\Psi_n({\bf r})^*\,{\bf{f}}_{\rm a,L}({\bf r})\Psi_n({\bf r})}\right]\label{fmnaxL_expanded}
\end{eqnarray}
and similarly for the (R) components.
In the following, we successively discuss the approximations we propose to evaluate the three terms in the right-hand side of Eq.(\ref{fmnaxL_expanded}).

{\it First term.} In this term, ${\bf f}_a({\bf r})$ represents the screened force between a nucleus $a$ with a test electron that is located at a position ${\bf r}$ lying outside all ionic cores (see red cross in Fig.~\ref{Fig:cartoon}).
At such a location, the core electrons perfectly shield the bare nuclei, which appear to the electron as a point-like charged particle with charge $Z_v=Z-Z_c$.
The remaining electronic screening is due to the valence electrons.
The situation is similar to that described in the previous section on local ECE pseudopotentials and we apply the same approximation.
Namely, we assume that the screening due to valence electrons in this interionic region (light blue area in Fig.~\ref{Fig:cartoon}) can be described by the dielectric function of the homogeneous electron gas model.
This yields
\ben
\lefteqn{\int_\Omega{d{\bf r}\,\Pi^{\rm out}({\bf r})\tilde{\Psi}_n({\bf r})^*\,{\bf f}_{\rm a,L(R)}({\bf r}) \tilde{\Psi}_m({\bf r})}}&&\\
&&\approx\int_\Omega{d{\bf r}\,\Pi^{\rm out}({\bf r})\tilde{\Psi}_n({\bf r})^*\,{\bf f}_{\rm a}^{\rm eg}({\bf r})\tilde{\Psi}_m({\bf r})}
\een
where ${\bf f}_{\rm a}^{\rm eg}({\bf r})$ is defined as in Eq.(\ref{fIeg}) with $\varepsilon_{\rm eg}$ the dielectric function of the homogeneous electron gas with density $Z_vN_i/\Omega$.

{\it Second term.} In this term, ${\bf r}$ lies inside the core of atom $a$ (see black cross in Fig.~\ref{Fig:cartoon}) where the core electrons do not fully screen the bare nucleus, the electron density varies widely and the homogoneous electron gas model is expected to fail.
This is illustrated in Fig.~(\ref{Fig:10}) that shows the components of the radial electron density surrounding an iron nucleus at solid density and melting temperature and a copper nucleus at liquid density and $T=0.2$ $\rm eV$ (see caption).
In order to deal with this term, we appeal to the exact sum rules 
\bes
\begin{align}
  &\sum_{\rm a=1}^{N_i}{\bf f}_{\rm a,L}({\bf r})=\int_{\Omega}{d{\bf r}'\nabla_{{\bf r}'}V_{ei}({\bf r}')\varepsilon_L({\bf r}',{\bf r},\omega=0)^{-1}}\nn\\
  &\hspace{1.5cm}=\nabla_{{\bf r}}V_{KS}[\rho_e]({\bf r})\label{sum_rule_L}\\
  &\sum_{\rm a=1}^{N_i}{\bf f}_{\rm a,R}({\bf r})=\int_{\Omega}{d{\bf r}'\varepsilon_R({\bf r},{\bf r}',\omega=0)^{-1}\nabla_{{\bf r}'}V_{ei}({\bf r}')}\nn\\
  &\hspace{1.5cm}=\nabla_{{\bf r}}V_{KS}[\rho_e]({\bf r})\label{sum_rule_R}
\end{align}
\label{sum_rule_L_and_R}
\ees
where $V_{KS}$ is given by Eq.(\ref{V_KS_PAW}); a proof of these sum rules can be found in the companion paper \cite{Daligault_Simoni_2019}.
In order to use Eq.(\ref{sum_rule_L_and_R}), we make two observations.
First, at the position ${\bf r}$ in core ${\rm a}$, the effect of other nuclei is perfectly shielded by their own core electrons and by their surrounding valence electrons.
This is illustrated in Fig.~\ref{Fig:10} that shows that every nucleus is typically surrounded by a neutralizing electronic sphere of radius $r_n$, whose magnitude is typically of the order of or smaller than half the average distance between two ions, which is denoted by $r_g$.
At solid density and above, the distance $r_g$, which we set equal to half the distance $r^*$ to the first peak of the ion-ion pair distribution function $g_{ii}(r^*)$  (not shown), is typically of the order of $\simeq 0.65 a$, where $a=(3\Omega/4\pi N_i)^{1/3}$ is the average interparticle distance.
As a consequence, the KS potential $V_{KS}$ is equal to the ``partial'' KS potential $V_{KS}^{\rm a}({\bf r})=-Ze^2/|{\bf r}-{\bf R}_{\rm a}| +v_{Hxc}^{\rm a}[\rho_e]({\bf r})$ with $\rho_e({\bf r})=\rho_v({\bf r})+\rho_{c}^{\rm a}({\bf r})$.
Second, we find intuitively reasonable to assume that all the cores are identical and contribute equally to the sum rules (\ref{sum_rule_L}) and (\ref{sum_rule_R}), i.e. the state of a given ion is only weakly dependent on the instantaneous configuration ${\bf R}$.
Overall, by applying both observations in Eq.(\ref{sum_rule_L_and_R}), we obtain
\begin{align*}
  &\Pi_{\rm a}^{\rm in}({\bf r}){\bf f}_{\rm a,L}({\bf r})\approx\nabla_{{\bf r}}\left[\frac{-Ze^2}{|{\bf r}-{\bf R}_{\rm a}|} +v_{Hxc}^{\rm a}[\rho_e]({\bf r})\right]\Pi_{\rm a}^{\rm in}({\bf r})\\
  &\Pi_{\rm a}^{\rm in}({\bf r}){\bf f}_{\rm a,R}({\bf r})\approx\nabla_{{\bf r}}\left[\frac{-Ze^2}{|{\bf r}-{\bf R}_{\rm a}|} +v_{Hxc}^{\rm a}[\rho_e]({\bf r})\right]\Pi_{\rm a}^{\rm in}({\bf r})\,.
\end{align*}
With this approximation, the second term in Eq.(\ref{fmnaxL_expanded}) reads as
\ben
\lefteqn{\int_\Omega{d{\bf r}\,\Pi_{\rm a}^{\rm in}({\bf r})\Psi_n({\bf r})^*{\bf f}_{\rm a,L(R)}({\bf r}) \Psi_n({\bf r})}}&&\\
&&\hspace{1.5cm}\simeq\int_\Omega{d{\bf r}\,\Pi_{\rm a}^{\rm in}({\bf r})\Psi_n({\bf r})^* \nabla_{{\bf r}}V_{KS}^{\rm a}({\bf r})\Psi_n({\bf r})}\,.
\een

{\it Third term.} The position ${\bf r}$ lies in the core of an ion ${\rm b}$ distinct from ${\rm a}$.
Due to the presence of the neutralizing sphere surrounding ${\rm a}$ discussed above, it is legitimate to assume that the strength of the force ${\bf f}_{\rm a,L(R)}({\bf r})$ is generally negligibly small.
We accordingly neglect the third term in Eq.(\ref{fmnaxL_expanded}).

In summary, we propose the following final approximation
\ba 
&f_{nm}^{\alpha={\rm ax,L(R)}}\approx{\bf e}_x\cdot\left[\int_\Omega{d{\bf r}\,\Pi^{\rm out}({\bf r})\tilde{\Psi}_n({\bf r})^*{\bf f}_{\rm a}^{\rm eg}({\bf r})\tilde{\Psi}_m({\bf r})}\right. \nn\\
&\hspace{1.1cm}\left.+\int_\Omega{d{\bf r}\,\Pi_{\rm a}^{\rm in}({\bf r})\Psi_n({\bf r})^* \nabla_{{\bf r}}V_{KS}^{\rm a}({\bf r})\Psi_n({\bf r})}\right] \label{fmnPAW}
\end{align}
For convenience, we give a list of mathematical relations that are useful to implement Eq.(\ref{fmnPAW}) in appendices \ref{F} and \ref{I}.

Before going further, we feel that a word of caution is in order regarding the use of the sum rules (\ref{sum_rule_L}) and (\ref{sum_rule_R}).
It may indeed be tempting to use the latter to approximate all the individual screened forces ${\bf f}_{\rm a,L}({\bf r})$ and ${\bf f}_{\rm a,R}({\bf r})$ by their average $\frac{1}{N_i}\sum_{\rm a=1}^{N_i}{\bf f}_{\rm a,L}({\bf r})$ and $\frac{1}{N_i}\sum_{\rm a=1}^{N_i}{\bf f}_{\rm a,R}({\bf r})$ for all positions ${\bf r}$, namely setting
\be
{\bf f}_{\rm a,L}({\bf r})={\bf f}_{\rm a,R}({\bf r})=\frac{1}{N_i}\nabla_{{\bf r}}V_{KS}[\rho_e]({\bf r})
\ee
for all ${\rm a}$.
This apparently reasonable assumption, however, yields a goofy result for the electron-ion coupling factor, namely $G_{ie}=0$.
This can be understood as a consequence of the exact sum rule $\sum_{\alpha,\beta}\gamma_{\alpha,\beta}^{[{\bf R}]}=0$ derived in \cite{Daligault_Simoni_2019} that is physically related to the conservation of the total linear momentum.
We will see in Sec.~\ref{Sec:3.3} that the individual friction coefficients depend on the position of the ions in the configuration ${\bf R}$ (see the dispersion of values around the average friction illustrated in Fig.~\ref{Fig:07}).

\subsection{First-principles calculation with a plane wave basis set}\label{sec:PW}

In simulations of bulk systems, the Kohn-Sham equations are often solved by imposing periodic boundary conditions and the wave functions are conveniently expanded over a plane-waves basis set.
In this section, we recast the previous results for the friction coefficients and the force matrix elements when periodic conditions are imposed.
This allows us to define some key numerical parameters that are varied in the next section in order to see the dependence of the temperature relaxation, including the number of ${\bf k}$ points and the number of bands.

\paragraph{Definitions and notations.}

We recall some useful notions needed when dealing with periodic systems.
A neutral system consisting $N_i$ ions and $N_e$ electrons (see Sec.~\ref{sec:2-2} on pseudopotentials for the meaning of ion and of $N_e$) is placed in a parallelipedic cell of volume $\Omega={\bf a}_1\cdot({\bf a}_2\cross {\bf a}_3)$ with primitive vectors $\{{\bf a}_1,{\bf a}_2,{\bf a}_3\}$.
We assume that the system is replicated periodically along the three primitive directions.
Calculations for liquid metals or plasmas are typically done assuming a cubic cell $\Omega=L^3$ and ${\bf a}_x=L{\bf e}_x$.
Yet, since the theory applies to a solid metal with $T_i$ significantly larger than its Debye temperature \cite{Daligault_Simoni_2019}, we here consider the case of a general Bravais lattice.

The Bloch theorem allows one to write the Kohn-Sham eigenstates in the form 
\ben
\Psi_{n{\bf k}}({\bf r}) = u_{n{\bf k}}({\bf r})e^{i{\bf k}\cdot{\bf r}},
\een
where ${\bf k}$ is a wave vector in the first Brillouin zone, $u_{n{\bf k}}({\bf r})$ is a function with the periodicity of the Bravais lattice, $n$ is the band index.
In practice, it is convenient to limit the number of allowed vectors ${\bf k}$ by imposing the Born-von Karman boundary conditions of macroscopic periodicity, namely
\be
\Psi_{n{\bf k}}({\bf r}+N_x{\bf a}_x)=\Psi_{n{\bf k}}({\bf r})\quad,\quad x=1,2,3\,,
\ee
where the $N_x$ are integers of order ${\cal{N}}_{{\bf k}}^{1/3}$, where ${\cal{N}}_{{\bf k}}=N_1N_2N_3$ is the total number of primitive cells.
Indeed, the Bloch theorem implies $u_{n{\bf k}}({\bf r})=\sum_{\bf G}{c_{n{\bf k}}({\bf G})e^{i{\bf G}\cdot{\bf r}}}/\sqrt{V}$ with $V={\cal{N}}_{{\bf k}}\Omega$, the Bloch vectors are restricted to the form 
\ben
{\bf k}=\sum_{x=1}^3{\frac{m_x}{N_x}{\bf  b}_x}\,,
\een
where the $m_x$ are all integers in the range $0\le m_x<N_x$, the ${\bf b}_x$ are the primitive vector of the reciprocal lattice, and ${\bf G}=\sum_x{n_x{\bf b}_x}$ are vectors of the reciprocal lattice.

The ground-state single particle density is written as
$n_{\rm e}({\bf r}) = 2\sum_n\sum_{{\bf k}\in{\rm BZ}}p_n({\bf k})|\tilde{\Psi}_{n{\bf
    k}}({\bf r})|^2$ where the sum is done over the first Brillouin
zone {\bf k} points. $p_n({\bf k})$ is now the Fermi-Dirac occupation probability of the KS state.

\paragraph{The friction coefficients.}
With these definitions, the friction coefficients (\ref{Eq:friction_tensor_KS_DFT}) read (see appendix \ref{A} for a complete derivation)
%
\begin{align}\label{Eq:14b}
\tilde{\gamma}_{\alpha\beta}^{[{\bf R}]} =&-\frac{\pi\hbar}{M}\sum_{n\neq m}^{\rm val}\sum_{{\bf k}\in{\rm IBZ}}W_{\bf k}\frac{p_n({\bf k})-p_m({\bf k})}{\epsilon_n({\bf k})-\epsilon_m({\bf k})}f_{nm}^\alpha({\bf k})f_{mn}^\beta({\bf k})\nonumber\\
&\times\delta(\epsilon_n({\bf k})-\epsilon_m({\bf k}))\,,
\end{align}
where the summation is performed only over the valence states and the ${\bf k}$ vectors belonging to the irreducible Brillouin zone (IBZ) --- i.e., the first Brillouin zone reduced by all of the symmetries in the point group of the lattice ---  $W_{\bf k}$ is the weight of each ${\bf k}$ points.

The form of the force matrix elements in Eq.~(\ref{Eq:14b}) depends on the type of pseudopotential used.
\paragraph{Matrix elements for a local ECE pseudopotential.} 
In this case the matrix elements read (see appendix \ref{F} for the complete derivation)
\begin{equation}\label{Eq:matr_el_feg}
  f_{nm}^{\alpha={\rm a,x}}(\mathbf{k}) = \frac{1}{\Omega}\int_\Omega
  d{\bf r}\,u_{n\mathbf{k}}^*(\mathbf{r})f_{\rm a,x}^{\rm eg}(\mathbf{r})u_{m\mathbf{k}}(\mathbf{r}),
\end{equation}
$\mathbf{f}_{\rm a}^{\rm eg}({\bf r})$ is the gradient of the screened electron ion potential from Eq.~(\ref{fIeg}). Thanks to the periodicity of the system this quantity can be easily computed in reciprocal space
\begin{equation}
  {\bf f}_{\rm a}^{\rm eg}(\mathbf{G}) = i {\bf G} e^{-i{\bf G}\cdot{\bf R}_{\rm a}}v_{ie}(|\mathbf{G}|)\tilde{\varepsilon}_0^{-1}(|\mathbf{G}|,\omega=0),
\end{equation}
where $v_{\rm ie}(|{\bf G}|)$ is the Fourier transform of
the local ECE pseudo potential. $f_{\rm a}^{\rm eg}(|{\bf G}|)$ can be then inverse Fourier transform back to real space by using a Fast Fourier Transform algorithm
\begin{equation}
{\bf f}_{\rm a}^{\rm eg}({\bf r}) = \frac{1}{\Omega}\sum_{{\bf G}}{\bf f}_{\rm a}^{\rm eg}({\bf G}) e^{i{\bf G}\cdot{\bf r}}\,,
\end{equation}
and used into Eq.~(\ref{Eq:matr_el_feg}) to complete the calculation of the matrix elements.

\paragraph{Matrix elements for a PAW pseudo-potential.}

By using the approximation (\ref{fmnPAW}) for the screened interaction together with the property (\ref{C13}), valid inside the core region of atom ${\rm a}$, we easily obtain the following final decomposition
\begin{align}\label{Eq:14}
& f_{nm}^{\alpha={\rm a,x}}({\bf k}) =
  \frac{1}{\Omega}\int_\Omega d{\bf r}\,\Pi_{\rm out}({\bf r}) u_{n{\bf k}}({\bf r})^* f_{\rm a,x}^{\rm eg}({\bf r})u_{m{\bf k}}({\bf r}) +\nonumber \\
&+ \sum_{i,j}\ip*{\tilde{\Psi}_{n{\bf k}}}{\tilde{p}_{{\rm a}i}}\int_{\Omega}d{\bf r}\Pi_a^{\rm in}({\bf r})\phi_{{\rm a}i}^*({\bf r})\nabla_{r_x}V_{\rm KS}^{\rm a}({\bf r})\phi_{{\rm a}j}({\bf r})\nonumber\\
&\times\ip*{\tilde{p}_{{\rm a}j}}{\tilde{\Psi}_{m{\bf k}}}\,.
\end{align}
The evaluation of the first term on the right hand side of Eq.~(\ref{Eq:14}) proceeds analogously to the computation of Eq.~(\ref{Eq:matr_el_feg}), the determination of the atomic core contribution requires instead the evaluation of a set of one-dimensional integrals described in appendix \ref{I}.


\section{Real-time calculation of the Kubo formulas}\label{sec:3}

In this section, we describe the method we used to evaluate the Kubo formulas for the friction coefficients and the electron-ion coupling, and we discuss how this differs from the approach that is generally used for the calculation of electronic conductivities.

For convenience, we introduce the following notations.
The ensemble averaged friction coefficient $\Gamma$ in Eq.(\ref{Eq:Temp_rel_rate}) is written as
\be
\Gamma=\bigg\langle\frac{1}{3N_{\rm i}}\sum_{\rm a=1}^{N_{\rm i}}\sum_{x=1}^3\gamma_{\rm ax,ax}^{[{\bf R}]}\bigg\rangle=\big\langle \Gamma^{[{\bf R}]}\big\rangle \label{Gamma}
\ee
where
\be
\Gamma^{[{\bf R}]}=\frac{1}{N_{\rm i}}\sum\limits_{\rm a=1}^{N_{\rm i}}\Gamma_{\rm a}^{[{\bf R}]}
\ee
is the average friction coefficient felt by any ion when the system is in the configuration ${\bf R}$ and $\ds\Gamma_{\rm a}^{[{\bf R}]}=\sum_{x=1}^3\tilde{\gamma}_{\rm ax,ax}^{[{\bf R}]}/3$ is the spatially average friction felt by ion ${\rm a}$ in this configuration.
Using the formulas of Sec.~\ref{sec:2}, the friction coefficient $\Gamma_{\rm a}^{[{\bf R}]}$ reads
\begin{align}\label{gammaabfrequency}
\Gamma_{\rm a}^{[{\bf R}]} =&-\frac{\pi\hbar}{3M}\sum_{\rm x=1}^{3}\sum_{n\neq m}^{\rm val}\sum_{{\bf k}\in{\rm IBZ}}W_{\bf k}\frac{p_n({\bf k})-p_m({\bf k})}{\epsilon_n({\bf k})-\epsilon_m({\bf k})}|f_{nm}^{\rm ax}({\bf k})|^2\nonumber\\
&\times\delta(\epsilon_n({\bf k})-\epsilon_m({\bf k}))\,,
\end{align}
where, as mentioned above, we neglect the correction term $\delta\tilde{\gamma}_{\alpha\beta}$ in Eq.(\ref{Eq:MB_friction_tensor}).

In the following, we first discuss the calculation of the frictions $\Gamma_{\rm a}^{[{\bf R}]}$ and $\Gamma^{[{\bf R}]}$.
We then discuss in Sec.~\ref{Sec:3.2} the self-averaging character of the ensemble average $\big\langle.\big\rangle$ in Eq.(\ref{Gamma}).
Finally, in Sec.~\ref{Sec:3.3}, we discuss the statistical distribution of friction coefficients $\Gamma_{\rm a}^{[{\bf R}]}$.

\subsection{Calculation of $\Gamma_{\rm a}^{[{\bf R}]}$ and $\Gamma^{[{\bf R}]}$ in the time domain}

In this section, we omit the superscript $[{\bf R}]$ on all quantities.
In principle, in order to compute $\Gamma_{\rm a}$, one could follow the method generally used for evaluating the Kubo-Greenwood formulas for the electrical and thermal conductivities, which is a direct evaluation of the expression (\ref{gammaabfrequency}).
\myFig{1}{1}{true}{0}{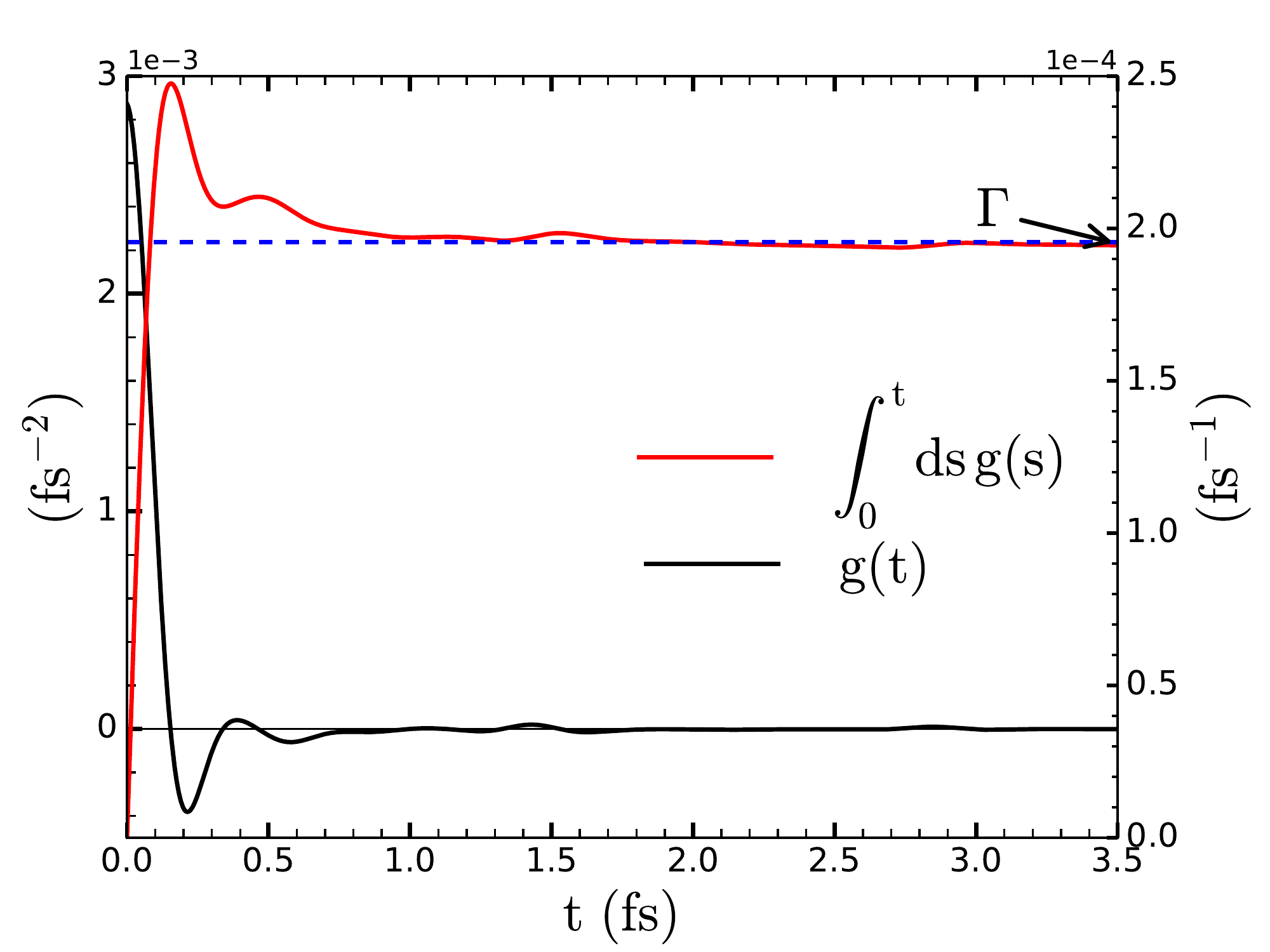}{(Color online) Illustration of the method used to determine the average friction $\Gamma$.
The black line shows the time correlation function $g(t)$ defined in Eqs.(\ref{gIoft},\ref{calGoft}) and the red line shows $\int_0^t{ds\,g(s)}$ as obtained for warm dense aluminum at solid density and $T=1160.4$ $\rm K$.
$g(t)$ decays rapidly to zero over a time scale of $\sim 1$ ${\rm fs}$.
Beyond this time scale, the cumulative sum remains constant at a plateau value corresponding to $\Gamma$ according to Eq.(\ref{cumulativegoft}).
Figure \ref{Fig02_2} shows another example in the presence of a higher level of numerical noise.}{Fig:02}
However, since the finite simulation volume results in a discrete spectrum, the $\delta$-function in Eq.(\ref{gammaabfrequency}) must be broadened; this is typically achieved by replacing the $\delta$-function with a Gaussian or a Lorentzian distribution with a finite width chosen ad-hoc.
To avoid introducing this extraneous parameter that needs to be
determined for each calculation, we return to the fundamental formulas (\ref{tildegammaab}) and evaluate them in stage by first calculating the force correlation function in time domain and then by integrating it over time; this approach is analogous to the standard method used in classical physics to calculate the transport coefficients from the positions and velocities calculated in a classical molecular dynamics simulation \cite{FrenkelSmit2001}.
\begin{figure}[t]
\begin{center} 
\includegraphics[width=\columnwidth]{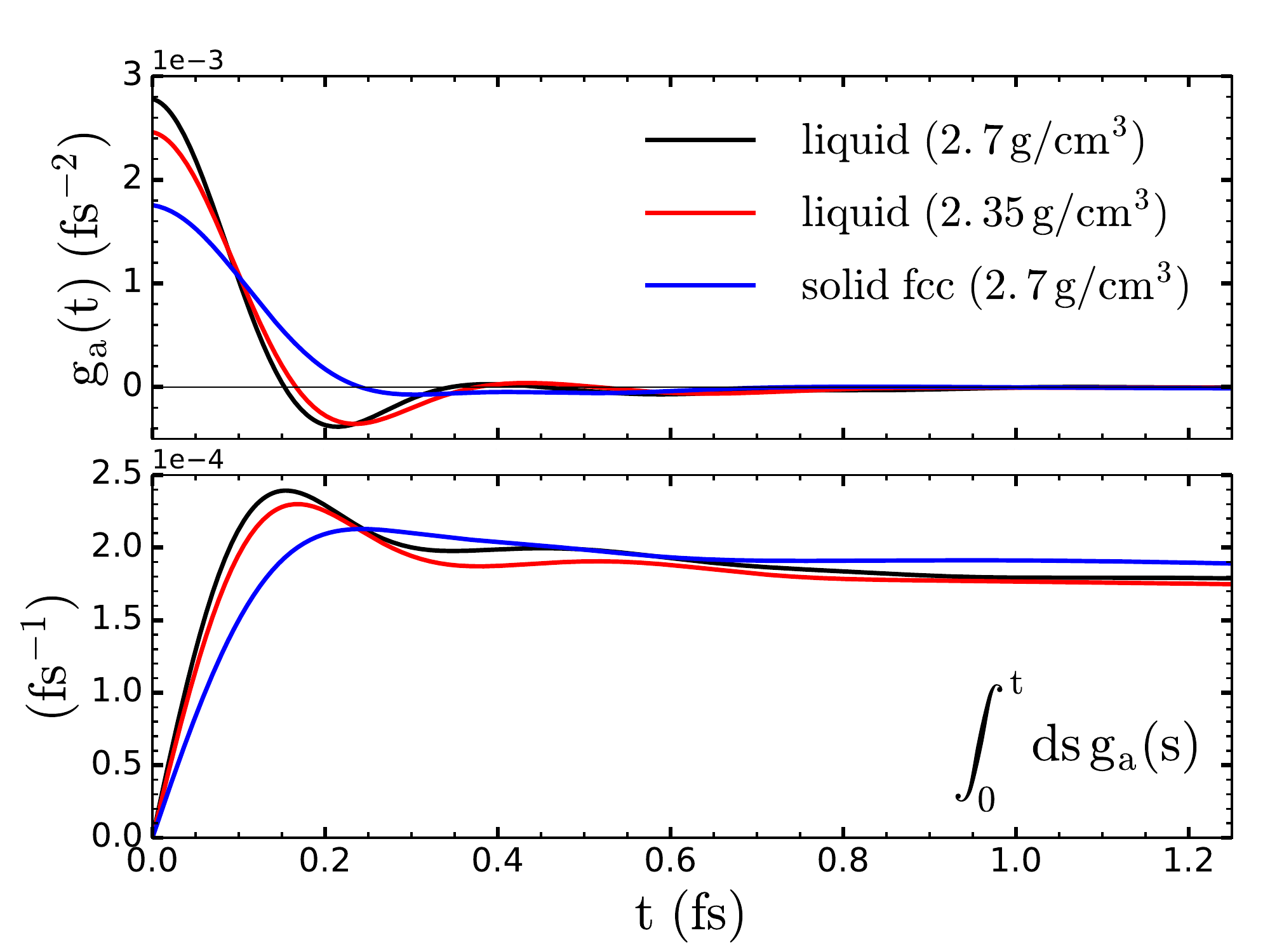}
\caption{(Color online) Another illustration of the method used to evaluate the friction coefficient $\Gamma_{\rm a}$ of an ion in aluminum under the conditions indicated in the legend and for $T=1160.4\,{\rm K}$.
The time correlation function $g_{\rm a}(t)$ defined in Eq.(\ref{gIoft}) and the corresponding cumulative sum are shown in the upper and lower panels, respectively.
In all three cases, beyond a correlation time scale of the order of $1$ $\rm fs$, the cumulative sums reach a plateau value equal to the friction coefficient $\Gamma_{\rm a}$ according to Eq.(\ref{gammaIcumulativesum}).}\label{Fig:14}
\end{center}
\end{figure}
By using Eq.(\ref{tildegammaab}), Eq.(\ref{gammaabfrequency}) is replaced by the following time integral
\be
\Gamma_{\rm a}&=&\lim_{T\to\infty}\int_0^T{dt\,g_{\rm a}(t)}\,, \label{gammaIcumulativesum}
\ee
where
\be
g_{\rm a}(t)&=&\frac{\beta_e}{6M}\sum_{x=1}^3{K_{\rm ax,ax}(t)} \label{gIoft}\\
&=&\frac{\hbar}{3M}\sum_{\rm x=1}^3\sum_{n\neq m}\sum_{\mathbf{k}\in\mathrm{IBZ}}W_\mathbf{k}\,p_n({\bf k})\big[1 - p_m({\bf k})\big]\big|f_{nm}^{\rm ax}(\mathbf{k})\big|^2\nonumber\\
&&\hspace{0.5cm}\times\frac{e^{\beta_e(\epsilon_n({\bf k})-\epsilon_m({\bf k}))}-1}{\epsilon_n({\bf k})-\epsilon_m({\bf k})}\cos\bigg(\frac{\epsilon_n(\mathbf{k})-\epsilon_m(\mathbf{k})}{\hbar}t\bigg)\,.\nn
\ee
Similarly, for the average friction, we use
\be
\Gamma&=&\lim_{T\to\infty}\int_0^T{dt\,g(t)}\,, \label{cumulativegoft}
\ee
where
\be
g(t)=\frac{1}{N_{\rm i}}\sum_{\rm a=1}^{N_{\rm i}}{g_{\rm a}(t)} \label{calGoft}
\ee
is the average over the ions of the temporal force correlation function.
In both Eqs.~(\ref{gammaIcumulativesum}) and (\ref{cumulativegoft}), the cumulative sum is expected to reach a plateau beyond the correlation time scale of the screened electron-ion force correlation function, i.e. the time beyond which $g_{\rm a}(t)$ and $g(t)$ vanish or are negliglibly small.
The method is illustrated in Figs.~(\ref{Fig:02}) and (\ref{Fig:14}).

Figure (\ref{Fig:02}) shows the average correlation function $g(t)$ (black line) and its cumulative sum $\int_0^t{ds\,g(s)}$ (red line) for liquid aluminum at solid density and $T=1160.4$ $\rm{K}$ (the details of the simulations are discussed in Sec.~\ref{sec:4}).
The correlation function decays rapidly to zero over a time scale of $\sim 1$ ${\rm fs}$.
As a result, its cumulative sum reaches a stable plateau beyond this correlation time scale that, according to Eq.(\ref{cumulativegoft}), corresponds to the average friction coefficient.

Similarly, Fig.~(\ref{Fig:14}) shows the correlation function $g_{\rm a}(t)$ (upper panel) and its cumulative sum $\int_0^t{ds\,g_{\rm a}(s)}$ (lower panel) of a randomly chosen atom ${\rm a}$ in aluminum at melting density $\rho=2.35$ $\rm g/cm^3$ and at solid density $2.7$ $\rm g/cm^3$, with $T=1160.4$ $\rm{K}$ in all cases; for $\rho=2.7$ $\rm g/cm^3$, two ionic structures are considered: a disordered, liquid structure and a superheated fcc crystal structure.
We again note the rapid decay to zero of the correlation functions and the concomitant evolution of their integrals to a plateau value corresponding to $\Gamma_{\rm a}$.

\begin{figure}[t] 
\begin{center} 
\includegraphics[width=\columnwidth]{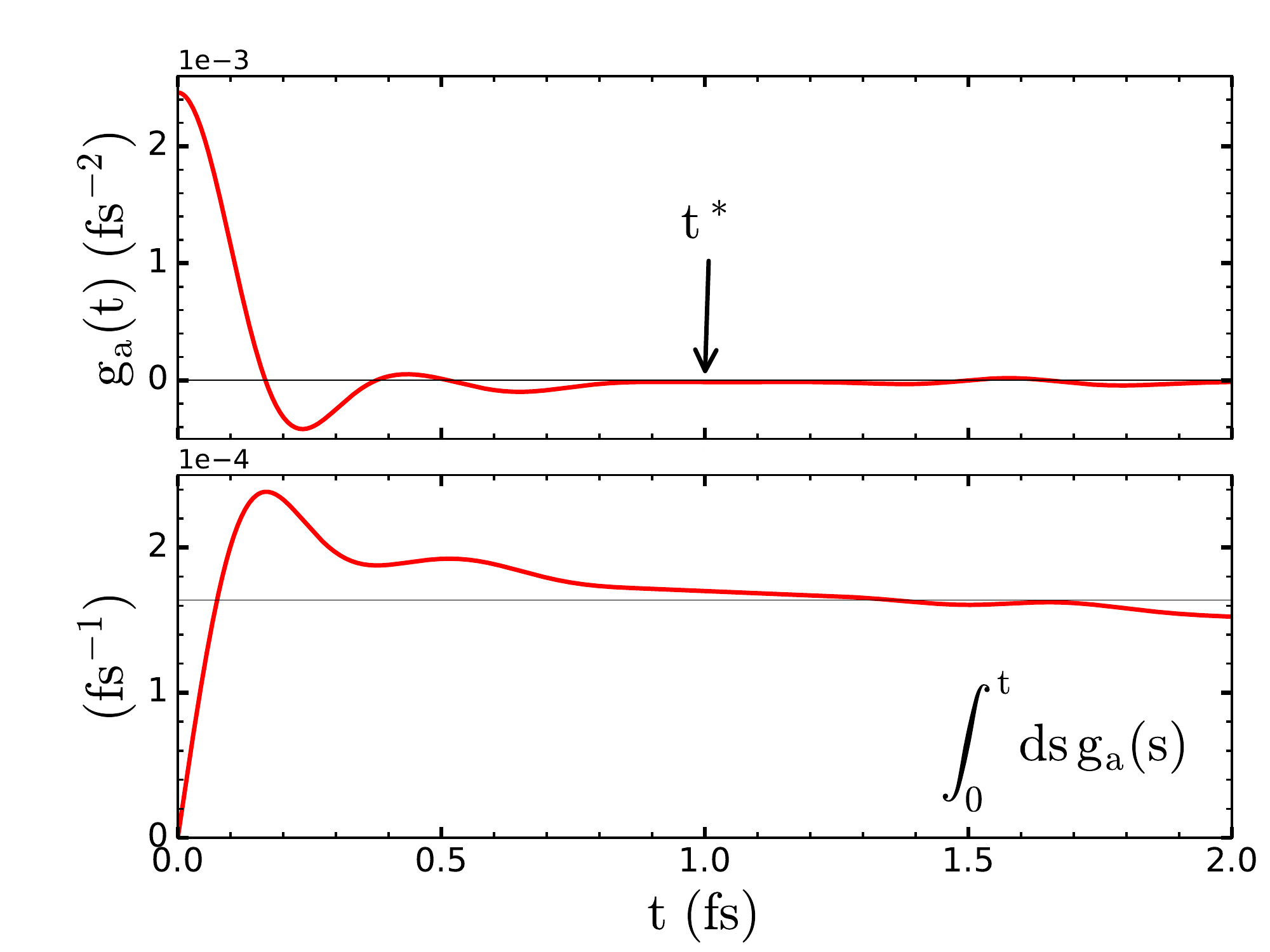}
\caption{(Color online) Same as in Fig.~\ref{Fig:02} in the presence of a higher level of numerical noise.}\label{Fig02_2}
\end{center}
\end{figure}

The calculations shown in Figs.~(\ref{Fig:02}) and (\ref{Fig:14}) are well converged and plateau values can be identified unambiguously.
In general, it may happen that, due to numerical inaccuracies, the force correlation function does not perfectly vanish and remains sligtly above or slighty below the zero line, resulting in positive or negative drifts in its cumulative sum.
This is illustrated in Fig.~\ref{Fig02_2}.
In all cases we studied, a very good estimate of the desired plateau value could be obtained by setting it equal to $\int_0^{t^*}{ds\,g(s)}$ where $t^*$ is the earliest time beyond which the correlation function has nearly vanished, e.g. $t^*=1$ ${\rm fs}$ in the case shown in Fig.~\ref{Fig02_2}.
An alternative to this educated guess consists in eliminating the apparent noise by multiplying the correlation function with an exponential that does not affect the short time behavior but forces the correlation function to vanish at later times.
This approach is equivalent to the standard method based on Eq.(\ref{gammaabfrequency}) combined with the broadening of the delta function.
We remark that we never had to use this approach for the calculations shown in this paper and in Ref.~\cite{Simoni_2019}.

\subsection{Self-averaging property.} \label{Sec:3.2}

In practice, the ensemble average in Eq.~(\ref{Gamma}) becomes an average over a finite number $\mathcal{N}_c$ of configurations $\{{\bf R}_c\}$ selected along the trajectory followed by the ions during a QMD simulation,
\be
\Gamma\simeq\frac{1}{\mathcal{N}_c}\sum_{c=1}^{\mathcal{N}_c}{\Gamma^{[{\bf R}_c]}}\,.
\ee
According to the published litterature, quantum molecular dynamics calculations of the electrical and thermal conductivities typically require $\mathcal{N}_c=10-20$ configurations in order to get good estimates for these quantities.
However, the friction $\Gamma$ differs from the conductivities in the fact that, for each ionic configuration $[{\bf R}_c]$, $\Gamma^{[{\bf R}_c]}$ is already an average property: it describes a single particle property, namely the average over the ions of the individual friction coeffcients.
By contrast, the electrical conductivity is a collective property: for each ionic configuration $[{\bf R}_c]$, the electrical current is a non-averaged sum over electrons.
\myFig{1}{1}{true}{0}{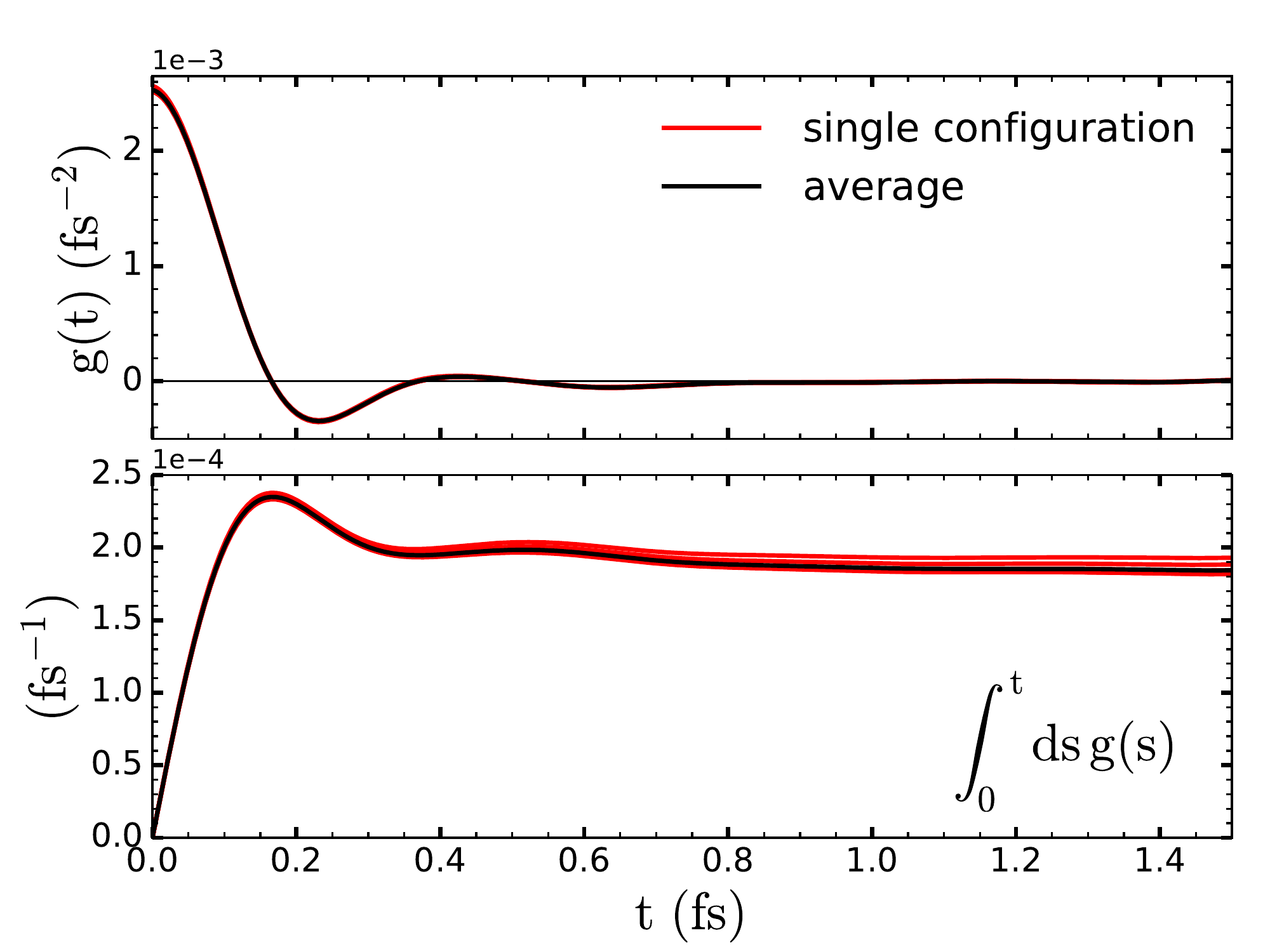}{(Color online) Illustration of the self-averaging character of $\Gamma$, Eq.(\ref{Gamma}).
The figure shows the correlation functions $g^{[{\bf R}_c]}(t)$ (red dashes, upper panel) defined by Eq.(\ref{calGoft}), and their cumulative sums $\int_0^t ds\,g^{[{\bf R}_c]}(s)$ (red dashes, lower panel) for $\mathcal{N}_c=10$ configurations taken during a QMD simulation of aluminum at $2.35$ $\rm g/cm^3$ and $T=0.1$ $\rm eV$; the configurations are separated in time by $0.67$ $\rm ps$. The averages over the ${\cal{N}}_c$ configurations are shown in black in both panels.}{Fig:self-averaging}
It is reasonable to expect that, if the system size, $N_i$, is sufficiently large, a single ionic configuration $\mathcal{N}_c=1$ is sufficient to accurately determine $\Gamma$; in other words, $\Gamma$ is self-averaging property.

As illustrated in Fig.~(\ref{Fig:self-averaging}), our calculations confirm the self-averaging character of $\Gamma$.
The figure shows the correlation function $g^{[{\bf R}_c]}(t)$ and its cumulative sum $\int_0^t{ds\,g^{[{\bf R}_c]}(s)}$ for $10$ configurations equidistant in time taken during a $6$ $\rm ps$ long QMD simulation of aluminum at $2.35$ $\rm g/cm^3$ and $T=1160.4$ $\rm K$.
The correlation functions (upper panel) are quite alike and it is hard to notice any change in shape.
The small differences are more apparent in the long-time values $\Gamma^{[{\bf R}_c]}$ reached by the cumulative sums.
Yet, the dispersion of different $\Gamma^{[{\bf R}_c]}$ remains quite small
\be
\Delta\Gamma\equiv\sqrt{\frac{1}{\mathcal{N}_c-1}\sum_{c=1}^{\mathcal{N}_c}{(\Gamma-\Gamma^{[{\bf R}_c]})^2}}\approx 4\times 10^{-6}\,{\rm fs^{-1}}\,,
\ee
with $\Delta\Gamma/\Gamma=0.0217$.

All the illustrative calculations discussed in the remaining of the paper use ${\cal{N}}_c=1$ configuration.

\myFig{1}{1}{true}{0}{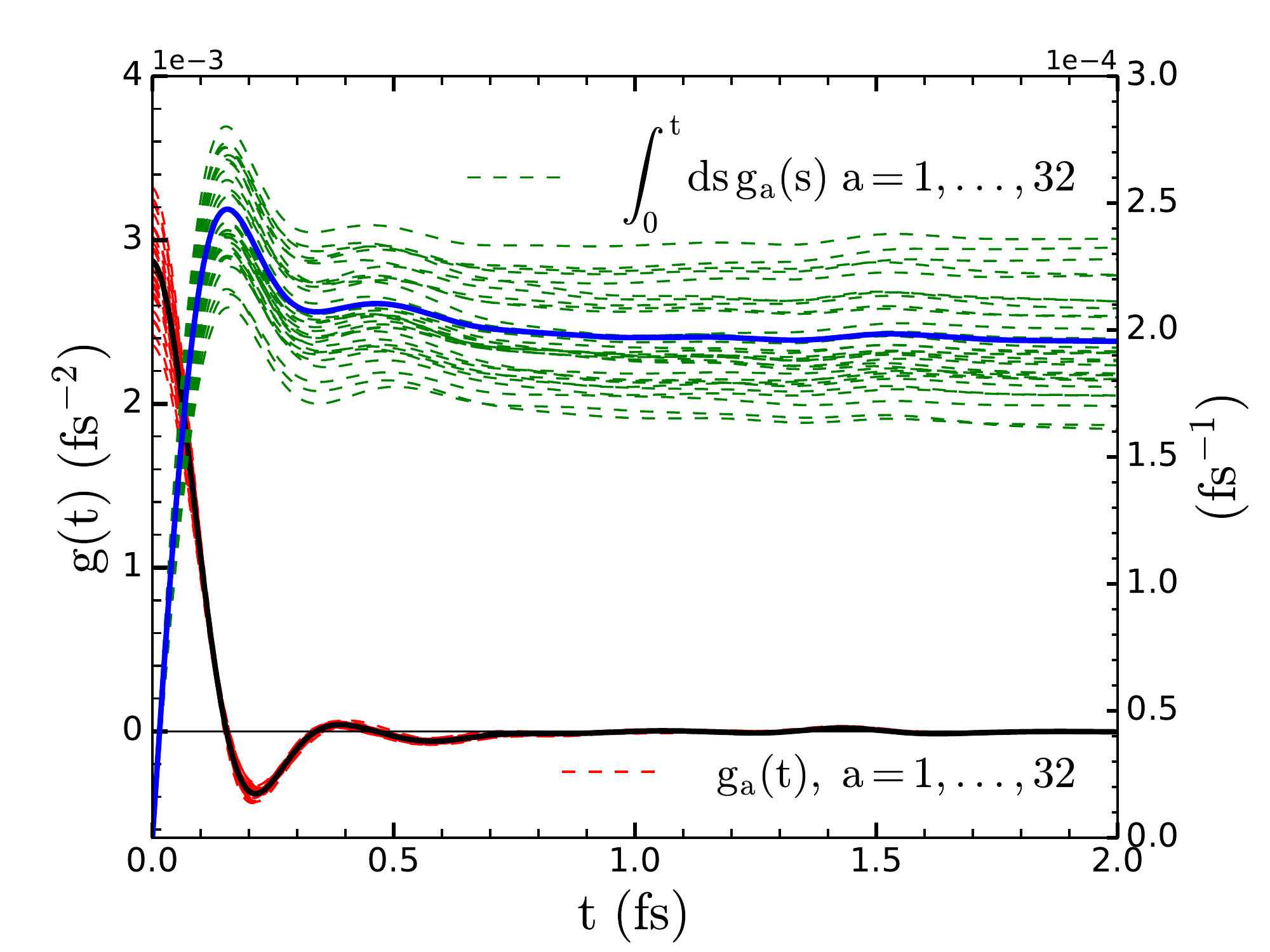}{(Color online) Distribution of friction coefficients $\Gamma_{\rm a}$ in an instantaneous ionic configuration in liquid aluminum at \SI{2.7}{g/cm^3} and $T=1160.4\,\rm{K}$. The figure shows the time correlation function $g_{\rm a}(t)$ (dashed red) and the corresponding cumulative sum $\int_0^t{ds\,g_{\rm a}(s)}$ (dashed green) for $32$ randomly selected ions. The full black line shows the average $g(t)$ of the individual functions $g_{\rm a}(t)$, the full blue line shows its cumulative sum $\int_0^t{ds\,g(s)}$.}{Fig:07}

\subsection{Distribution of friction coefficients} \label{Sec:3.3}

We now discuss the statistical distribution of individual friction coefficients $\Gamma_{\rm a}$ around their average $\Gamma$ .
Figure~(\ref{Fig:07}) shows the correlation functions $g_{\rm a}(t)$ (red dashes) and the corresponding cumulative sums $\int_0^t{ds\,g_{\rm a}(s)}$ (dashed green) for $32$ randomly chosen atom in a QMD simulation of liquid aluminum at \SI{2.7}{g/cm^3} and $T=1160.4$ $\rm{K}$ (the simulation contained $N_i=64$ ions, we show results for only $32$ of them in the figures for clarity).
The figure also shows the average correlation function $g(t)$ (black full line) and its cumulative sum $\int_0^t{ds\,g(s)}$ (blue full line).

All the correlation functions $g_{\rm a}(t)$ and cumulative sums show similar variations in time.
We note, however, a non-negligible spread in the initial value, which result in a non-negligible dispersion of the individual friction coefficients (see later time behavior of dashed green lines).
Quantitatively, here, the average friction is $\Gamma=1.955\times 10^{-4}$ $\rm fs^{-1}$, whereas the standard deviation of individual frictions is $\sigma=\sqrt{\sum_{\rm a= 1}^{N_i}\left[\Gamma_{\rm a}-\Gamma\right]^2/(N_i-1)}=1.94\times 10^{-5}$ ${\rm fs^{-1}}$, that gives $\sigma/\Gamma=0.0994$.

\section{Parametric and convergence study}\label{sec:4}

In this section we study the dependence of the calculation of friction coefficients discussed above both on the numerical parameters involved in plane-wave-based QMD calculations and on the approximations used for quantities such as the dielectric function and the exchange-correlation energy.

All the QMD calculations presented in this paper were performed by using the Quantum Espresso package, an open-source plane-wave DFT code \cite{QE2}.
Given the system's chemical composition, its mass density $\rho$ and its ionic and electronic temperatures $T_i$ and $T_e$, a typical calculation consists of two main parts. 
The first part is a standard QMD simulation within the Born-Oppenheimer approximation.
Given a number of ions $N_i$ and electrons $N_e$ in a cell of volume $\Omega$, the electronic ground state is computed by assuming Born-von Karman periodic conditions for every ionic configuration and by solving the set of KS equations at temperature $T_e$.
The ionic positions are then updated by using the instantaneous Born-Oppenheimer force in the Newton equations.
A thermostat is used to ensure that during the dynamics the ionic
temperature $T_i$ does not change, in all the calculations the
Andersen thermostat was employed.
The system is first carefully equilibrated and then it is evolved for a sufficiently long time (few picoseconds) in order to collect enough ionic configurations.
The number of electrons, $N_e$, per atom that are directly accounted for in the simulation depends on the pseudopotential used.
For all calculations shown in this paper and in \cite{Simoni_2019}, only the valence electrons enter explicitely the calculation with the exception of the iron atom for which also semi-core states ($3s$ and $3p$ atomic shells) contribute to the self-consistent evaluation of the electronic structure.
At that stage, only the $\Gamma$ point is used for the representation of the Brillouin zone (calculations with higher order ${\bf k}$-point sets were examined with no significant effect on the transport coefficients).
Following the molecular dynamics simulation, a number of instantaneous ionic configurations are selected.
However, as we discussed in Sec.~\ref{Sec:3.2}, while tens of configurations may be necessary to determine collective properties like electrical conductivity, one single configuration is often enough for the calculation of the average friction provided the number of atoms $N_i$ in the simulation box is large enough.\\
In the second part of the calculation we extract the average friction coefficient $\Gamma$ from the knowledge of the temporal correlation function $g(t)$, which is calculated using Eqs.~(\ref{gIoft},\ref{calGoft})
For a given selected ionic configuration, a refined calculation of the electronic structure is performed where the number of bands, the cut-off energy for the plane waves expansion together with the number of {\bf k}-points are increased to ensure convergence of the summations (\ref{gIoft}) and (\ref{calGoft}).
The details of all the parameters used for the calculation of $\Gamma$ in the case of different material systems are shown in table~(\ref{Tab:1}), while a study of the convergence of $g(t)$ and $\Gamma$ with respect to the choice of these parameters is given in Sec.~(\ref{sec:4-1}).
\begin{table}[H]
  \centering
  \caption{Typical values for the number of bands, number of {\bf k}-points, cut-off energy and number of atoms used in the calculations for aluminum at solid and liquid density and for melted iron and copper. If $T_e$ is changed, $\mathcal{N}_b$ needs also to be changed in the way explained in the main text.}
  \label{Tab:1}
  \begin{tabular}{ |p{1.8cm}||p{1.5cm}|p{1.5cm}|p{1.5cm}|p{1.5cm}|  }
    \hline
    ${\rm material}$ & ${\rm Al}$ & ${\rm Al}$ & ${\rm Cu}$ & ${\rm Fe}$ \\ 
    $ \rho\,(g/cm^3)$ & {\bf 2.7} & {\bf 2.35} & {\bf 8.02} &
    {\bf 7.87} \\
    $ T_{\rm e,i}\,(eV)$ & {\bf 0.1} & {\bf 0.1} & {\bf 0.2} &
    {\bf 0.156} \\
    \hline\hline
    $\mathcal{N}_b$ & $250$ & $250$ & $620$ & $750$ \\
    $\mathcal{N}_{\bf k}$ & $8$ & $8$ & $8$ & $8$ \\
    $E_{\rm cut}\,(Ry)$ & $150$ & $150$ & $150$ & $150$ \\
    $N_i$ & $64$ & $64$ & $64$ & $64$ \\
    \hline
  \end{tabular}
\end{table}
%


\subsection{Convergence with the number of bands and of ${\bf k}$-points.} \label{sec:4-1}

We consider the dependence of the force correlation functions and their cumulative sums on the number of band $\mathcal{N}_b$ and of ${\bf k}$-points ${\cal{N}}_{{\bf k}}=N_1\times N_2\times N_3$ used when evaluating Eq.~(\ref{gIoft}).
For this discussion, we consider simulations of liquid aluminum at \SI{2.7}{g/cm^3} and $T=1160.4\,\rm{K}$ using the numerical parameters listed in table~(\ref{Tab:1}) where either $\mathcal{N}_b$ or ${\cal{N}}_{\bf k}$ is varied.
\myFig{1}{1}{true}{0}{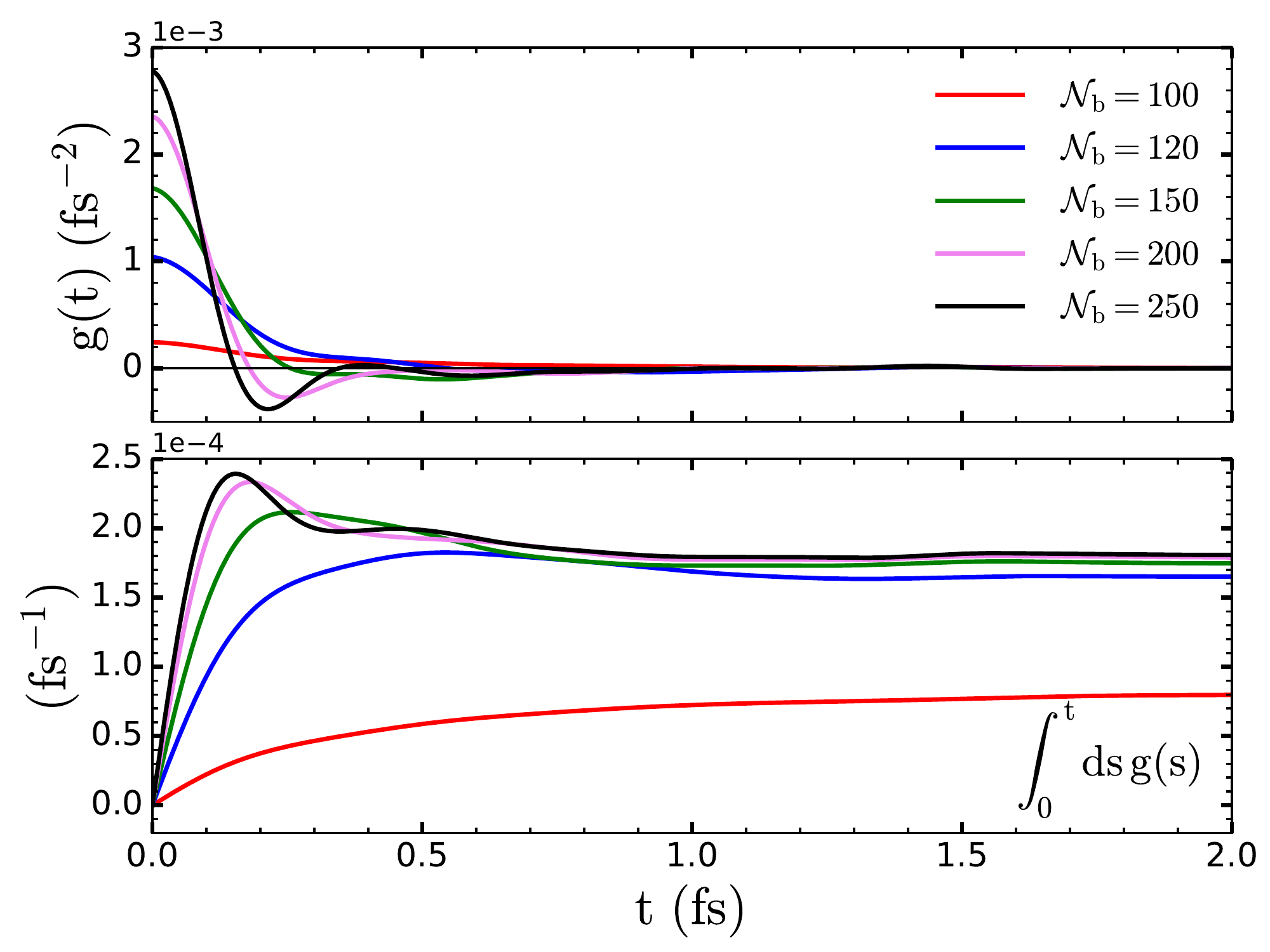}{(Color online) 
Effect of the number of bands.
The figure shows the time correlation function $g(t)$  (upper panel) and the corresponding cumulative sum $\int_0^t{ds\,g(s)}$  (bottom panel) for liquid aluminum at \SI{2.7}{g/cm^3} and $T=1160.4\,\rm{K}$ using the number of bands indicated in the legend. The other parameters are fixed and listed in table~(\ref{Tab:1}).}{Fig:12}
Figure~(\ref{Fig:12}) shows $g(t)$ (upper panel) and its cumulative sum (lower panel) obtained by using different number of bands varying between $\mathcal{N}_b=100$ and $250$.
By increasing the number of bands, we notice the convergence of the results towards a stable limit.
With a closer look at the calculation, we observe that, with only $100$ bands (red line), there are states with a non-negligible occupation number $p_n=0.1$ or even greater that are not included in the calculation.
As can be seen in the figure, these states, which are then included as the number of bands increases, significantly contribute to the correlation function.
Interestingly, here, as a result of a fortunate cancellation of errors in the cumulative sums, the plateau values with $\mathcal{N}_b\ge 200$ are less sensitive to the number of bands than the correlation functions.
\myFig{1}{1}{true}{0}{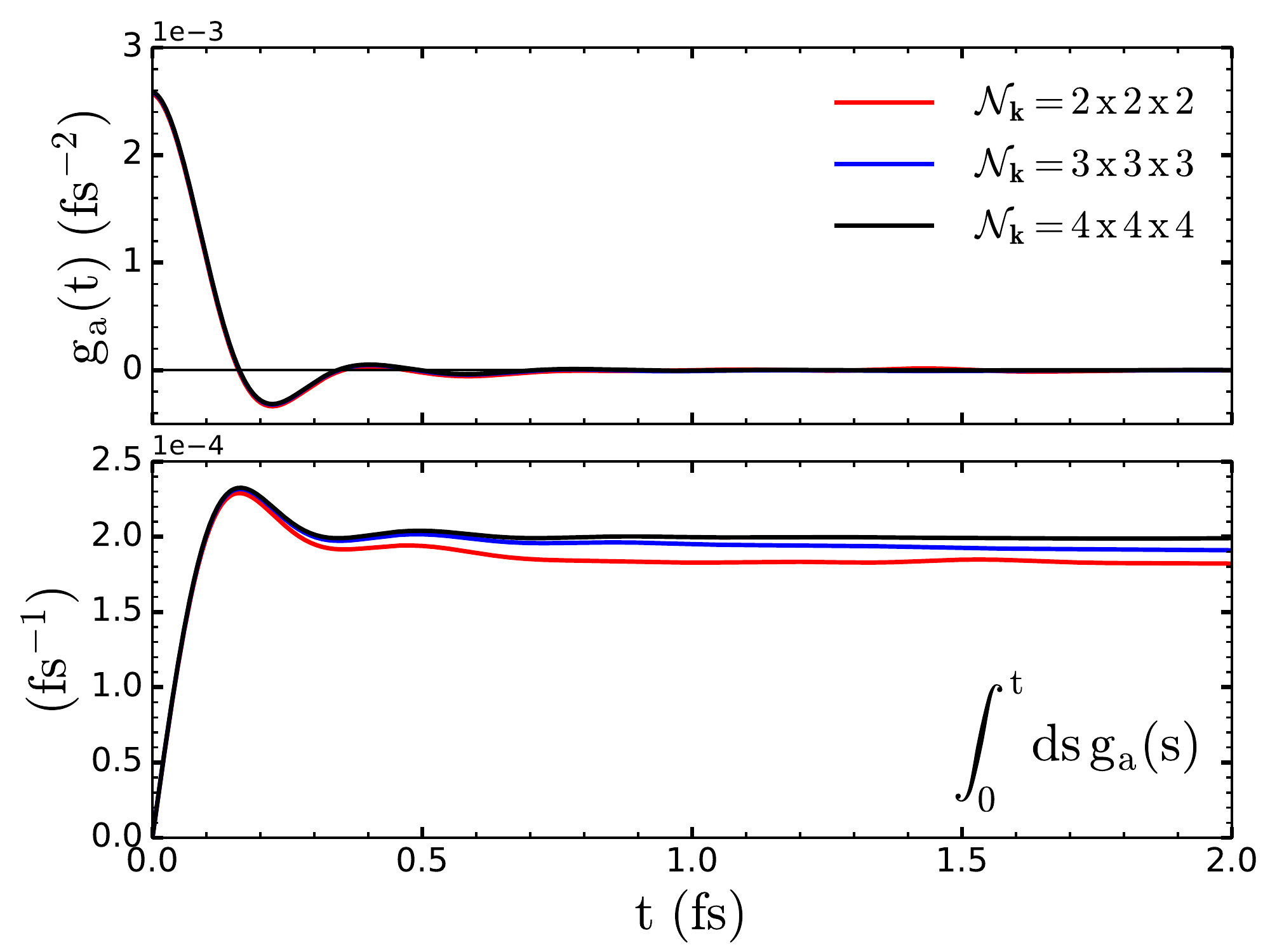}{(Color online) Effect of the number $N_1\times N_2\times N_3$ of ${\bf k}$-points.
The figure shows the time correlation function $g_{\rm a}(t)$ (upper panel) and the corresponding cumulative sum $\int_0^t{ds\,g_{\rm a}(s)}$ (bottom panel) for a single atom randomly selected in liquid aluminum at \SI{2.7}{g/cm^3} and $T=1160.4\,\rm{K}$ using the number of ${\bf k}$-points indicated in the legend while the other parameters are fixed and indicated in table~(\ref{Tab:1}).}{Fig:13}
Figure~(\ref{Fig:13}) shows the time correlation function $g_{\rm a}(t)$ (upper panel) for a randomly chosen ion ${\rm a}$ and its cumulative sum (lower panel) for different numbers $\mathcal{N}_{\bf k}=N_1\times N_2\times N_3$ of ${\bf k}$-points.
The convergence with increasing $\mathcal{N}_{\bf k}$ is much faster than with the number of bands that we have previously discussed.
We see that the $\mathcal{N}_{\bf k}=2\times 2\times 2$ result may be considered already converged given that its variation with respect to the fully converged $3\times 3\times 3$ one is lower than the spread due to the different atomic contributions (see Fig.~\ref{Fig:07}).

The previous analysis is valid for all the calculations considered here.
At lower densities it may be necessary to lower the number of atoms $N_i$ in order to do not increase too much the size of the simulation box; in this case, a higher value of $\mathcal{N}_{\bf k}$ may be required.
At higher $T_e$ (see below), we need instead to increase $\mathcal{N}_b$ given that more states at higher energy will have non zero occupations.

\subsection{Effect of the electronic structure in two-temperature $T_e\ne T_i$ calculations}

At a given ionic temperature $T_i$, the friction coefficients (\ref{Eq:14b}) and the temperature relaxation $G_{\rm ie}$ will depend on $T_e$ explicitly through the Fermi-Dirac occupations $p_n=\left(1+e^{-(\mu(T_e)-\epsilon_n)/k_BT_e}\right)^{-1}$ and implicitly through the KS spectrum $\{\Psi_{n},\epsilon_{n}\}$ since the KS Hamiltonian depends on $T_e$ via the density $\rho_e=2\sum_n{p_n|\Psi_n|^2}$.
Given that the determination of the KS spectrum and the calculation of the force matrix elements are computationally demanding, we wonder here about the importance of accounting for the implicit dependence on $T_e$ of these quantities.
We shall refer to a calculation as self-consistent when the KS spectrum is recalculated for each $T_e$ (by increasing the number of bands with the temperature).
By contrast, we refer to a non self-consistent calculation when the equilibrium KS spectrum at $T_e=T_i$ is used and the electronic temperature is only varied in the populations (the number of bands is therefore fixed and given in table~(\ref{Tab:1})).
\myFig{1}{1}{true}{0}{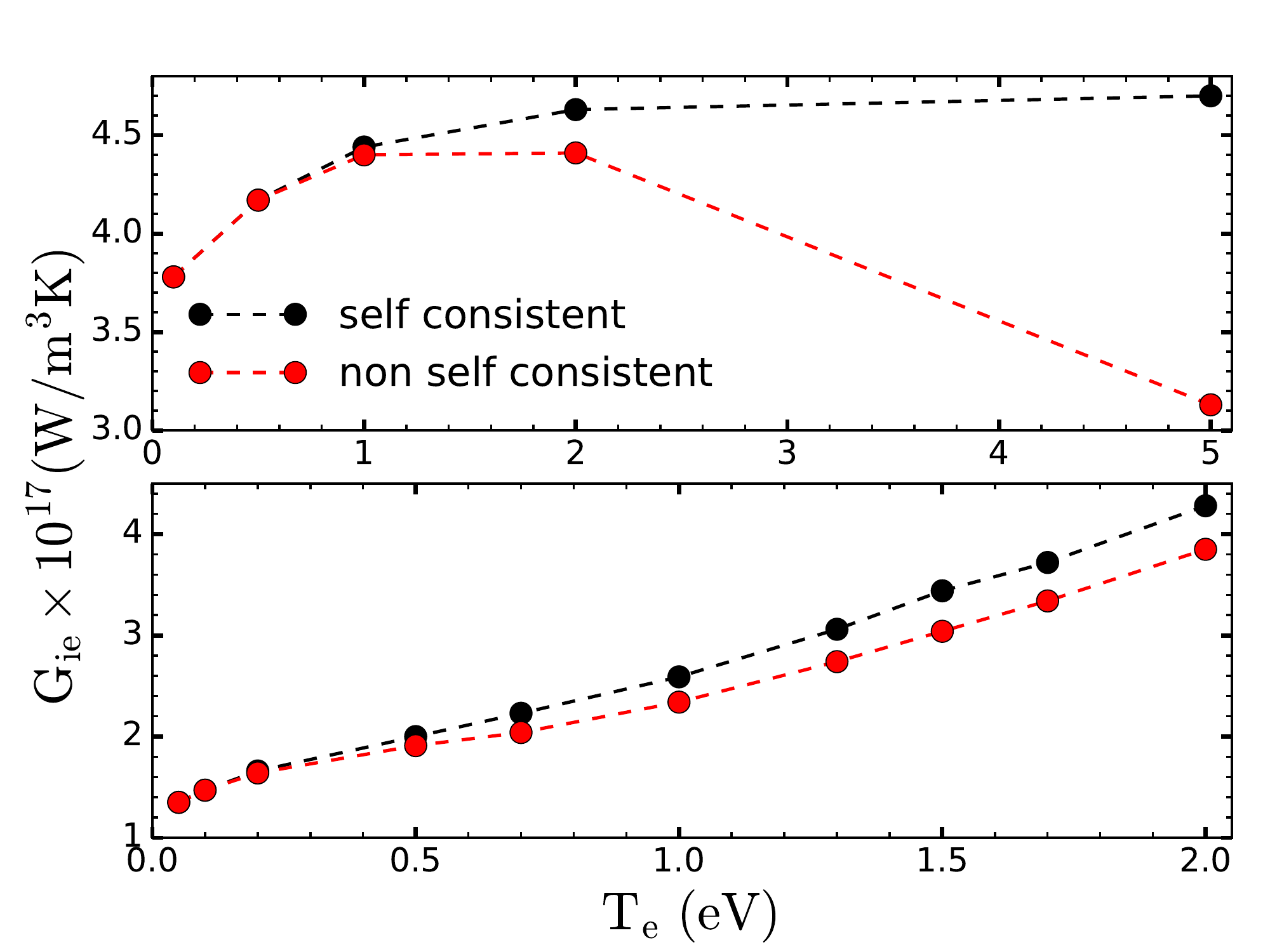}{(Color online) Effect of the self-consistency of the calculation on the temperature relaxation rate for aluminum (\SI{2.35}{g/cm^3}, $T_{\rm i}=$\SI{0.1}{eV}) (upper panel) and copper (lower panel) at \SI{8.02}{g/cm^3}, $T_{\rm i}=$\SI{0.2}{eV}.}{Fig:18}
Figure~(\ref{Fig:18}) shows both self-consistent and non self-consistent calculations of the electron-ion coupling constant $G_{\rm ie}(T_{\rm e},T_{\rm i})$ for aluminum at liquid density $\rho=2.35$ $\rm g/cm^3$ with $0.1\le T_{\rm e}\le 5$ $\rm eV$ and $T_i=0.1$ $\rm eV$ (upper panel), and for copper at liquid density $\rho=8.02$ $\rm g/cm^3$ with $0.2\le T_e\le 2$ $\rm eV$ and $T_i=0.2$ $\rm eV$.
For aluminum, the non self-consistent calculations start to significantly differ from the self-consistent ones beyond $T_e=10 T_i$.
In copper, the calculations differ at a lower temperature $T_e\approx 2 T_i$ but the difference remains of the same magnitude up to $T_e=10 T_i$, the non self-consistent calculation being $\sim 6 \%$ lower between $0.6$ and $2$ $\rm eV$.
\myFig{1}{1}{true}{0}{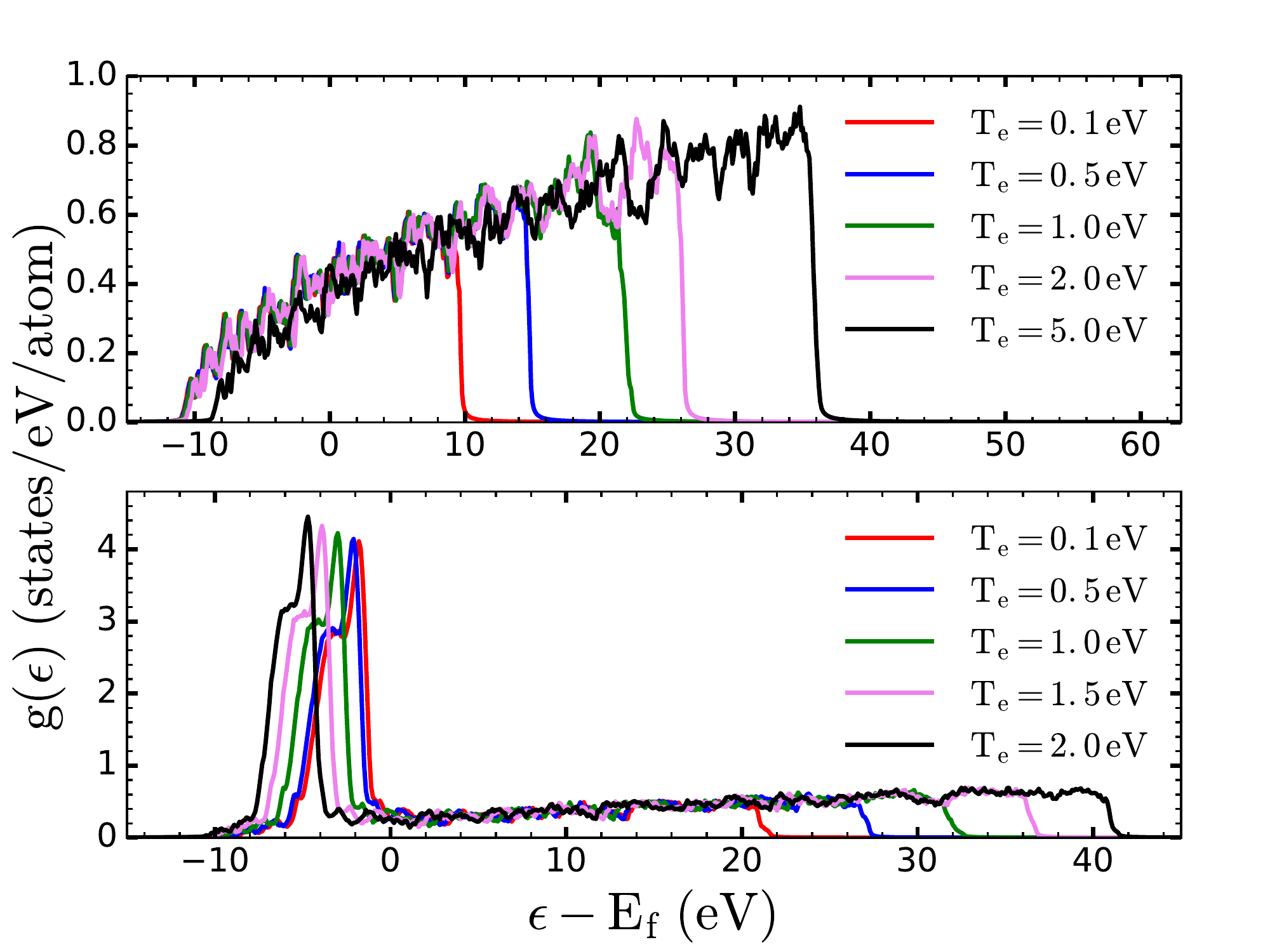}{(Color online) Density of states ($g(\epsilon)$) for aluminum (\SI{2.35}{g/cm^3}, $T_{\rm i}=$\SI{0.1}{eV}) at different $T_e$ (upper panel) and copper (lower panel) at \SI{8.02}{g/cm^3}, $T_{\rm i}=$\SI{0.2}{eV}.}{Fig:24}
These findings can be understood from the dependence on $T_e$ of the electronic density of states $g(\epsilon)$ shown in Fig.~(\ref{Fig:24}).
In the case of aluminum, since $g(\epsilon)$ increases in magnitude with the energy of the state, a non self consistent calculation using the density of states obtained at $T_e=0.1\,eV$ at higher temperatures will become a bad approximation quite soon given that more and more states contributing to the sum (\ref{gIoft}) will be neglected.
In the case of copper instead the number of states increases only slightly with the energy of the state with the consequence that a non self-consistent calculation does not affect so drastically the final $G_{ie}$ value as in the case of aluminum. 

\subsection{Importance of screening} \label{sec:4-3}
\begin{figure*}[t] 
\begin{center} 
\includegraphics[width=\columnwidth]{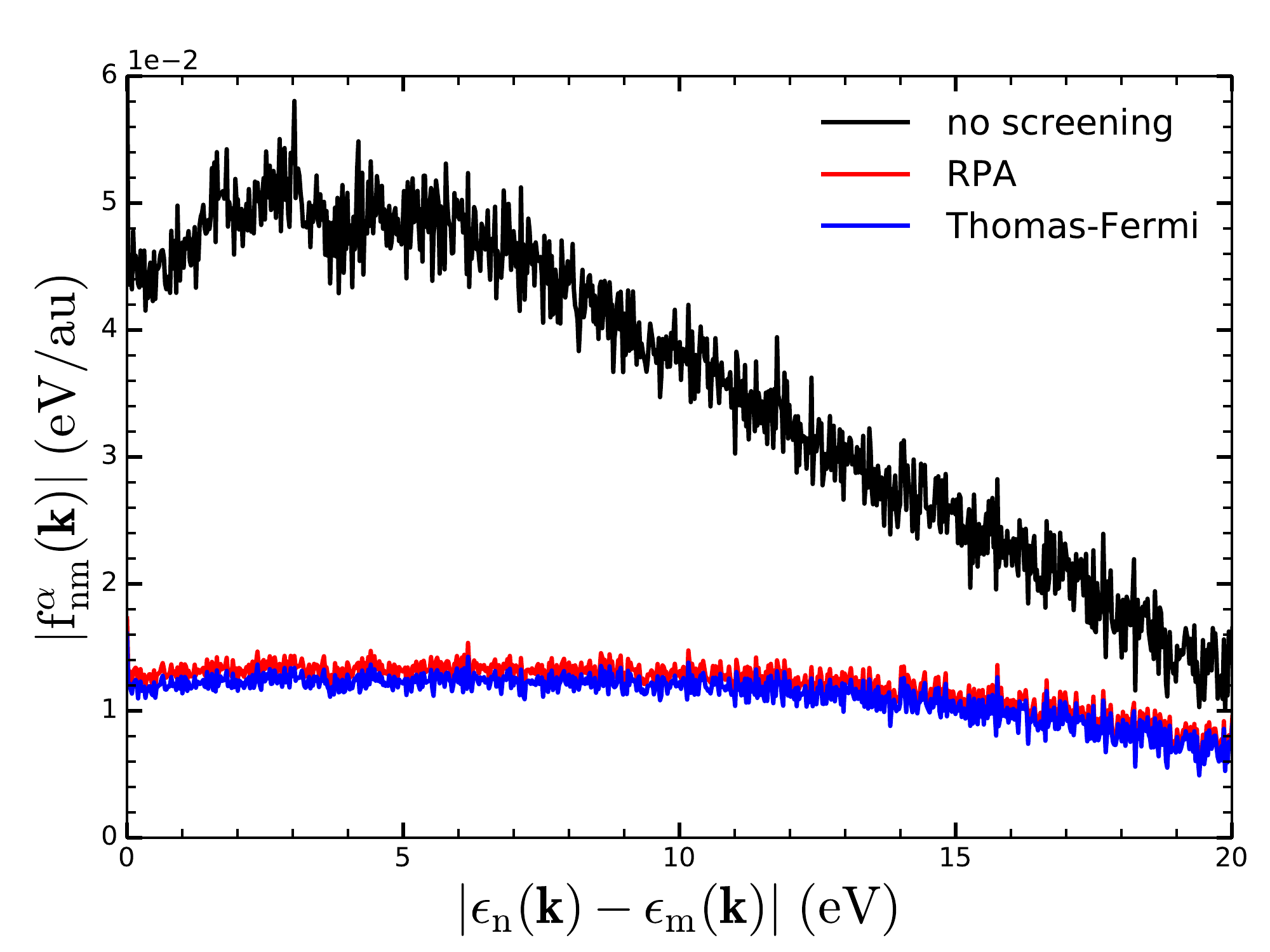}
\includegraphics[width=\columnwidth]{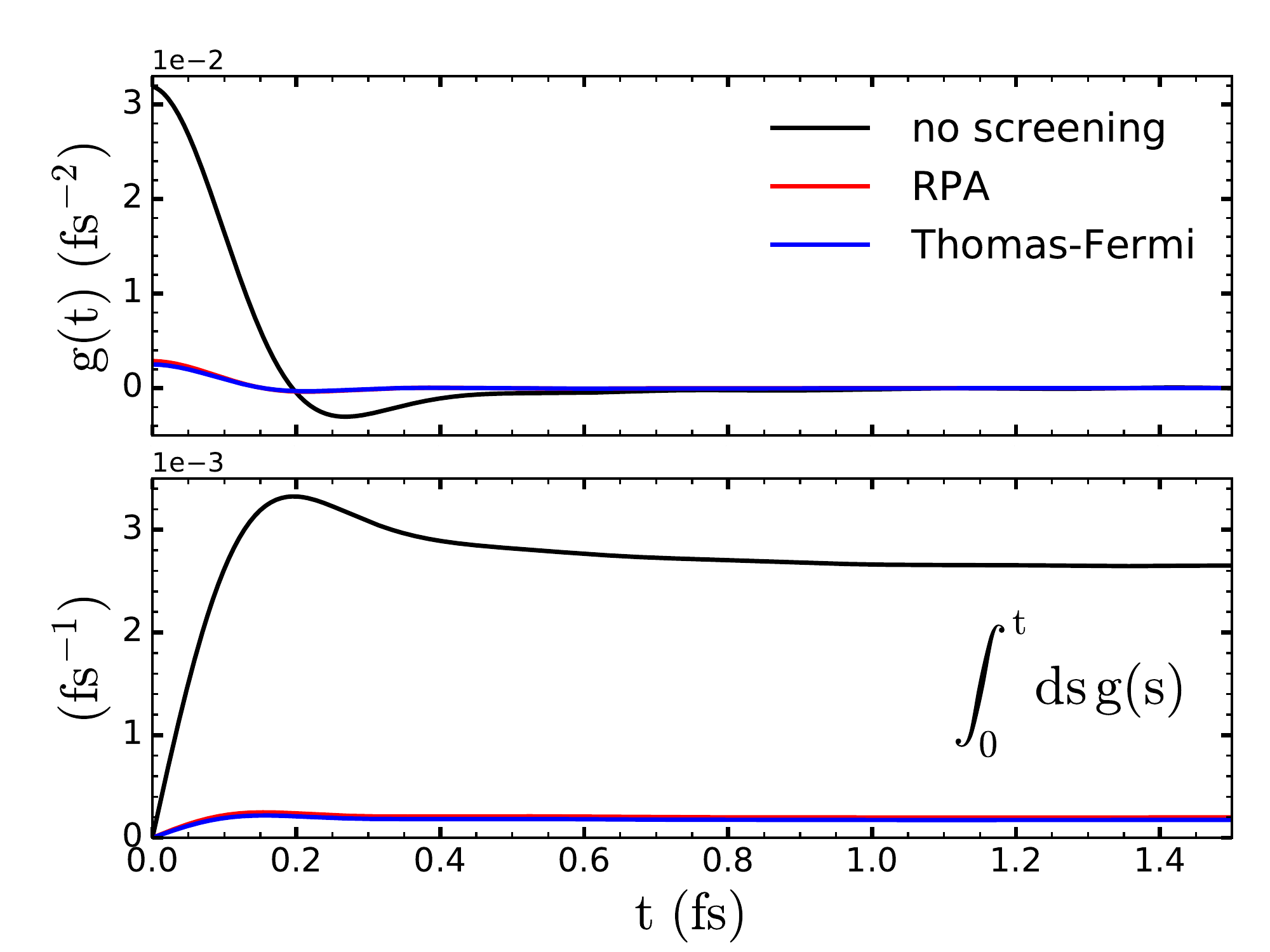}
\caption{(Color online) Left: Dependence of the force matrix elements $f_{nm}^{\mathrm{ax}}(\mathbf{k})$ from the dielectric function. The red curve employs the Lindhard RPA screening, the blue curve uses the Thomas-Fermi screened interaction and the black curve neglects screening effects.\\
Right: Effect of the dielectric function on $g(t)$ (upper panel) and its cumulative sum (lower panel). The colors correspond to the same type of screening shown in the left figure.}\label{Fig:0305}
\end{center}
\end{figure*}
\begin{figure}[h] 
\begin{center} 
\includegraphics[scale=0.35]{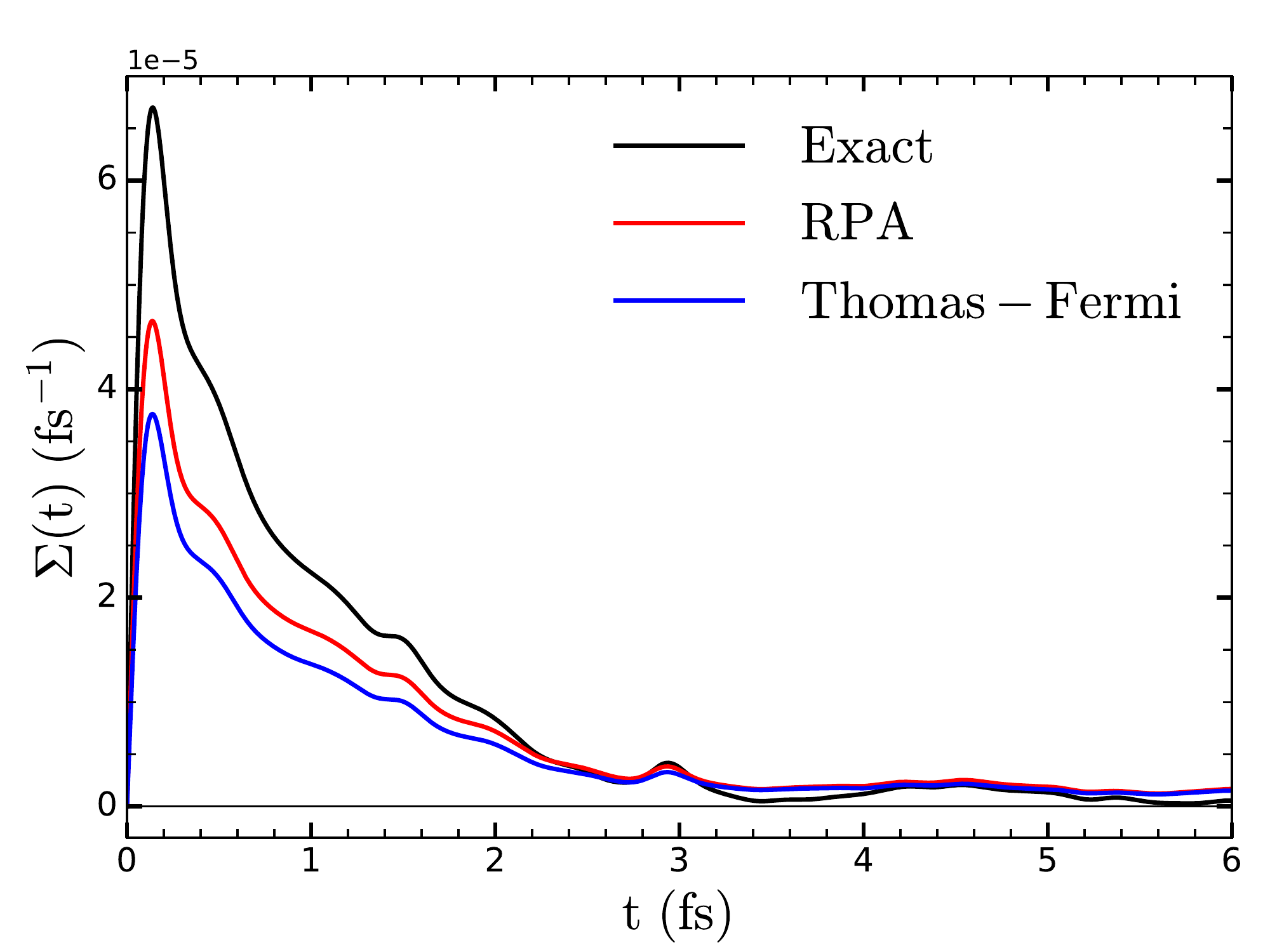}\\
\includegraphics[scale=0.35]{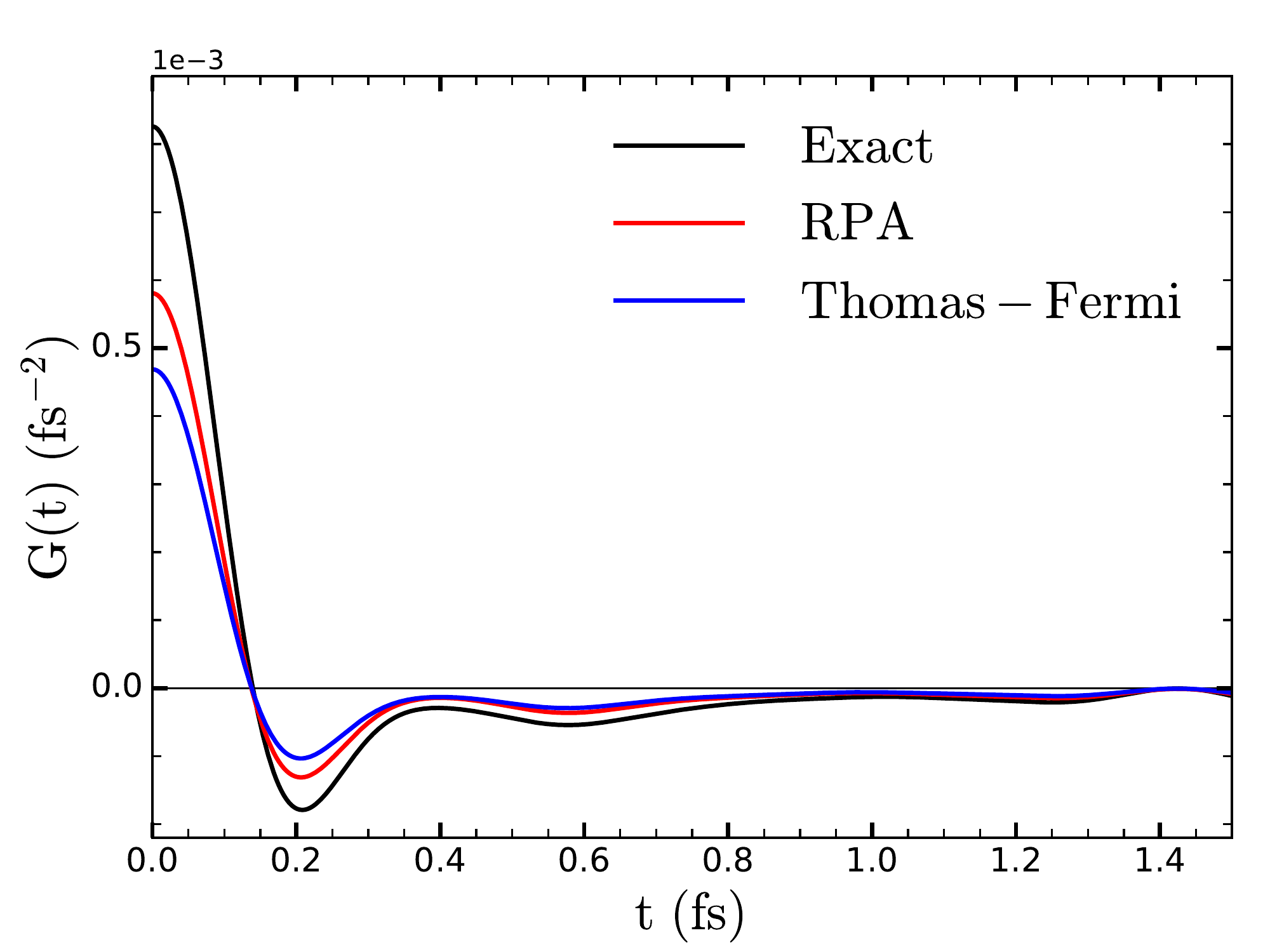}\\
\includegraphics[scale=0.35]{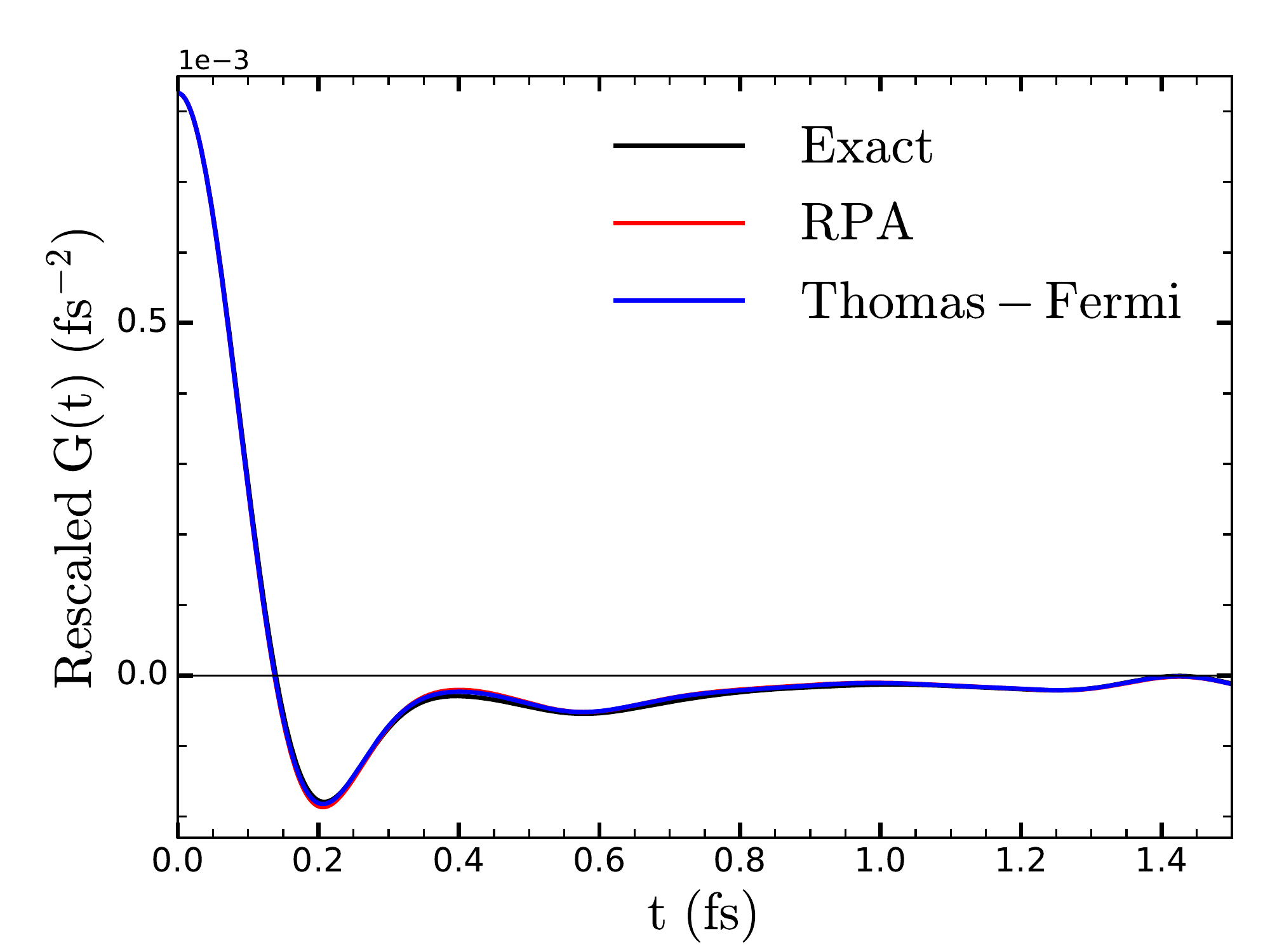}
\caption{(Color online) Effect of the dielectric function on the sum rule (\ref{sum_ab_gamma_ax_by}). The upper and middle panels show $G(t)$, Eq.(\ref{Eq:g-sr2}), and the corresponding cumulative sums $\Sigma(t)=\int_0^t{ds\,G(s)}$ obtained by using $\sum_{\rm a=1}^{N_i}\hat{f}_{\rm ax,L(R)}=\nabla_{\bf r}V_{KS}$ (black lines), the RPA dielectric function (\ref{epsilon_RPA}) (red line) and the Thomas-Fermi approximation (\ref{epsilon_TF}) (blue line). In the bottom panel, the RPA and Thomas-Fermi data shown in the upper panel are re-scaled to match the initial value of the 'exact' case; the resulting three curves are nearly indistinguishable.}\label{Fig:08}
\end{center}
\end{figure}
In Sec.~\ref{sec:2}, we pointed out the importance of accounting for the shielding of the electron-ion interaction due to the other electrons.
Moreover, we suggested to approximate the inhomogeneous dielectric functions $\varepsilon_{L,R}^{[{\bf R}]}$ with that of the electron gas $\varepsilon_{\rm eg}$.
In this section, we illustrate quantitatively the effect due to the screening by using different models for the dielectric function $\varepsilon_{eg}$ in Eq.(\ref{fIeg}).

The left panel of Fig.~(\ref{Fig:0305}) shows the matrix elements $f_{nm}^\alpha({\bf k})$ as a function of the energy differences $\epsilon_n({\bf k})-\epsilon_m({\bf k})$ for liquid aluminum at $\rho=2.7$ $\rm g/cm^3$ and $T=0.1$ $\rm eV$ (for the parameters used in the calculation see table~(\ref{Tab:1})), as obtained for three popular models of the static dielectric function $\varepsilon_{\rm eg}$, including: (1) $\varepsilon_{\rm eg}({\bf k})\simeq 1$, i.e. the screening effect of valence electrons is neglected and, as discussed in Sec.~\ref{sec:2-2}, this choice corresponds to the popular Kubo-Greenwood approximation; (2) the RPA approximation
\be
\varepsilon_{\rm eg}({\bf k})\simeq\varepsilon_{\rm RPA}({\bf k})=1-\frac{4\pi e^2}{k^2}\chi_0(\mathbf{k},\omega=0)\,, \label{epsilon_RPA}
\ee
where $\chi_0$ is the free electron (Lindhard) density-density response function at finite temperature;
(3) the Thomas-Fermi (TF) approximation
\be
\varepsilon_{\rm eg}({\bf k})\simeq\varepsilon_{\rm TF}({\bf k})&=&1-\frac{4\pi e^2}{k^2}\chi_0(\mathbf{k}=0,\omega=0)\nn\\
&=&1+\frac{k_{TF}^2}{k^2}\,, \label{epsilon_TF}
\ee
where $1/k_{TF}$ is the (finite-temperature) Thomas-Fermi screening length, corresponding to the asymptotic limit of $\varepsilon_{\rm eg}$ as $k$ goes to zero.
The effect of electron screening is evident.
The unscreened matrix elements are significantly larger than the screened ones, by a factor $5$ at the lowest energy excitations, which are incidentally the most important ones since low energy transitions dominate the sum (\ref{gIoft}).
The Thomas-Fermi and Lindhard screening give similar results for the matrix elements, supporting the idea that the chief effect of the valence electrons is to shield the Coulomb tail of the bare electron-ion interactions.

The effect on the corresponding correlation functions and their cumulative sums is shown in the right panel of Fig.~\ref{Fig:0305}.
The magnitude of the unscreened correlation function and of the corresponding friction is an order of magnitude larger than the others; in particular we find $\Gamma_{\rm no\,screen}=25.9$ , $\Gamma_{\rm RPA}=1.93$ and $\Gamma_{\rm TF}=1.69$ $\times 10^{-4}$ $\rm fs^{-1}$.
It can be shown that the unscreened calculation actually diverges but at a slow, logarithmic rate; the finitude of our calculation results from the numerical truncation of the KS spectrum at large energies.
The screened calculations differ instead by only $12\%$, a difference that we found for all the elements and conditions that we have considered.

In order to try to quantify the error made in replacing $\varepsilon_{L,R}^{[{\bf R}]}$ by $\varepsilon_{\rm eg}$, we appeal to the following exact sum rule satisfied by the set of friction coefficients $\tilde{\gamma}_{\rm ax,by}^{[{\bf R}]}$
\be
\sum_{{\rm a,b}=1}^{N_i}\tilde{\gamma}_{\rm ax,by}^{[{\bf R}]}=0\,, \label{sum_ab_gamma_ax_by}
\ee
for all directions $x$ and $y$.
As discussed in Ref.~\cite{Daligault_Simoni_2019}, this sum rule is a direct consequence of the following relation between the matrix elements of the screened forces and of the single-particle momentum $\hat{\bf p}$, 
\bes
\be
\sum_a f_{nm}^{ax,L}&=&\langle n|\nabla_x V_{KS}|m\rangle=\frac{1}{i\hbar}\langle n|\hat{p}_x|m\rangle(\epsilon_n-\epsilon_m)\nn\\\\
\sum_a f_{nm}^{ax,R}&=&\langle n|\nabla_x V_{KS}|m\rangle=\frac{1}{i\hbar}\langle n|\hat{p}_x|m\rangle(\epsilon_n-\epsilon_m)\nn\\
\ee
\label{sum_a_fnm_ax_L_R}
\ees
These relations are themselves a consequence of the properties (\ref{sum_rule_L_and_R}) discussed in Sec.~\ref{sec:2-2}.

When approximating the inhomogeneous dielectric functions $\varepsilon_{L,R}^{[{\bf R}]}$ with $\varepsilon_{\rm RPA}$ or $\varepsilon_{\rm TF}$ in the expression of the screened forces, the properties (\ref{sum_a_fnm_ax_L_R}) and, in turn, the sum rules (\ref{sum_ab_gamma_ax_by}) are not expected to hold perfectly.
In order to investigate the effect of these approximations on the sum rule, we introduce the following quantity
\begin{align}\label{Eq:g-sr2}
&G(t)=\nonumber\\
&\frac{\hbar}{3MN_i}\sum_{x=1}^3\sum_{n\neq m}\sum_{{\bf k}\in{\rm IBZ}}W_{\bf k}p_n({\bf k})\big[1-p_m({\bf k})\big]\big|F_{nm}^x({\bf k})\big|^2\times\nonumber\\
&\times\frac{e^{\beta_e(\epsilon_n({\bf k})-\epsilon_m({\bf k}))}-1}{\epsilon_n({\bf k})-\epsilon_m({\bf k})}\cos\Big(\frac{\epsilon_n({\bf k})-\epsilon_m({\bf k})}{\hbar}t\Big)
\end{align}
defined such that its cumulative sum $\Sigma(t)=\int_0^t{ds\,G(s)}$ satisfies 
\be
\lim_{t\to\infty}\Sigma(t)=\sum_{{\rm a,b}=1}^{N_i}{\sum_{x=1}^3{\tilde{\gamma}_{\rm ax,bx}}} = 0\,.
\ee
In Eq.~(\ref{Eq:g-sr2}) we also introduced the total force matrix elements $F_{nm}^x$
\begin{align}
F_{nm}^x({\bf k})=\sum_{{\rm a}=1}^{N_i}{f_{nm}^{\rm ax}({\bf k})}=\mel{\Psi_{n{\bf k}}}{\sum_{\rm a=1}^{N_i}\hat{f}_{\rm ax,L(R)}}{\Psi_{m{\bf k}}}\,.\label{hatfLR}
\end{align}
Figure~(\ref{Fig:08}) shows $G(t)$ (middle panel) and $\Sigma(t)$ (upper panel) obtained by using the usual three different models for the screening of the electron-ion interaction in the case of liquid aluminum at $2.7$ $\rm g/cm^3$ and $T=0.1$ $\rm eV$.
We first check numerically the validity of the sum rule (\ref{sum_ab_gamma_ax_by}), the black lines show results obtained by using relation (\ref{sum_a_fnm_ax_L_R}), i.e. setting $\sum_{\rm a=1}^{N_i}\hat{f}_{\rm ax,L(R)}=\nabla_{\bf r}V_{KS}$ in Eq.~(\ref{hatfLR}); we denote by $G_{exact}$ the resulting $G$.
In practice, due to the discrete character of calculations, the correlation function and its sum do not vanish perfectly at late times.
Yet, the sum rule (\ref{sum_ab_gamma_ax_by}) is satisfied to very good accuracy: the cumulative sum $\Sigma(t)$ converges towards zero at large times.
We remark that the correlation time scale of $G(t)$ is $\sim 3-4$ $\rm fs$ and it is larger than that for $g(t)$ discussed earlier, which is $\sim 1$ fs ( e.g., see Fig.~(\ref{Fig:02})).
We believe that this difference is indicative of the different physical nature of these quantities: $g(t)$ is related to energy exchanges while $G(t)$ is related to momentum exchange.
A detailed study is beyond the scope of this work and we hope to return to this effect in a future work.

Secondly, we consider the results obtained when using either the RPA (red lines) or the Thomas-Fermi (blue lines) dielectric functions to calculate $\hat{f}_{ax,L(R)}$ in Eq.(\ref{hatfLR}).
In both cases, the sum rule (\ref{sum_ab_gamma_ax_by}) is well satisfied, i.e. $\Sigma(t)$ tends toward zero a large times.
Errors caused by the approximation are clearly seen in the detailed temporal variations of $G(t)$, with the RPA case being closer to the exact one.
In fact, the main changes are less in the temporal behavior than in the initial value $G(t=0)$.
This is shown in the bottom panel of Fig.~(\ref{Fig:08}) through the rescaled quantities
\be \label{Eq:exact_SR}
G(t)=G(t)\frac{G_{exact}(t=0)}{G(t=0)}\,,
\ee
the rescaled $G(t)$ computed by using different models of screening overlap nearly perfectly over the entire time scale.
This suggests that the time correlation function could be renormalized by using an exact sum rule for $G(t=0)$. By starting from Eq.~(\ref{Eq:g-sr2}) and after some manipulation we can show that
\begin{equation}
G_{exact}(t=0) = \frac{\hbar}{3MN_i}\int_\Omega d{\bf r}\,\rho_e({\bf r})\nabla^2_{\bf r}V_{\rm KS}({\bf r}),
\end{equation}
that used into Eq.~(\ref{Eq:exact_SR}) provides a more accurate rescaling of $G(t)$.
\begin{figure}[t] 
\begin{center} 
\includegraphics[width=\columnwidth]{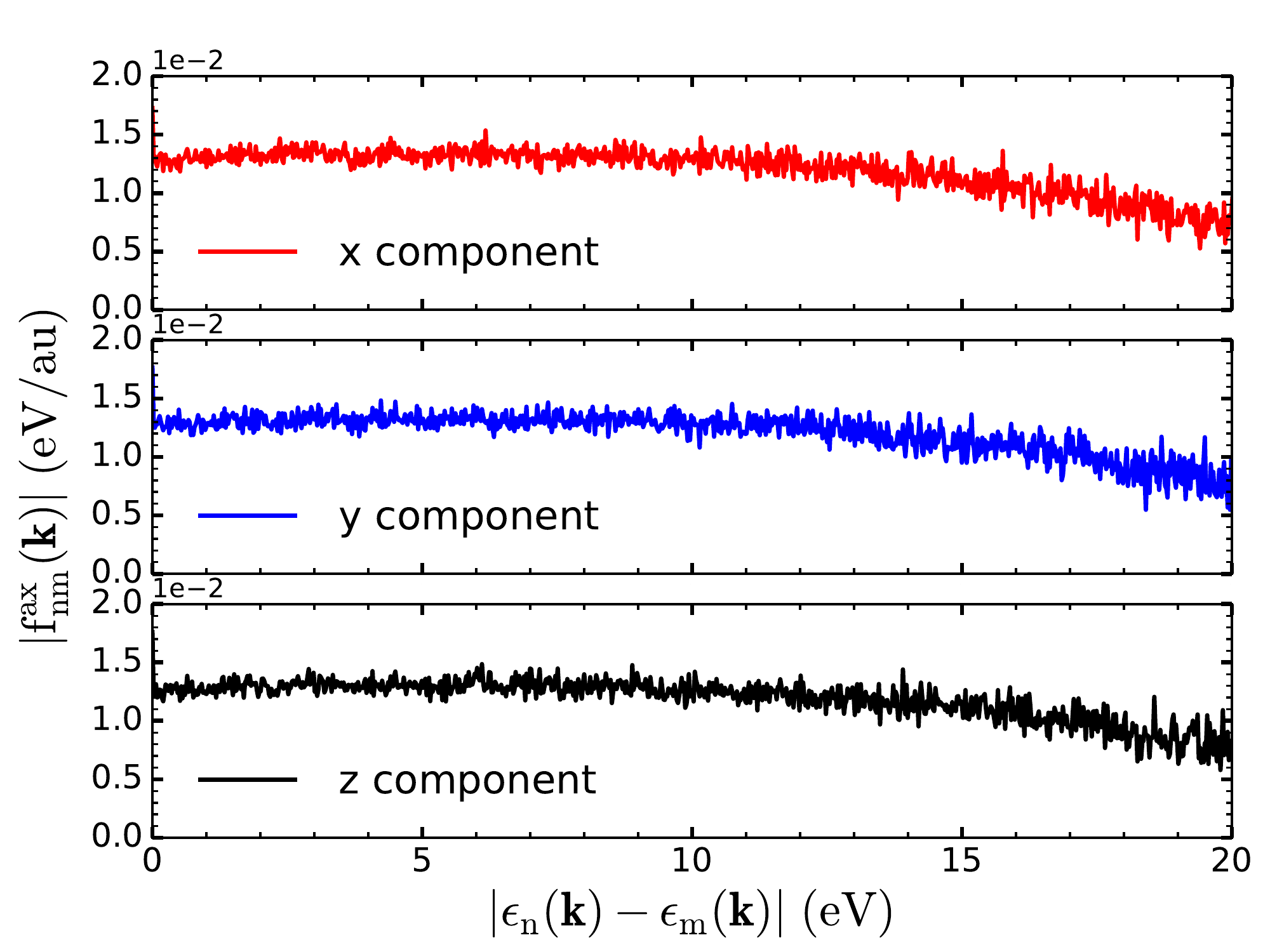}
\includegraphics[width=\columnwidth]{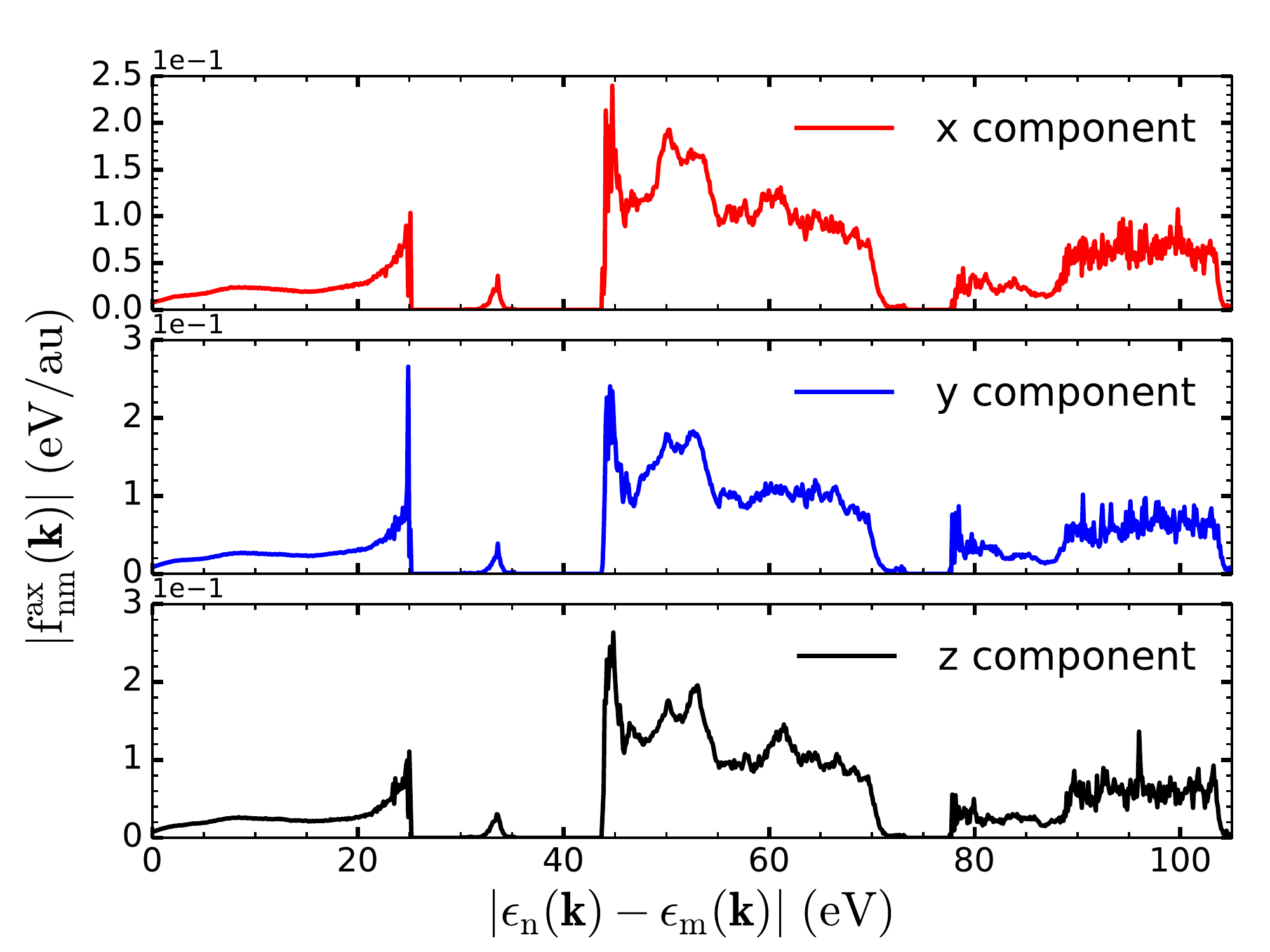}
\caption{(Color online) (top figure) Variation of the matrix elements $f_{nm}^{\rm ax}(\mathbf{k})$ with the excitation energies $|\epsilon_n({\bf k})-\epsilon_m({\bf k})|$  for a randomly selected ion ${\rm a}$ and along the three spatial directions in liquid aluminum. For convenience, the figure shows the absolute value of the matrix elements.\\
(bottom figure) Same as above for iron at melting temperature $T=1811\,\rm{K}$ and \SI{7.874}{gr/cm^3} (see table~(\ref{Tab:1}) for the parameters used in the calculation).} \label{Fig:04} 
\end{center}
\end{figure}
\subsection{Analysis of screened electron-force matrix elements}

In this section, we discuss the dependence of the force matrix elements,
$f_{nm}^{\mathrm{a}x}(\mathbf{k})$, Eq.~(\ref{Eq:matr_el_feg}), on the transition energies,
$\epsilon_n(\mathbf{k})-\epsilon_m(\mathbf{k})$.
Figure (\ref{Fig:04}) shows  $f_{nm}^{\mathrm{a}x}(\mathbf{k})$
versus $\epsilon_n(\mathbf{k})-\epsilon_m(\mathbf{k})$ for two distinct cases, name solid density aluminum at
$T=1160.4$ $\rm K$ (top panel) and solid density iron at $1811$ $\rm K$
(bootom panel).
In both cases, $|f_{nm}^{\rm ax}({\bf k})|$ is computed for a single atom $\mathrm{a}$ along the three different directions $x, y, z$.

In both cases, the force matrix elements are almost identical in all the three directions, this isotropic nature of the matrix elements suggests that the three different $x$ directions in the sum (\ref{gIoft}) contribute equally to the final correlation function $g_{\rm a}(t)$.
However, the variations with
$\epsilon_n(\mathbf{k})-\epsilon_m(\mathbf{k})$ differ widely.
For aluminum, the matrix elements appear to
be approximately constant up to approximately \SI{13}{eV}).
On the contrary, for iron, we note significant variations with the
transition energies.
It is interesting to think about these findings in the light of the
reference model due to Wang et al. \cite{Wang1994} and popularized by Lin et
al. \cite{Lin_2008} for
the temperature relxation rate in hot solids due to electron-phonon
scattering.
As discussed in Ref.~\cite{Daligault_Simoni_2019}, this simplified model also
results from our theory (\ref{Eq:Temp_rel_rate}), which potentially extends the
original model to liquid metals and plasmas.
As described in \cite{Lin_2008} and \cite{Daligault_Simoni_2019}, in this
simplified model,  the detailed of the electron-phonon or electron-ion
matrix elements are factorized out of the sum of electronic
transitions and are lumped together into a single prefactor to be
determined.
Clearly, such a factorization is justified for the aluminum system
shown in Fig.~(\ref{Fig:04}) as the low-energy matrix elements, which
contribute the most to Eq.(\ref{Eq:friction_tensor_KS_DFT}), are
nearly equal to one another.
The suitability of the approximation is much more questionable in the case of iron.

\subsection{Dependence on the exchange-correlation energy.} \label{sec:4-6}

The exchange-correlation effects affect the friction coefficients $\gamma_{\alpha\beta}$ in Eq.(\ref{Eq:MB_friction_tensor}) in two main ways: through the exchange-correlation potential $v_{\rm xc}$ in the KS Hamiltonian, which affect the spectrum, and through the exchange-correlation kernel $f_{\rm xc}$, which enters in both the dielectric functions $\varepsilon_{\rm L,R}$ and the correction term $\delta\tilde{\gamma}_{\alpha\beta}$, Eq.(\ref{Eq:MBcorrection}).
As discussed in Sec.~\ref{sec:2-2}, this work neglects the effects of $f_{xc}$.
In this section, we only consider the dependence of the temperature relaxation rate on the exchange-correlation energy potential $v_{\rm xc}$.
\myFig{1}{1}{true}{0}{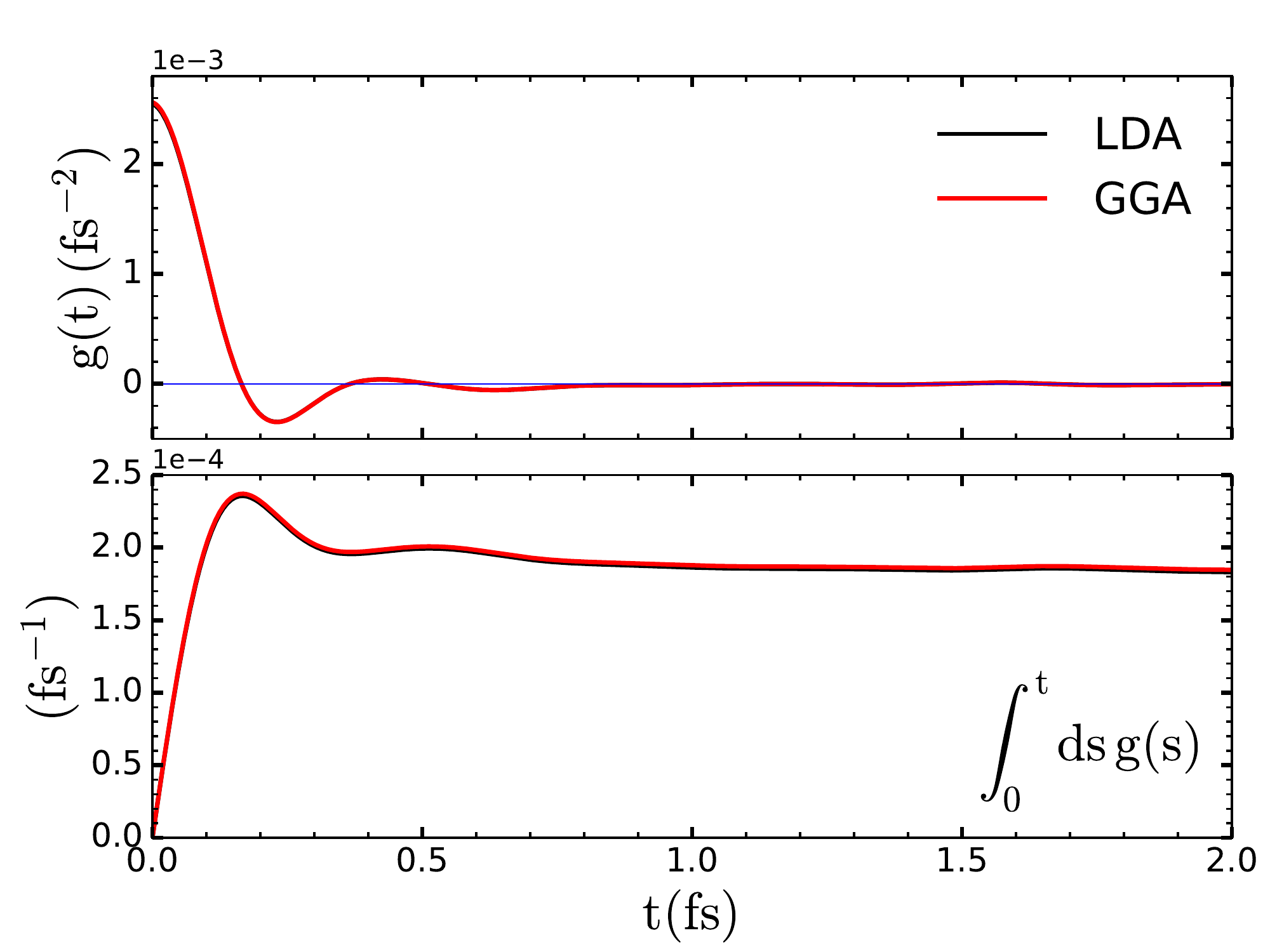}{(Color online) Correlation function $g(t)$ (upper panel) and its cumulative sum (lower panel) for Aluminum (\SI{2.35}{g/cm^3}, $T_{\rm i,e}=$\SI{0.1}{eV}), obtained using two popular approximations for the exchange-correlation functional (see table~(\ref{Tab:1}) for the calculation's details).}{Fig:17}
For illustration, Fig.~(\ref{Fig:17}) shows the correlation function $g(t)$ and its cumulative sum $\int_0^t{ds\,g(s)}$ obtained with two standard approximations for $v_{\rm xc}[\rho_e]$, namely the local-density approximation (LDA) of Perdew and Zunger~\cite{PZ}, and the generalized-gradient approximation (GGA) approximation of Perdew, Burke and Ernzerhof~\cite{PBE}, for liquid aluminum at melt density $2.35$ $\rm g.cm^{-3}$ and $T=0.1$ $\rm eV$.
Both functionals are strictly speaking zero-temperature approximations, which is reasonable here since $T_e/T_F\sim 0.01$. The exchange-correlation functional has a negligible effect, a result which we found also for the other calculations we have done so far.
This is supported by the density of states obtained from the two calculations. 
As shown in Fig.~(\ref{Fig:23}), both exchange-correlation functionals generate essentially the same $g(\epsilon)$.
Overall, in all our present calculations, the frictions coefficients, and more precisely the force matric elements, are generally more sensitive on the choice  of the screening model than on $v_{\rm xc}$.
\myFig{1}{1}{true}{0}{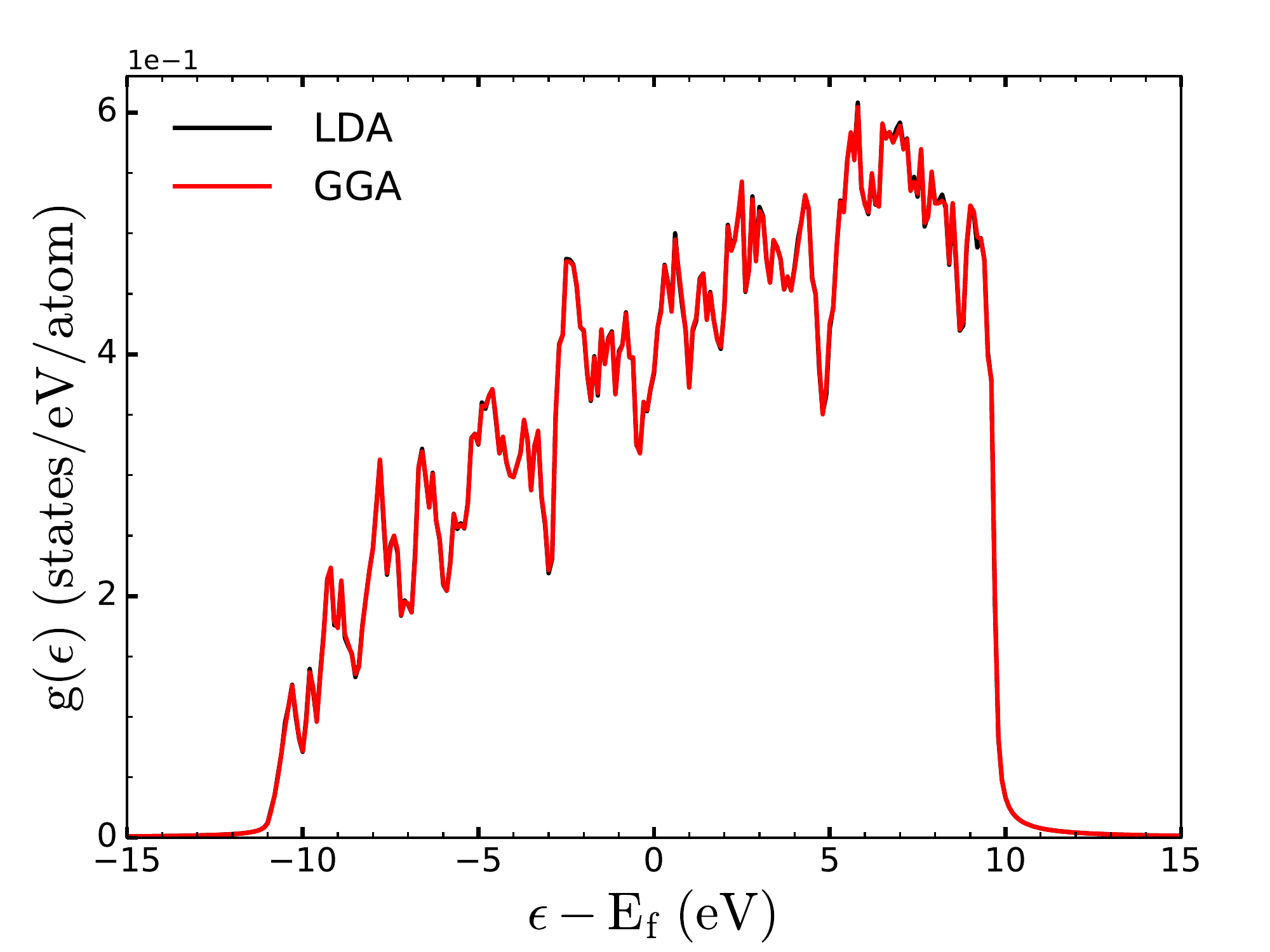}{(Color online) Comparison between the density of states, $g(\epsilon)$, obtained from the same two calculations of Fig.~(\ref{Fig:17}).}{Fig:23}
\section{Conclusions}\label{sec:5}
In this paper we have shown how to compute the temperature relaxation
rate $G_{\rm ei}(T_{\rm e},T_{\rm i})$ and frictions coefficients of
plasmas and liquid metals by means of QMD sumulations.
Specific calculations of $G_{\rm ei}(T_{\rm e},T_{\rm i})$ for
different materials were presented in a previous paper \cite{Simoni_2019} and others will be published elsewhere in the future.

The practical calculation presents difficulties that are unlike those
encountered with the Kubo formulas for the electrical and thermal conductivities.
In particular, the widely used Kubo-Greenwood approximation is
inapplicable here and the screening of electron-ion interactions by
all electrons must be carefully taken into account.
We have discussed the approximations we applied
to deal with these complications in pseudopotential calculations based
on either local of plane-augmented wave potentials.
We have presented a detailed parametric and convergence study with the numerical param- eters, including the system size, the number of bands and k-points, and the physical approximations for the dielectric function and the exchange-correlation energy.
Future useful extensions of this work should include a better
description of screening effects, e.g., using a self-consistent
calculation of the dielectric functions (\ref{Eq:12a}-\ref{Eq:12b}),
and the inclusion of dynamical many-body correlation effects modeled
by the exchange-correlation kernel $f_{xc}$.

\acknowledgments

This work was supported by the US Department of Energy through the Los Alamos National Laboratory through the LDRD Grant No.20170490ER and the Center of Non-Linear Studies (CNLS).
Los Alamos National Laboratory is operated by Triad National Security, LLC, for the National Nuclear Security Administration of U.S. Department of Energy (Contract No. 89233218CNA000001).


\begin{appendix}


\section{Lehmann representation for the Kohn-Sham friction tensor, Eq.~(\ref{Eq:14b})}\label{A}
The Eq.~(\ref{Eq:friction_tensor_KS_DFT}) in presence of periodic boundary conditions may
be straightforwardly rewritten as
\begin{align}
  \tilde{\gamma}_{\alpha\beta}^{[{\bf R}]} &= -\frac{\pi\hbar}{MN_{\bf
      k}}\sum_{n\neq m}\sum_{{\bf k},{\bf k'}\in{\rm
      BZ}}\frac{p_n({\bf k})-p_m({\bf k'})}{\epsilon_n({\bf
      k})-\epsilon_m({\bf k'})}f_{n{\bf k},m{\bf
      k'}}^{\alpha,{\rm L}}\times\nonumber\\
&\times f_{m{\bf k'},n{\bf k}}^{\beta,{\rm
      R}}\delta(\epsilon_n({\bf k})-\epsilon_m({\bf k'})),
\end{align}
For both ECE and PAW pseudopotentials, the force matrix elements satisfies (see next appendix)
\begin{equation}
  f_{n{\bf k},m{\bf k'}}^{\alpha,{\rm L(R)}} \simeq
  \delta_{{\bf k},{\bf k'}} f_{nm}^\alpha({\bf k}),
\end{equation}
where we neglected the effect of the inhomogeneous ionic background on the dielectric function, leading to
\begin{align}
  \tilde{\gamma}_{\alpha\beta}^{[{\bf R}]} &=
  -\frac{\pi\hbar}{M}\sum_{n\neq m}\sum_{{\bf k}\in{\rm IBZ}}W_{\bf k}\frac{p_n({\bf
      k})-p_m({\bf k})}{\epsilon_n({\bf k})-\epsilon_m({\bf
      k})}f_{nm}^\alpha({\bf k})f_{mn}^\beta({\bf
    k})\nonumber\\
&\times\delta(\epsilon_n({\bf k})-\epsilon_m({\bf k})),
\end{align}
where the sum is done only over the {\bf k}-points of the irreducible
Brillouin zone weighted with the factor $W_{\bf k}$.
From the previous expression we may reduce the friction
tensor to three different types of contributions
\begin{equation}
  \tilde{\gamma}_{\alpha\beta}^{[{\bf R}]} =
  \tilde{\gamma}_{\alpha\beta}^{{\rm v}\rightarrow{\rm v}[{\bf R}]} +
  \tilde{\gamma}_{\alpha\beta}^{{\rm c}\rightarrow{\rm v}[{\bf R}]} +
  \tilde{\gamma}_{\alpha\beta}^{{\rm c}\rightarrow{\rm c}[{\bf R}]},
\end{equation}
these correspond to electronic transitions between respectively valence electrons, valence and core electrons and only core electrons. In particular the last term,
$\tilde{\gamma}_{\alpha\beta}^{{\rm c}\rightarrow{\rm c}[{\bf R}]}$,
is exactly zero given that core states are always fully
occupied ($p_n({\bf k})=p_m({\bf k})$). The term
$\tilde{\gamma}_{\alpha\beta}^{{\rm c}\rightarrow{\rm v}[{\bf R}]}$
can be neglected by assuming that the electron-ion scattering
potential is weak enough to do not induce transitions between core and valence states. We are then left with only the first term and the following final expression for the friction tensor
\begin{align}
  \tilde{\gamma}_{\alpha\beta}^{[{\bf R}]} &=
  -\frac{\pi\hbar}{M}\sum_{n\neq m}^{\rm val}\sum_{{\bf k}\in{\rm IBZ}}W_{\bf k}\frac{p_n({\bf
      k})-p_m({\bf k})}{\epsilon_n({\bf k})-\epsilon_m({\bf
      k})}f_{nm}^\alpha({\bf k})f_{mn}^\beta({\bf
    k})\nonumber\\
&\times\delta(\epsilon_n({\bf k})-\epsilon_m({\bf k})),
\end{align}
that gives Eq.~(\ref{Eq:14b}).

\section{Proof of Eq.~(\ref{Eq:matr_el_feg}) for the force matrix elements}\label{F}
Under the assumption of weak inhomogeneity of the ionic background we can write the following expression
\begin{align}\label{G:0}
  f_{n\mathbf{k},m\mathbf{k'}}^{\alpha,\mathrm{L(R)}}\simeq f_{n\mathbf{k},m\mathbf{k'}}^\alpha &= \mel*{\tilde{\Psi}_{n{\bf k}}}{\hat{f}_\alpha^{\rm eg}}{\tilde{\Psi}_{m{\bf k'}}}\nonumber\\
  &=\int_Vd{\bf r}\,\tilde{\Psi}_{n\mathbf{k}}^*(\mathbf{r})f_\alpha^{\rm eg}(\mathbf{r})\tilde{\Psi}_{m\mathbf{k'}}(\mathbf{r}),
\end{align}
where $\tilde{\Psi}_{n\mathbf{k}}(\mathbf{r})$ is the KS wave
function and $V={\cal N}_{\bf k}\Omega$.
\begin{equation}\label{G:2}
  f_{n\mathbf{k},m\mathbf{k'}}^\alpha =
  \frac{1}{V}\int_V\,d{\bf r}\,e^{-i(\mathbf{k}-\mathbf{k'})\cdot\mathbf{r}}u_{n\mathbf{k}}^*(\mathbf{r})f_\alpha^{\rm eg}(\mathbf{r})u_{m\mathbf{k'}}(\mathbf{r})\,.
\end{equation}
The term $f(\mathbf{r})=u_{n\mathbf{k}}^*(\mathbf{r})f_\alpha^{\rm eg}(\mathbf{r})u_{m\mathbf{k'}}(\mathbf{r})$, that has the same periodicity of the Bravais lattice,
$f(\mathbf{r}+n_x\boldsymbol{a}_x)=f(\mathbf{r})$, we can expanded it in Fourier series over the reciprocal lattice vectors
\begin{equation}
  f(\mathbf{r}) = \frac{1}{\Omega}\sum_\mathbf{G}\tilde{f}(\mathbf{G})e^{i\mathbf{G}\cdot\mathbf{r}},
\end{equation}
where $\{\mathbf{G}\}$ defines a set of reciprocal space vectors, i.e. ${\bf G}=\sum_x m_x{\bf b}_x$, for the periodic Bravais lattice of primitive vectors $\{\boldsymbol{a}_x\}$. By using this expansion into (\ref{G:2})
\begin{align}\label{G:1}
  f_{n\mathbf{k},m\mathbf{k'}}^\alpha &=
  \frac{1}{\Omega}\sum_\mathbf{G}\tilde{f}(\mathbf{G})\cdot\frac{1}{V}\int_Vd{\bf r}\,
  e^{-i(\mathbf{k}-\mathbf{k'}-\mathbf{G})\cdot\mathbf{r}}\nonumber\\
  &= \delta_{{\bf k},{\bf k'}}\cdot\frac{1}{\Omega}\int_\Omega d{\bf r}\,u_{n{\bf k}}^*({\bf
    r})f_\alpha^{\rm eg}({\bf r})u_{m{\bf
      k'}}({\bf r})\nonumber\\
  &= \delta_{{\bf k},{\bf k'}}f_{nm}^\alpha({\bf k}),
\end{align}
where we have used the following result
\begin{align}
  \frac{1}{V}\int_Vd{\bf r}
  e^{-i(\mathbf{k}-\mathbf{k'}-\mathbf{G})\cdot\mathbf{r}} &=
  \prod_{x=1}^3\int_0^L\frac{dy}{L}\,e^{-\frac{2\pi i}{L}(n_x-n'_x-N_x m_x)y}\nonumber\\
  &= \frac{V}{(2\pi)^3}\prod_{x=1}^3\delta_{n_x,n'_x+N_x m_x}\nonumber\\
  &= \delta_{\mathbf{k},\mathbf{k'}} \delta_{\mathbf{G},{\bf 0}}.
\end{align}
$N_x$ was defined in Sec.~(\ref{sec:PW}) as the number of unit cells along the $x$ direction of total length $L=N_x|{\bf a}_x|$ and in the
last step we have used the fact that $0 \le n_x,n'_x < N_x$.

\section{Quick review of the PAW formalism} \label{appendix_definition_PAW}
In Sec.~(\ref{subsub:PAW}) we have briefly introduced the fundamental equations of the PAW method, here we look in more detail at the PAW all-electron wave functions and show how to build it. In (\ref{H1}) we explain how to compute the screened electron-ion interaction in the core region and in (\ref{H2}) we show how to generalize the calculation of the force matrix elements to the PAW formalism by proving Eq.~(\ref{Eq:14}).

The most general expression for the all-electron PAW wave functions is
\begin{align}
  \ket{\Psi_{n{\bf k}}} &= \hat{\boldsymbol{\tau}}\ket*{\tilde{\Psi}_{n{\bf k}}}\nonumber\\
  &= \ket*{\tilde{\Psi}_{n{\bf k}}} +\sum_{\rm a=1}^{N_i}\sum_i(\ket{\phi_{{\rm a}i}}-\ket*{\tilde{\phi}_{{\rm a}i}})\ip*{\tilde{p}_{{\rm a}i}}{\tilde{\Psi}_{n{\bf k}}}
\end{align}
where $\ket*{\tilde{\Psi}_{n{\bf k}}}$ is the smooth function solution of the modified KS equations $\hat{\boldsymbol{\tau}}^\dagger\hat{h}_{KS}\hat{\boldsymbol{\tau}}\ket*{\tilde{\Psi}_{n{\bf k}}}=\epsilon_{n{\bf k}}\hat{\boldsymbol{\tau}}^\dagger\hat{\boldsymbol{\tau}}\ket*{\tilde{\Psi}_{n{\bf k}}}$. The second term on the right hand side corresponds instead to an expansion over the atomic wave functions centered around the different nuclei. It allows to correctly reconstruct the nodal structure of the all-electron wave function $\ket{\Psi_{n{\bf k}}}$ inside the core regions.

The symbol $i$ is used here to label the set of quantum numbers for the atomic functions $\ket{\phi_{{\rm a}i}}$ and $\ket*{\tilde{\phi}_{{\rm a}i}}$. These wave functions are the eigenstates respectively of the isolated KS-DFT all-electron atom
\begin{align}
&\bigg[-\frac{\hbar^2}{2m}\nabla_{\bf r}^2 + v_{\rm KS}^{\rm a}(r)\bigg]\phi_{{\rm a}i}({\bf r})=\epsilon^{\rm a}_i\phi_{{\rm a}i}({\bf r}),\nonumber\\
&v_{\rm KS}^{\rm a}(r) = -\frac{Z_{\rm a} e^2}{r} + v_{\rm Hxc}[\rho_e^{\rm a}](r),\,\,\rho_e^{\rm a}(r)=\rho_v^{\rm a}(r) + \rho_c^{\rm a}(r), 
\end{align}
where $Z_{\rm a}$ is the atomic number of the element and $\rho_e^{\rm a}(r)$ is the total electron density of atom ${\rm a}$ and of the isolated KS-DFT pseudo-atom that accounts only for the valence electrons through an effective pseudopotential $v_{\rm ps}^{\rm a}(r)$ that coincides with the all-electron potential outside the atomic core radius 
\begin{align}
&\bigg[-\frac{\hbar^2}{2m}\nabla_{\bf r}^2 + \tilde{v}_{\rm KS}^{\rm a}(r)\bigg]\tilde{\phi}_{{\rm a}i}({\bf r})=\epsilon^{\rm a}_i\tilde{\phi}_{{\rm a}i}({\bf r}),\nonumber\\
&\tilde{v}_{\rm KS}^{\rm a}(r) = v_{\rm ps}^{\rm a}(r) + v_{\rm Hxc}[\rho_v^{\rm a}](r). 
\end{align}
The KS potential of both the systems is spherically symmetric and therefore the principal quantum number $n$, the orbital numbers $l=0,\ldots,n-1$ and the magnetic numbers $m=-l,\ldots,l$ define a good set $i=\{n, l, m\}$ of quantum numbers (the system is assumed to be spin unpolarized, as always throughout the paper, therefore the spin is neglected). 

We may rewrite the mapping $\hat{\boldsymbol{\tau}}$ between the all-electron and the smooth wave functions as follows
\begin{equation}
\hat{\boldsymbol{\tau}}=\hat{1}+\sum_{\rm a=1}^N\hat{\boldsymbol{\tau}}_{\rm a},
\end{equation}
this allows to simplify the notation since we can write the action of the operators $\hat{\boldsymbol{\tau}}_{\rm a}$ on the smooth functions as
\begin{equation}
  \hat{\boldsymbol{\tau}}_{\rm a}\ket*{\tilde{\Psi}_{n{\bf k}}} =
  \ket{(\phi_{n{\bf k}})_{\rm a}} - \ket*{(\tilde{\phi}_{n{\bf k}})_{\rm a}},
\end{equation}
here $\ket{(\phi_{n{\bf k}})_{\rm a}}$ and $\ket*{(\tilde{\phi}_{n{\bf
      k}})_{\rm a}}$ represent the expansion of the smooth state,
$\ket*{\tilde{\Psi}_{n{\bf k}}}$, over, respectively, the all-electron and the
pseudo valence atomic functions localized around the atom ${\rm a}$
\begin{align}
  \ket{(\phi_{n{\bf k}})_{\rm a}} &=
  \sum_{i}\ket{\phi_{{\rm a}i}}\ip*{\tilde{p}_{{\rm a}i}}{\tilde{\Psi}_{n{\bf
        k}}}, \\
  \ket*{(\tilde{\phi}_{n{\bf k}})_{\rm a}} &=
  \sum_{i}\ket*{\tilde{\phi}_{{\rm a}i}}\ip*{\tilde{p}_{{\rm a}i}}{\tilde{\Psi}_{n{\bf
        k}}}.
\end{align}
The sum is performed over the complete set of quantum numbers $i$. $\ket*{\tilde{p}_{{\rm a}i}}$ are some fixed set of functions termed smooth projector functions and they satisfy the following duality condition
\begin{equation}
  \sum_{i}\ket*{\tilde{\phi}_{{\rm a}i}}\bra*{\tilde{p}_{{\rm a}i}}
  = \hat{1},
\end{equation}
inside each augmentation sphere implying also that
\begin{equation}
\ip*{\tilde{p}_{{\rm a}i}}{\tilde{\phi}_{{\rm a}j}} = \delta_{i,j},\,\,\,\mathrm{for }\,|{\bf r}-{\bf R}_{\rm a}|< r_c^{\rm a}.
\end{equation}
\subsection{The screened electron-ion interaction $V_{\rm KS}^{\rm a}$ in the core region} \label{H1}
Here we want to explain how to compute the screened electron-ion potential, $V^{\rm a}_{\rm KS}$, appearing in Eq.~(\ref{fmnPAW}) and entering the expression (\ref{Eq:14}) for the PAW force matrix elements. We should notice that although the definition of $V_{\rm KS}^{\rm a}$ introduced in Sec.~\ref{subsub:PAW} looks identical to the expression of $v_{\rm KS}^{\rm a}$ from the previous section the two quantities should not be confused. While $v_{\rm KS}^{\rm a}$ is the all-electron KS potential of the isolated atom and $\rho_e^{\rm a}$ represents here the electron density of the isolated atom, $V_{\rm KS}^{\rm a}$ is computed by using the electron density of the full many atoms system in the vicinity of atom ${\rm a}$
\begin{equation} \label{Eq:VKSa}
V_{\rm KS}^{\rm a}({\bf r}) = -\frac{Z e^2}{|{\bf r}-{\bf R}_{\rm a}|} + v_{Hxc}^{\rm a}[\rho_e]({\bf r}).
\end{equation}
The first term on the right hand side of Eq.~(\ref{Eq:VKSa}) is the all-electron Coulomb potential of atom ${\rm a}$, the second one is the Hartree plus exchange-correlation potential that depends on the all-electron density around atom ${\rm a}$, i.e. $\rho_e=\rho_v+\rho_c^{\rm a}$, where $\rho_v$ is the valence density of the system and $\rho_c^{\rm a}$ is the core electron density of atom $a$. In order to compute this second term we need to make a couple of assumptions 1) the electron density around each atom may be considered approximately spherical; 2) the atom becomes neutral at a certain distance $r_n^{\rm a}$ smaller than the average inter-atomic distance. At a distance $r$ from the atomic center ${\bf R}_{\rm a}$ less or equal to the core radius $r_c^{\rm a}$, the potential becomes
\begin{align}
v_{Hxc}^{\rm a}&[\rho_e](r)\simeq v_{xc}[\rho_v+\rho_c^{\rm a}](r) +\nonumber\\
&+ 4\pi e^2\bigg[\int_0^{r_c^{\rm a}}dr'\,r'^{2}\frac{\rho_c^{\rm a}(r')}{|r-r'|}+\int_0^{r_n^{\rm a}}dr'\,r'^{2}\frac{\rho_v(r')}{|r-r'|}\bigg].
\end{align}
%
\subsection{Calculation of the PAW force matrix elements Eq.~(\ref{Eq:14})} \label{H2}
In this appendix we prove the expression (\ref{Eq:14}) for the PAW force matrix elements, $f_{n{\bf k},m{\bf k'}}^\alpha$. 
We start by noticing that the projectors $\ket*{\tilde{p}_{{\rm a}i}}$ are non zero only inside the augmentation sphere of the
atom, while at distances from the atomic center greater than the
core radius $r_c^{\rm a}$ pseudo and all-electron atomic functions coincide, leading to
\begin{align} \label{C12}
  \ip{{\bf r}}{(\phi_{n{\bf k}})_{\rm a}} = \ip*{{\bf
      r}}{(\tilde{\phi}_{n{\bf k}})_{\rm a}},\qquad & |{\bf r}-{\bf R}_{\rm a}| >
  r_c^{\rm a} \nonumber \\
  & {\rm a} = 1, \ldots, N_i
\end{align}
while inside each sphere we easily obtain
\begin{align} \label{C13}
  \Psi_{n{\bf k}}({\bf r}) = \ip{{\bf r}}{(\phi_{n{\bf k}})_{\rm a}},\qquad &
  |{\bf r}-{\bf R}_{\rm a}|\le r_c^{\rm a}\nonumber \\
  & {\rm a} = 1, \ldots, N_i.
\end{align}
The previous relations are always satisfied given that we have a
complete basis of atomic functions allowing for an accurate
representation of $\Psi_{n{\bf k}}({\bf r})$ inside the core regions. In terms of these quantities the all-electron wave functions may be rewritten as follows
\begin{align*}
  \ket{\Psi_{n{\bf k}}} =& \ket*{\tilde{\Psi}_{n{\bf k}}} +
  \sum_{\rm a=1}^{N_i}\big[\ket*{(\phi_{n{\bf k}})_{\rm a}} - \ket*{(\tilde{\phi}_{n{\bf
          k}})_{\rm a}}\big],\\
  =& \ket*{\tilde{\Psi}_{n{\bf k}}} +
  \sum_{\rm a=1}^{N_i}\ket*{(\Delta\phi_{n{\bf k}})_{\rm a}},
\end{align*}
by using the previous definitions the force matrix elements become (where we have omitted the ${\rm L(R)}$ labels)
\begin{align*}
  &f^\alpha_{n{\bf k},m{\bf k'}} = \mel*{\tilde{\Psi}_{n{\bf
        k}}}{\hat{f}_\alpha}{\tilde{\Psi}_{m{\bf k'}}} +
  \sum_{\rm a=1}^{N_i}\mel*{(\Delta\phi_{n{\bf
        k}})_{\rm a}}{\hat{f}_\alpha}{\tilde{\Psi}_{m{\bf k'}}} +\\
  &+ \sum_{\rm a=1}^{N_i}\mel*{\tilde{\Psi}_{n{\bf
        k}}}{\hat{f}_\alpha}{(\Delta\phi_{m{\bf k'}})_{\rm a}} +
  \sum_{\rm a=1}^{N_i}\mel*{(\Delta\phi_{n{\bf
        k}})_{\rm a}}{\hat{f}_\alpha}{(\Delta\phi_{m{\bf k'}})_{\rm a}} +\\
  &+ \sum_{\rm a\neq b}\mel{(\Delta\phi_{n{\bf
        k}})_{\rm a}}{\hat{f}_\alpha}{(\Delta\phi_{m{\bf k'}})_{\rm b}}
\end{align*}
where the last term is exactly zero since the spheres do not overlap
and $\ket*{(\Delta\phi_{n{\bf k}})_{\rm a}}=0$ outside each atomic sphere,
after some rearrangements we arrive to
\begin{align*}
  f_{n{\bf k},m{\bf k'}}^\alpha &= \mel*{\tilde{\Psi}_{n{\bf
        k}}}{\hat{f}_\alpha}{\tilde{\Psi}_{m{\bf k'}}} +
  \sum_{\rm a=1}^{N_i}\mel*{(\Delta\phi_{n{\bf
        k}})_{\rm a}}{\hat{f}_\alpha}{(\phi_{m{\bf k'}})_{\rm a}} +\\
  &+ \sum_{\rm a=1}^{N_i}\mel*{(\tilde{\phi}_{n{\bf
        k}})_{\rm a}}{\hat{f}_\alpha}{(\Delta\phi_{m{\bf k'}})_{\rm a}}\\
  &= \mel*{\tilde{\Psi}_{n{\bf
        k}}}{\hat{f}_\alpha}{\tilde{\Psi}_{m{\bf k'}}} +
  \sum_{\rm a=1}^{N_i}\Delta f_{n{\bf k},m{\bf k'}}^{\rm a,\alpha}
\end{align*}
where
\begin{equation}
  \Delta f_{n{\bf k},m{\bf k'}}^{\rm a,\alpha} = \mel*{(\phi_{n{\bf
        k}})_{\rm a}}{\hat{f}_\alpha}{(\phi_{m{\bf k'}})_{\rm a}} -
  \mel*{(\tilde{\phi}_{n{\bf
        k}})_{\rm a}}{\hat{f}_\alpha}{(\tilde{\phi}_{m{\bf k'}})_{\rm a}},
\end{equation}
in addition, by following an analogous procedure to that outlined in
(\ref{F}) we obtain $f_{n{\bf k},m{\bf k'}}^\alpha=\delta_{{\bf
    k},{\bf k'}}f_{nm}^\alpha({\bf k})$. Finally, from the properties of the all-electron wave functions (\ref{C12}) and (\ref{C13}) it is easy to show that the previous expression is equivalent to Eq.~(\ref{Eq:14}).

\begin{widetext}
\section{Evaluation of the atomic sphere's contribution to the force matrix elements in Eq.~(\ref{Eq:14})}\label{I}
In this appendix we show how to compute the atomic sphere contribution to the force matrix elements in Eq.~(\ref{Eq:14})
\begin{equation}
{\bf f}_{ij}^{\rm a} = \int_\Omega d{\bf r}\,\Pi_{\rm a}^{\rm in}({\bf r})\phi_{{\rm a}i}^*({\bf r})\nabla_{\bf r}V_{\rm KS}^{\rm a}({\bf r})\phi_{{\rm a}j}({\bf r}),
\end{equation}
by integrating explicitly over the spatial projector $\Pi_{\rm a}^{\rm in}({\bf r})$ we obtain 
\begin{equation}
  {\bf f}_{ij}^{\rm a} =
  \int_{\mathcal{S}_{\rm a}}d{\bf r}\,\phi_{nlm}^{\rm a}({\bf
    r})^*\nabla_{\bf r}V_{\rm KS}^{\rm a}({\bf r})\phi_{n'l'm'}^{\rm a}({\bf r})
\end{equation}
where now the integral is computed over the
atomic sphere of radius $r_c^{\rm a}$ centered on atom ${\rm a}$. Since both the wave functions $\phi_{nlm}^{\rm a}({\bf r})$ and the potential $V_{\rm KS}^{\rm a}$, as explained in \ref{appendix_definition_PAW} and \ref{H1}, are centered around atom ${\rm a}$ we can make a change of variables ${\bf r}\rightarrow {\bf r}'={\bf r}+{\bf R}_{\rm a}$ and switch to spherical coordinates.
By using for the potential gradient
\begin{equation}
  \nabla_{\bf r'}V_{\rm KS}^{\rm a}(|{\bf r}'-{\bf R}_{\rm a}|) = \frac{{\bf r}'-{\bf R}_{\rm a}}{|{\bf r}'-{\bf
      R}_{\rm a}|}\cdot\frac{d}{dr}V_{\rm KS}^{\rm a}(r)
\end{equation}
where $r=|{\bf r}'-{\bf R}_{\rm a}|$ the integral becomes
\begin{align*}
   {\bf f}_{ij}^{\rm a} &= \int_{\mathcal{S}_{\rm a}}d{\bf r}'\,\phi^{\rm a}_{nlm}({\bf
    r}'-{\bf R}_{\rm a})^*\frac{{\bf r}'- {\bf R}_{\rm a}}{|{\bf r}'-
      {\bf R}_{\rm a}|}\frac{dV_{\rm KS}^{\rm a}(r)}{dr}\phi_{n'l'm'}^{\rm a}({\bf r}'-{\bf R}_{\rm a})\\
  &= \int_{\mathcal{S}_{\rm a}}d{\bf r}\,R^{\rm a}_{nl}(
    r)^* Y_{lm}(\hat{\bf n})^*\frac{\bf r}{r}\frac{dV_{\rm KS}^{\rm a}(r)}{dr}R_{n'l'}^{\rm a}(r)Y_{l'm'}(\hat{\bf n})
\end{align*}
where we have rewritten the atomic functions in terms of the spherical harmonics $\phi_{nlm}^{\rm a}({\bf r}'-{\bf R}_{\rm a})=R_{nl}^{\rm a}(r)Y_{lm}(\hat{\bf n})$. 
The previous integral can be very naturally computed in spherical coordinates only one time for every atom ${\rm a}$, direction and atomic transition. We first consider the direction $x$
\begin{align}
\hat{\bf e}_x\cdot{\bf f}_{ij}^{\rm a} &= \int_0^{r_c^{\rm a}}dr\,r^2\int_{-1}^1 d\cos\theta\int_0^{2\pi}d\varphi R_{nl}^{\rm a}(r)^*Y_{lm}(\theta,\varphi)^* \sin\theta \cos\varphi\frac{d V_{\rm KS}^{\rm a}(r)}{dr}R_{n'l'}^{\rm a}(r) Y_{l'm'}(\theta,\varphi)\nonumber\\
&= C_{lm}C_{l'm'}\int_0^{r_c^{\rm a}}dr\,r^2 R_{nl}^{\rm a}(r)^* \frac{d V_{\rm KS}^{\rm a}(r)}{dr}R_{n'l'}^{\rm a}(r)\int_{-1}^1d\cos\theta\, \sin\theta P_{lm}(\cos\theta) P_{l'm'}(\cos\theta)\int_0^{2\pi}d\varphi\,\cos\varphi e^{-i(m-m')\varphi}
\end{align}
where in the last step we have rewritten the spherical harmonics in terms of Legendre polynomials $Y_{lm}(\theta,\varphi)=C_{lm}P_{lm}(\cos\theta)e^{im\varphi}$. The integral over $\varphi$ can be exactly computed reducing the expression to a simpler double integral
\begin{equation}
\hat{\bf e}_x\cdot{\bf f}_{ij}^{\rm a} = \pi C_{lm} C_{l'm'} (\delta_{m,m'+1} + \delta_{m,m'-1})\int_0^{r_c^{\rm a}}dr\,r^2 R_{nl}^{\rm a}(r)^* \frac{d V_{\rm KS}^{\rm a}(r)}{dr}R_{n'l'}^{\rm a}(r)\int_{-1}^1 dx\,\sqrt{1-x^2}P_{lm}(x)P_{l'm'}(x).
\end{equation}
Similarly along the $y$ direction we find
\begin{align}
\hat{\bf e}_y\cdot{\bf f}_{ij}^{\rm a} &= \int_0^{r_c^{\rm a}}dr\,r^2\int_{-1}^1 d\cos\theta\int_0^{2\pi}d\varphi R_{nl}^{\rm a}(r)^*Y_{lm}(\theta,\varphi)^* \sin\theta\sin\varphi\frac{d V_{\rm KS}^{\rm a}(r)}{dr}R_{n'l'}^{\rm a}(r) Y_{l'm'}(\theta,\varphi)\nonumber\\
&= C_{lm}C_{l'm'}\int_0^{r_c^{\rm a}}dr\,r^2 R_{nl}^{\rm a}(r)^* \frac{d V_{\rm KS}^{\rm a}(r)}{dr}R_{n'l'}^{\rm a}(r)\int_{-1}^1d\cos\theta\, \sin\theta P_{lm}(\cos\theta) P_{l'm'}(\cos\theta)\int_0^{2\pi}d\varphi\,\sin\varphi e^{-i(m-m')\varphi}
\end{align}
leading to the final result
\begin{equation}
\hat{\bf e}_y\cdot{\bf f}_{ij}^{\rm a} = \frac{\pi}{i} C_{lm} C_{l'm'} (\delta_{m,m'+1} - \delta_{m,m'-1})\int_0^{r_c^{\rm a}}dr\,r^2 R_{nl}^{\rm a}(r)^* \frac{d V_{\rm KS}^{\rm a}(r)}{dr}R_{n'l'}^{\rm a}(r)\int_{-1}^1 dx\,\sqrt{1-x^2}P_{lm}(x)P_{l'm'}(x).
\end{equation}
Along the $z$ direction we have instead
\begin{align}
\hat{\bf e}_z\cdot{\bf f}_{ij}^{\rm a} &= \int_0^{r_c^{\rm a}}dr\,r^2\int_{-1}^1 d\cos\theta\int_0^{2\pi}d\varphi R_{nl}^{\rm a}(r)^*Y_{lm}(\theta,\varphi)^* \cos\theta\frac{d V_{\rm KS}^{\rm a}(r)}{dr}R_{n'l'}^{\rm a}(r) Y_{l'm'}(\theta,\varphi)\nonumber\\
&= C_{lm}C_{l'm'}\int_0^{r_c^{\rm a}}dr\,r^2 R_{nl}^{\rm a}(r)^* \frac{d V_{\rm KS}^{\rm a}(r)}{dr}R_{n'l'}^{\rm a}(r)\int_{-1}^1d\cos\theta\,\cos\theta P_{lm}(\cos\theta) P_{l'm'}(\cos\theta)\int_0^{2\pi}d\varphi\,e^{-i(m-m')\varphi}
\end{align}
that by integrating over $\varphi$ gives
\begin{equation}
\hat{\bf e}_z\cdot{\bf f}_{ij}^{\rm a} = 2\pi C_{lm} C_{l'm'}\delta_{m,m'}\int_0^{r_c^{\rm a}}dr\,r^2 R_{nl}^{\rm a}(r)^* \frac{d V_{\rm KS}^{\rm a}(r)}{dr}R_{n'l'}^{\rm a}(r)\int_{-1}^1 dx\,x P_{lm}(x)P_{l'm'}(x).
\end{equation}
In conclusion the entire calculation reduces to the evaluation of only three different types of one dimensional integrals, namely $\int_0^{r_c^{\rm a}}dr\,r^2 R_{nl}^{\rm a}(r)^*\,dV_{\rm KS}^{\rm a}(r)/dr\,R_{n'l'}^{\rm a}(r)$, that needs to be computed numerically from the knowledge of the radial all-electron wave functions $R_{nl}^{\rm a}(r)$ and of the gradient of the KS potential $V_{\rm KS}^{\rm a}(r)$, and $\int_{-1}^1 dx\,\sqrt{1-x^2}P_{lm}(x)P_{l'm'}(x)$ together with $\int_{-1}^1 dx\,x P_{lm}(x)P_{l'm'}(x)$ that can be instead computed analytically. 

\end{widetext}

\end{appendix}

%
\end{document}